\documentclass[twocolumn,twocolappendix]{emulateapj}

\usepackage{amssymb}
\usepackage{amsmath}
\usepackage{subfigure}
\usepackage{multirow}
\usepackage{graphicx}
\usepackage{times}
\usepackage{longtable}
\usepackage{array}
\usepackage[colorlinks,dvipdfm,linkcolor=blue,anchorcolor=,citecolor=blue]{hyperref}

\usepackage{color}

\def\note #1]{{\bf #1]}}

\slugcomment{Submitted to AAS journal November 2, 2018; accepted July 3, 2019}
\shorttitle{Solar models with overshoot, the solar wind, and PMS accretion}
\shortauthors{Zhang, Li \& Christensen-Dalsgaard}

\begin{document}

\title{Solar models with convective overshoot, the solar-wind mass loss, and a PMS disk accretion: helioseismic quantities, Li depletion and neutrino fluxes}
\author{Qian-Sheng Zhang \altaffilmark{1,2,3,4}, Yan Li \altaffilmark{1,2,3,4}, \& J{\o}rgen Christensen-Dalsgaard \altaffilmark{5,6}}
\email{zqs@ynao.ac.cn(QSZ)}

\altaffiltext{1}{Yunnan Observatories, Chinese Academy of Sciences, 396 Yangfangwang, Guandu District, Kunming 650216, China}
\altaffiltext{2}{Center for Astronomical Mega-Science, Chinese Academy of Sciences, 20A Datun Road, Chaoyang District, Beijing 100012, China}
\altaffiltext{3}{Key Laboratory for the Structure and Evolution of Celestial Objects, Chinese Academy of Sciences, 396 Yangfangwang, Guandu District, Kunming 650216, China}
\altaffiltext{4}{University of Chinese Academy of Sciences, Beijing 100049, China}
\altaffiltext{5}{Stellar Astrophysics Centre and Department of Physics and Astronomy, Aarhus University, DK-8000 Aarhus C, Denmark}
\altaffiltext{6}{Kavli Institute for Theoretical Physics, University of California Santa Barbara, CA 93106-4030, USA}

\begin{abstract}

Helioseismic observations have revealed many properties of the Sun: the depth and the helium abundance of the convection zone, the sound-speed and the density profiles in the solar interior. Those constraints have been used to judge the stellar evolution theory. With the old solar composition (e.g., GS98), the solar standard model is in reasonable agreement with the helioseismic constraints. However, a solar model with revised composition (e.g., AGSS09) with low abundance $Z$ of heavy elements cannot be consistent with those constraints. This is the so-called ``solar abundance problem", standing for more than ten years even with the recent upward revised Ne abundance. Many mechanisms have been proposed to mitigate the problem. However, there is still not a low-$Z$ solar model satisfying all helioseismic constraints. In this paper, we report a possible solution to the solar abundance problem. With some extra physical processes that are not included in the standard model, solar models can be significantly improved. Our new solar models with convective overshoot, the solar wind, and an early mass accretion show consistency with helioseismic constraints, the solar Li abundance, and observations of solar neutrino fluxes.

\end{abstract}


\section{Introduction} \label{SecIntro}

Observations have revealed many properties of the Sun with high accuracy. Element abundances can be determined from the absorption line analyses of the solar atmosphere. The information of the solar interior can be extracted from helioseismology. The properties of the solar core can be probed from observations of the solar neutrino fluxes. Therefore the Sun is the best target to benchmark the stellar evolutionary theory in detail.

\subsection{The solar abundance problem}

\begin{figure}
\centering
\includegraphics[scale=0.5]{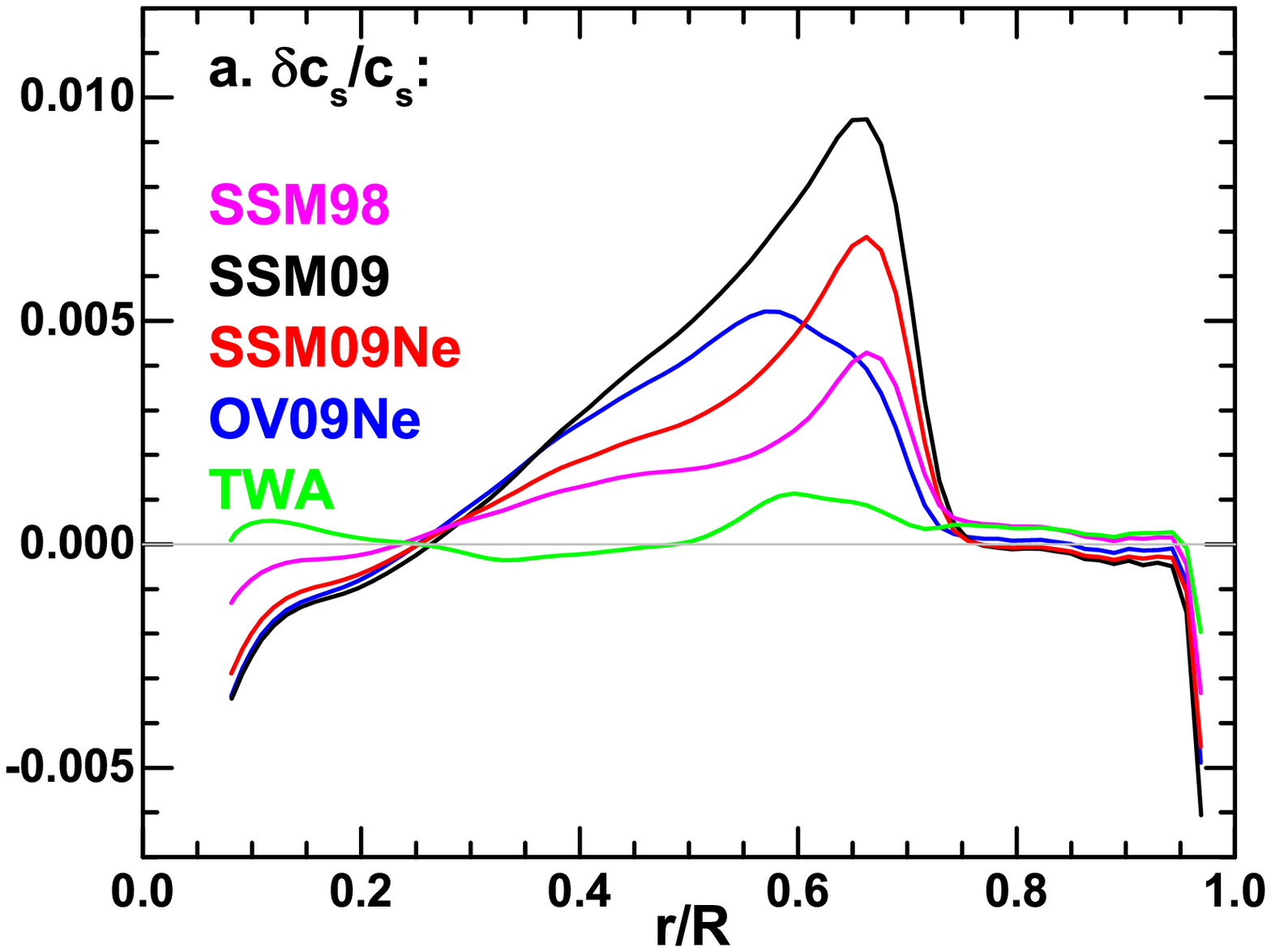}
\includegraphics[scale=0.5]{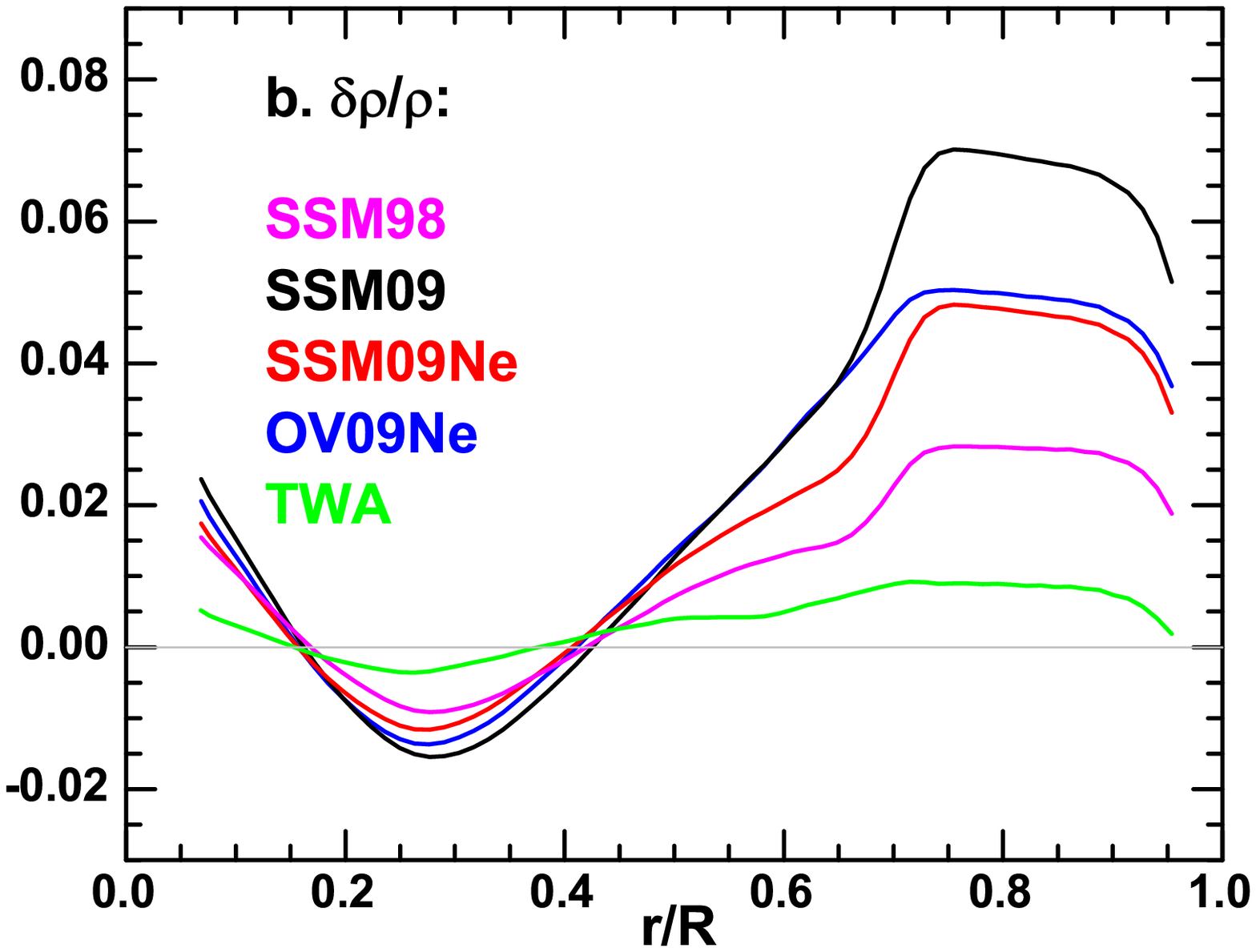}
\caption{
	Relative differences of sound speed and density in the solar interior between helioseismic inferences and solar models (see Table~\ref{tab:basmod}). $\delta c_{\rm{s}} / c_{\rm{s}} = (c_{{\rm{s}},\odot}-c_{\rm{s}})/c_{{\rm{s}},\odot}$ and $\delta \rho / \rho = (\rho_{\odot}-\rho)/\rho_{\odot}$ obtained from helioseismic inversion (see Section~\ref{Secresult} for details). The black line corresponds to the AGSS09 SSM. The red line corresponds to the AGSS09Ne (i.e., AGSS09 composition and revised Ne abundance) SSM. The purple line corresponds to the GS98 SSM. The blue line corresponds to the solar model with AGSS09Ne and a helioseismically based overshoot model. The green line corresponds to a typical improved solar model (see Section \ref{Secresult}).
} \label{std_c}
\end{figure}

The standard solar models (SSMs) based on old solar composition (e.g., with $Z/X = 0.0245$ for GN93 \citep{GN93} or 0.0229 for GS98 \citep{GS98}) are in reasonable agreement with the helioseismic inferences on the solar sound-speed profile, the location of the base of the convection zone $R_{\rm{bc}}$ and the helium abundance in the convection zone $Y_{\rm{S}}$ \citep[e.g.,][]{JCD93,JCD96,Basu1997,Bah01}. However, using three-dimensional hydrodynamic model atmospheres \citep{SN98,AN00,FSD02} and relaxing the assumption of local thermodynamic equilibrium in the spectral line formation \citep{AGS04} the resulting value of the solar metallicity has been significantly revised downwards; for example, the AGSS09 composition shows $Z/X = 0.0181$ \citep{AGSS09}. Compared with the SSMs with GN93 or GS98 compositions, the SSM with AGSS09 compositions shows significant deviations from helioseismic inferences \citep[e.g.,][]{Bah05,AGSS09,ser09}. For example, the properties of the SSM with AGSS09 composition (Model SSM09) are shown in Table~\ref{modelinfo}. Comparing with the GS98 SSM (Model SSM98), the base of the convection zone (BCZ) is shallower and $Y_{\rm{S}}$ is lower. The sound-speed and density deviations resulting from helioseismic inversions are shown in Fig.~\ref{std_c}. The deviations of sound speed of Model SSM98 are less than 0.2\% in most part of the solar interior and are significant (but also less than 0.5\%) only in the region $0.6<r/R_\odot<0.7$ below the BCZ. However, Model SSM09 shows overall significant deviations of sound speed in the solar interior. The density profiles also show that Model SSM09 is worse than Model SSM98. Those significant deviations on $R_{\rm{bc}}$, $Y_{\rm{S}}$ and the sound-speed and density profiles caused by the low-$Z$ composition in SSM are called the ``the solar abundance problem". More comprehensive comparisons between solar models with GS98 and AGSS09 compositions by using a Monte Carlo method taking into account the observations of solar neutrino fluxes and uncertainties of model physics confirmed the existence of the problem that the AGSS09 composition is excluded at a high confidence level when used in a standard solar model \citep[e.g.,][]{Villante14,Vinyoles17,Song18}.

\begin{table*}
\centering
\caption{ Basic information of solar models. }\label{modelinfo}
\begin{tabular}{lcccccc}
\hline\noalign{\smallskip}
                                      & SSM98       & SSM09       & SSM09Ne     & OV09Ne      & TWA         & the Sun \\
\hline\noalign{\smallskip}
$\alpha_{\rm{MLT}}$                   & 2.3396      & 2.3084      & 2.3413      & 2.3073      & 2.3708      & ------ \\
$X_{\rm{0}}$                          & 0.7089      & 0.7190      & 0.7164      & 0.7206      & 0.7096      & ------ \\
$Z_{\rm{0}}$                          & 0.01837     & 0.01478     & 0.01527     & 0.01466     & 0.01472     & ------ \\
$X_{\rm{C}}$                          & 0.3487      & 0.3592      & 0.3564      & 0.3609      & 0.3485      & ------ \\
$Z_{\rm{C}}$                          & 0.01962     & 0.01580     & 0.01630     & 0.01565     & 0.01572     & ------ \\
$\log T_{\rm{C}}$                     & 7.1936      & 7.1908      & 7.1915      & 7.1900      & 7.1925      & ------ \\
$\log  \rho_{\rm{C}}$                 & 2.1782      & 2.1734      & 2.1754      & 2.1743      & 2.1857      & ------ \\
$M_{\rm{acc}}/M_{\odot}$              & ------      & ------      & ------      & ------      & 0.0585      & ------ \\
$M_{\rm{L}}/M_{\odot}$                & ------      & ------      & ------      & ------      & 0.0028      & 0.001$-$0.03 $^{\rm{a}}$ \\
$Y_{\rm{S}}$                          & 0.2453      & 0.2381      & 0.2405      & 0.2458      & 0.2450      & 0.2485$\pm$0.0035 $^{\rm{b}}$ \\
$(Z/X)_{\rm{S}}$                      & 0.0229      & 0.0181      & 0.0188      & 0.0188      & 0.0188      & 0.0188$\pm$0.0012 $^{\rm{c}}$  \\
$ R_{\rm{bc}}/R_{\odot}$              & 0.7152      & 0.7239      & 0.7207      & 0.7155      & 0.7110      & 0.713$\pm$0.001 $^{\rm{d}}$  \\
A(Li)$^{\rm{e}}$                      & 2.44        & 2.73        &  2.60       & 0.84        & 0.82        & 1.05$\pm$0.10 $^{\rm{f}}$ \\
\hline
neutrino fluxes $^{\rm{g}}$ & & & & & & \\
in $({\rm{cm}}^{-2}{\rm{s}}^{-1})$ & & & & & & \\
\hline
$pp$ $(10^{10})$                      & 5.96(0.5\%) & 6.00(0.5\%) & 5.99(0.5\%) & 6.00(0.5\%) & 5.98(0.5\%) &5.97(0.5\%)  \\
$pep$ $(10^{8})$                      & 1.45(0.9\%) & 1.46(0.9\%) & 1.46(0.9\%) & 1.47(0.9\%) & 1.47(0.9\%) &1.45(0.9\%)  \\
$hep$ $(10^{3})$                      & 8.01(30\%)  & 8.19(30\%)  & 8.15(30\%)  & 8.23(30\%)  & 8.13(30\%)  &19(55\%)     \\
$^7$Be $(10^{9})$                     & 4.91(6\%)   & 4.63(6\%)   & 4.70(6\%)   & 4.57(6\%)   & 4.84(6\%)   &4.80(5\%)    \\
$^8$B $(10^{6})$                      & 5.35(12\%)  & 4.74(12\%)  & 4.89(12\%)  & 4.60(12\%)  & 5.13(12\%)  &5.16(2.2\%)  \\
$^{13}$N $(10^{8})$                   & 2.86(14\%)  & 2.18(14\%)  & 2.21(14\%)  & 2.04(14\%)  & 2.19(14\%)  &$\leq$13.7   \\
$^{15}$O $(10^{8})$                   & 2.14(16\%)  & 1.59(16\%)  & 1.62(16\%)  & 1.48(16\%)  & 1.63(16\%)  &$\leq$2.8    \\
$^{17}$F $(10^{6})$                   & 5.30(18\%)  & 3.47(18\%)  & 3.55(18\%)  & 3.22(18\%)  & 3.59(18\%)  &$\leq$85     \\
\hline
\end{tabular}
\label{tab:basmod}
\tablecomments{ Model SSM98 is the standard solar model with the GS98 \citep{GS98} composition, Model SSM09 is the standard solar model with the AGSS09 \citep{AGSS09} composition, Model SSM09Ne is the standard solar model with the AGSS09Ne composition (i.e., AGSS09 composition and the revised Ne abundance \citep{Young18}; see Table~\ref{tab:abund}), Model OV09Ne is the solar model with the AGSS09Ne and convective overshoot, and Model TWA is a typical improved solar model with convective overshoot, solar wind and pre-main-sequence (PMS) accretion (see Section~\ref{Secresult} for details). $\alpha_{\rm{MLT}}$ is the mixing-length parameter, $X_{\rm{0}}$ is the initial hydrogen abundance, $Z_{\rm{0}}$ is the initial metallicity, $X_{\rm{C}}$ is the center hydrogen abundance, $Z_{\rm{C}}$ is the center metallicity, $T_{\rm{C}}$ is the center temperature, $\rho_{\rm{C}}$ is the center density, $M_{\rm{acc}}$ is the accreted mass, $M_{\rm{L}}$ is the mass loss.
Notes:
a. \citet{windmass}. b. \citet{ba04}. c. see Table~\ref{AGSS09Ne}. d. \citet{JCD91} and \citet{ba97}. e. Lithium abundance index is defined by ${\rm{A(Li)}}=\log (n_{\rm{Li}}/n_{\rm{H}})+12$ where $n_{\rm{Li}}$ and $n_{\rm{X}}$ are number densities for lithium and hydrogen. f. \citet{AGSS09}. g. Observations of solar neutrino fluxes and their uncertainties are from \citet{Bergstrom16} and the uncertainties of neutrino fluxes for solar models are from \citet{Vinyoles17}. }
\end{table*}

\subsection{Some attempts to adjust solar models}

\citet{ser09} pointed out that an opacity enhancement of $12\%-15\%$ at the BCZ to $2\%-5\%$ in the solar core can improve the sound-speed profile of the AGSS09 SSM to the level of GS98 SSM. Similarly, an opacity enhancement of about 20\% from the BCZ to about 2\% in the solar core is required to improve solar model to the level of Model S \citep{JCD10}. Comparing with the widely used OPAL or OP \citep{OPtables} opacity tables, a recently revised opacity table OPAS \citep{OPAS} shows an opacity enhancement of about 6\% at BCZ. Although solar models with AGSS09 composition and OPAS opacity table show some improvements in the sound-speed profile and $R_{\rm{bc}}$, the discrepancies cannot be efficiently removed \citep{OPASSSM}. A main result of the OPAS opacity table is that the individual contribution of iron to opacity is 40\% higher than that in OP tables. This tendency is consistent with the measurements of iron opacity at physical conditions similar to the base of the solar convection zone \citep{Bai15} which yield values $30\%-400\%$ higher than calculations. It is urgent to extend the effects of the revision on opacity to a larger range of temperature and density and more kinds of elements, as well as to understand the physical origin of the differences \citep[e.g.,][]{Pradha2018}. Even so, it would be a remarkable coincidence if the errors in the opacity calculations exactly compensate the variation of solar composition.

An enhancement in the solar neon abundance could enlarge the opacity and improve solar models. It was found that the required enhancement of neon abundance is about $200\%-300\%$ \citep{ab05,dp06,Zaatri07}. However, a higher neon abundance enlarges the discrepancy in the adiabatic exponent in the region $(0.75-0.9)R_{\odot}$ \citep{lin07}. Investigation of low-activity stars has shown a correlation between their Ne/O ratio and stellar activity and suggested that a significantly enhanced solar Ne/O ratio seems unlikely \citep{RSF08}. Recently, \citet{Young18} inferred the Ne abundance in the solar photosphere by using the Mg/Ne and Ne/O ratios in the transition region of the quiet Sun and found a Ne abundance enhancement of $\sim 40\%$, which is significantly less than the level of enhancement required to restore the solar model.

Molecular diffusion leads to heavy-element settling which increases the opacity below BCZ. \citet{AGS04} suggested that the sound speed of solar models with low-$Z$ could be improved by an enhancement of molecular diffusion. The effects of enhanced diffusion in solar models were explored \citep{ba04,Montalban04,Guzik05,yb07,yang16,yang19}. It is found that $R_{\rm{bc}}$ and the sound-speed profile can be significantly improved when the diffusion is enhanced by a factor $1.5-2.5$. However, there is no justification for such a large enhancement. And, even though $R_{\rm{bc}}$ and the sound-speed profile are improved, the helium abundance in the convection zone $Y_{\rm{S}}$ and the $^7$Be and $^8$B neutrino fluxes become worse.

\citet{Guzik10} investigated low-$Z$ solar models with mass loss. They found that sound speed, $Y_{\rm{S}}$ and $R_{\rm{bc}}$ can be improved simultaneously and the sound-speed profile can be almost restored for a $0.3M_{\odot}$ mass loss solar model. However, $Y_{\rm{S}}$ and $R_{\rm{bc}}$ are still not in their acceptable ranges for that model. Also, such extensive mass loss strongly blows away the stellar envelope, exposing on the surface regions of the interior where lithium has been destroyed.

\citet{Guzik05} noted that the solar envelope may be diluted by a PMS low-$Z$ accretion. In this case the bulk metallicity of the Sun is higher than that of the envelope. A possible justification for the low-$Z$ accretion is that planet formation locks high-$Z$ elements in planets \citep{Melendez09,ser11}. \citet{Castro07} and \citet{Guzik10} calculated solar models with low-$Z$ accretion near the zero-age main sequence (ZAMS). They found that the sound speed in the high-$Z$ interior and $Y_{\rm{S}}$ is improved but the sound-speed deviations near BCZ remain and $R_{\rm{bc}}$ is not significantly improved. A more detailed analysis of solar models with PMS accretion was carried out by \citet{ser11}, who calculated solar models with different masses, metallicities, starting time and durations of accretion in the PMS stage. Surprisingly, they found that some models with high-$Z$ accretion show good agreements on sound-speed profiles and $R_{\rm{bc}}$. However, those models have worse $Y_{\rm{S}}$ and their neutrino fluxes of $^7$Be and $^8$B are too low due to the low-$Z$ cores.

\citet{Montalban06} and \citet{Guzik10} investigated the effect of convective overshoot below BCZ on low-$Z$ solar models. They did not find significantly improvements on the sound-speed profile below BCZ. \citet{bi11} and \citet{yang16} calculated the low-$Z$ solar models with rotational mixing and magnetic field and found that $Y_{\rm{S}}$ can be improved.

Recently, \citet{vSZ16} published a new solar metallicity $Z_{\odot}=0.0196\pm0.0014$ derived from in situ measurement of the solar-wind composition (vSZ16), which is much higher than the AGSS09 composition. \citet{ser16} and \citet{vfz17} calculated solar models with vSZ16 metallicity. They found that, although the solar models have correct $R_{\rm{bc}}$, they have excessively modified sound-speed profile, very high $Y_{\rm{S}}$, and neutrino fluxes of $^7$Be and $^8$B. \citet{ser16} argued that the vSZ16 metallicity based on the solar-wind measurement cannot be trusted as representative of the photosphere or the bulk sun because of the FIP effect.

\subsection{The contradiction of the structure of the solar convective envelope}

\citet{TKFZ14} investigated the solar convective envelope models with AGSS09 composition. Because the gravitational energy release can be ignored for the Sun and the abundances are determined by given $Y_{\rm{S}}$ and $(Z/X)_{\rm{S}}$, the structure of the solar convective envelope can be directly determined by integrating the stellar structure equations from the solar surface with given radius and luminosity downward to BCZ without calculations of the solar evolutionary models. For the standard model of the solar convection envelope with the old GN93 composition, the density profile, $Y_{\rm{S}}$ and $R_{\rm{bc}}$ are all in good agreement with helioseismic inferences. However, for the corresponding model with the AGSS09 composition, the density profile, $Y_{\rm{S}}$ and $R_{\rm{bc}}$ cannot be consistent with helioseismic inferences simultaneously. This is an inherent contradiction of the standard model of the solar convective envelope with AGSS09 composition, which could be a part of the reason causing the ``solar abundance problem". The profile of density $\rho$ determines the profile of pressure $P$ in the solar interior by integrating the hydrostatic equation and hence largely determines the sound-speed $c_{\rm s}$, with
\begin{equation} \label{soundspeed}
c_{\rm s}^2 = {\Gamma_1 P \over \rho} \; ,
\end{equation}
given that $\Gamma_1$ is nearly constant in most of the solar interior; here $\Gamma _1= (\partial \ln P / \partial \ln \rho)_S$ is the adiabatic index, the derivative being at constant specific entropy $S$. This contradiction could explain that the sound-speed profile, $Y_{\rm{S}}$ and $R_{\rm{bc}}$ cannot be improved simultaneously for solar models with extra physics which does not affect the input physics of the convective envelope, e.g., enhanced molecular diffusion, mass loss, accretion, and rotational mixing.

A successful AGSS09 solar model must eliminate this contradiction. In order to do that, beside opacity enhancement, a probable mechanism has to be taken into account, i.e., the turbulent kinetic energy flux $F_{\rm K}$ below BCZ caused by convective overshoot. The turbulent kinetic energy flux below BCZ is negative; therefore the equilibrium of the total flux requires a higher temperature gradient. The effect is similar to an opacity enhancement. Even if the contradiction is not completely caused by missing the turbulent kinetic energy flux, it would reduce the opacity enhancement required to bring the solar models in accordance with the observations.

For the standard model of the solar convection envelope with vSZ16 metallicity, in order to obtain the required density profile, $Y_{\rm{S}}$ and $R_{\rm{bc}}$, we require a positive $F_{\rm{K}}$ which is physically unacceptable, or, a reduction of opacity near the BCZ which is excluded by the opacity measurement \citep[e.g.,][]{Bai15} and calculations \citep[e.g.,][]{OPAS}.

\subsection{About this paper}

In this paper, we focus on the ``solar abundance problem" and attempt to find a possible solution. We investigate solar models with convective overshoot, solar wind and a PMS accretion. The details of those physical process are introduced and discussed in Section \ref{Secphysics}. The resulting solar models are described in Section \ref{Secresult}. A discussion of the results is presented in Section \ref{Secdisc}, while Section \ref{Secconc} provides a summary and conclusions.

\section{input physics in solar models} \label{Secphysics}

Solar models are calculated using the YNEV code \citep{YNEVZ15} with a revision to calculate accretion/mass-loss with specific composition for the accreted/lost mass. The adopted element abundances (except Ne) are based on the AGSS09 \citep{AGSS09} solar photosphere composition. The latest solar photosphere abundances of elements heavier than Ne are available \citep{Grevesse2015,Scott2015a,Scott2015b} but the revisions are generally slight. Neon abundance in the AGSS09 \citep{AGSS09} solar photosphere composition is not directly measured and is based on the Ne/O ratio of the quite Sun measured by \citep{Young05}. Recently, the Ne/O ratio of the quite Sun has been upward revised about $\sim 40\%$ due to the updated atomic data \citep{Young18}. Therefore the AGSS09 metal composition with the revised Ne abundance was assumed as the solar composition. This composition is denoted as AGSS09Ne in this paper. The abundances of the main heavy elements and the resulting ratio of metallicity to hydrogen at the solar surface are listed in Table~\ref{AGSS09Ne}. The equation of state is interpolated from the OPAL equation of state tables \citep{OPALEOS}. The opacities are interpolated from the OPAL tables \citep{OPAL} and low-temperature opacity tables \citep{OPLT}. The OPAL tables are used in the high temperature region with ${\rm{lg}}T\geq4.5$ and the low-temperature opacity tables are used in the low temperature region with ${\rm{lg}}T\leq4.3$. In the region with $4.3<{\rm{lg}}T<4.5$, the opacity is smoothly interpolated from the two tables by using the following formula:
\begin{eqnarray}
{\rm{log}} \kappa = f_{\kappa} {\rm{log}} \kappa_{1} + (1-f_{\kappa}) {\rm{log}} \kappa_{2}, \\ \nonumber
f_{\kappa} = \frac{1}{2}\left\{ 1 + \sin \left[\left(\frac{{{\rm{lg}}T - 4.3}}{{0.2}} - \frac{1}{2}\right)\pi \right]\right\},
\end{eqnarray}
where $\kappa_{1}$ is the opacity given by the OPAL tables and $\kappa_{2}$ is the given by the low-temperature tables. In this case, $\kappa$ in the super-adiabatic convective envelope of solar models are from the low-temperature opacity tables. It should be pointed out here that the interpolation scheme between high- and low-temperature opacity tables is not unique and it may be different in different stellar evolutionary codes. Because the integral of the temperature gradient in the solar super-adiabatic convective envelope is restricted by the calibration of the model radius and the temperature gradient profile in that region is determined by the opacity profile and the value of $\alpha_{\rm{MLT}}$, the interpolation of the opacity from two tables could affect the value of the mixing-length parameter $\alpha_{\rm{MLT}}$ for solar models. Therefore different interpolation schemes between high- and low-temperature opacity tables in different codes should lead to a variation of $\alpha_{\rm{MLT}}$. Nuclear reaction rates are adopted from SFII \citep{SFII}, enhanced by using a weak screening \citep{SCR}. Molecular diffusion is taking into account by solving Burgers equation with resistance coefficients in the screening case \citep{DIFSCRZ17}. The convective heat flux is calculated by using the standard mixing-length theory. The K-S relation between temperature and optical depth \citep{KSATM} in the solar atmosphere is adopted.

\begin{table}
\centering
\caption{ AGSS09Ne - the adopted solar composition. }\label{AGSS09Ne}
\begin{tabular}{lc}
\hline\noalign{\smallskip}
element & abundance index \\
\hline\noalign{\smallskip}
 C  & 8.43$\pm$0.05 \\
 N  & 7.83$\pm$0.05 \\
 O  & 8.69$\pm$0.05 \\
 Ne & 8.08$\pm$0.09 \\
 Na & 6.24$\pm$0.04 \\
 Mg & 7.60$\pm$0.04 \\
 Al & 6.45$\pm$0.03 \\
 Si & 7.51$\pm$0.03 \\
 S  & 7.12$\pm$0.03 \\
 Ar & 6.40$\pm$0.13 \\
 Ca & 6.34$\pm$0.04 \\
 Cr & 5.64$\pm$0.04 \\
 Mn & 5.43$\pm$0.04 \\
 Fe & 7.50$\pm$0.04 \\
 Ni & 6.22$\pm$0.04 \\
\hline
 $(Z/X)_{\rm{s}}$ & 0.0188$\pm$0.0012 \\
\hline
\end{tabular}
\label{tab:abund}
\tablecomments{ Abundance index is defined as ${\rm{A}}({\rm{W}})={\rm{log_{10}}}(n_{\rm{W}}/n_{\rm{H}})+12$ for element $\rm{W}$. Values of abundance indices for all elements (except for Ne) come from \citet{AGSS09}, abundance index of Ne comes from \citet{Young18}. The resulting $(Z/X)_{\rm{s}}$ is derived by using a Monte Carlo simulation with 1,000,000 samples with element abundances based on the suggested values and standard deviations. }
\end{table}

The initial hydrogen abundance $X_{\rm{0}}$, initial metallicity $Z_{\rm{0}}$, the mixing-length parameter $\alpha_{\rm{MLT}}$ of each solar model are iteratively adjusted to ensure that the solar model at the present age $\tau_{\odot}=4.57\,\rm{Gyr}$ has the correct luminosity $L_{\odot}=3.8418\times10^{33}\rm{erg}/\rm{s}$ \citep{Bah05}, radius  $R_{\odot}=6.9598\times10^{10}\rm{cm}$ and the ratio of metallicity to hydrogen at the surface $(Z/X)_{\rm{S}}=0.0188$. The solar models are evolved from a core temperature $T_{\rm{C}}=10^5\,{\rm{K}}$ in the PMS stage to the present solar age.

\subsection{The problem remains: the standard solar model with AGSS09Ne composition} \label{SecSSM}

The information of SSMs with three different compositions (Model SSM09Ne for AGSS09Ne, Model SSM09 for the original AGSS09, and Model SSM98 for  GS98) is listed in Table~\ref{modelinfo}. The sound-speed and density deviation derived from helioseismic inversions are shown in Fig.~\ref{std_c}. It is found that, although the revised Ne abundance leads to overall improvements on solar model, the ``solar abundance problem" remains. The depth and helium abundance of the convection zone and the sound-speed and density profiles are still not in agreement with helioseismic inferences. This is not surprise since the required Ne enhancement for solving the ``solar abundance problem" is 200\%-300\% \citep{ab05,dp06,Zaatri07} but the actual enhancement of the revised Ne abundance is only about 40\%.

In order to try to alleviate the ``solar abundance problem", we take into account some extra physical processes missed in the SSM: the convective overshoot (leads to mixing and turbulent kinetic energy flux) below the BCZ, the inhomogeneous mass-loss caused by the solar wind, and PMS accretion with inhomogeneous materials. The motivation for including those processes are as follows. Although it has been shown that the turbulent kinetic energy flux could be a possible mechanism to eliminate the contradiction of the structure of the solar convective envelope \citep{TKFZ14}, its effects on the whole solar interior model have not been investigated. Since that contradiction could be a part of the reason causing the ``solar abundance problem'', it is necessary to investigate the effects of turbulent kinetic energy flux in solar evolutionary models. Observations have shown that the composition of the solar wind is not same as of the solar photosphere. Although the solar wind is currently weak, it may therefore have affected the evolution of the solar photospheric composition, which, to our knowledge, has not been taken into account in solar evolutionary models. Here we do take into account the such effects. The effects of the inhomogeneous PMS accretion on solar models have been investigated \citep[e.g.,][]{Castro07,Guzik10,ser11}. They have mainly concerned the varied metallicity and have not found a satisfactory solar model. \citet{ser11} have pointed out that PMS accretion with varied helium abundance could improve some properties of solar models. However, there has been no detailed investigation of that. Although PMS accretion cannot solve the ``solar abundance problem" in solo because it cannot eliminate the contradiction of the structure of the solar convective envelope, those investigations of PMS accretion \citep[e.g.,][]{Castro07,Guzik10,ser11} have shown that PMS accretion significantly affects the structure of the solar models. Therefore it should be taken into account.

\subsection{The convective overshoot below the solar convection zone} \label{Secovershoot}

Convective overshoot leads to a turbulent kinetic energy flux $F_{\rm{K}}$, which may contribute to resolving the contradiction that the standard solar convective envelope structure is not consistent with helioseismic inferences \citep{TKFZ14}, and overshoot mixing below the BCZ, which is a possible mechanism for the solar Li depletion. Partial turbulent mixing caused by convective overshoot may also play a role in eliminating the bump in the sound-speed difference found for some models (cf.\ Fig.~\ref{std_c}) just below the convection zone \citep{JCD18}. The basic theory of convective overshoot is generally complicated. Although there are non-local mixing-length models for convective overshoot \citep[e.g.,][]{Shaviv73,Maeder75,Bressan81}, they are excessively simplified and have been excluded by helioseismic inferences \citep{JCD11}. These helioseismic inferences show that the temperature gradient $\nabla$ in the solar convective overshoot region smoothly changes between $\nabla_{\rm{R}}$ and $\nabla_{\rm{ad}}$. Non-local mixing-length overshoot models show nearly adiabatic $\nabla$ in the overshoot region, with $\nabla$ changing abruptly to $\nabla_{\rm{R}}$ at the boundary of the region. At present, the required smooth profile of $\nabla$ can be obtained only by using statistical turbulent convection models \citep[e.g.,][]{xio81,xio85,xio89,Deng06,LY07}. However, the closure models in those models introduce many parameters which cannot be determined by first principle. Although they are more reasonable than non-local mixing-length models, statistical turbulent convection models still cannot perfectly reproduce the numerical simulations. Here, we introduce a simple model to deal with the convective overshoot below the base of the solar convection zone.

\subsubsection{The turbulent kinetic energy flux} \label{Secovershoottkef}

Because buoyancy opposes fluid motion outside the convectively unstable region (i.e., the convection zone defined by the Schwarzchild criterion), the source of the energy to support convection in the overshoot region is the kinetic energy flux which transports kinetic energy from the convective unstable region to the overshoot region. Therefore the equation to describe the convective overshoot is the equation of transport of turbulent kinetic energy \citep[e.g.,][]{xio81,xio85,LY07,Meakin10}:
\begin{eqnarray} \label{OVequation}
\frac{{\partial L_{\rm{K}}}}{{\partial {m_r}}} =  - \frac{g}{\rho }\overline {{u_r}'\rho '}  - {\varepsilon _{\rm{turb}}} = \frac{{\delta g{F_{\rm{C}}}}}{{\rho {c_P}T}} - {\varepsilon _{\rm{turb}}},
\end{eqnarray}%
where $L_{\rm{K}}=4\pi {r^2}{F_{\rm{K}}}$ is the turbulent kinetic energy luminosity, $g=G m_r / r^2$ is the gravitational acceleration, $G$ is the gravitational constant, $m_r$ is the enclosed mass within radius $r$, $\delta=-(\partial \ln \rho /\partial \ln T)_P$, $\varepsilon _{\rm{turb}}$ is the dissipation rate of turbulent kinetic energy. The term $- g \overline {{u_r}'\rho '} / \rho$ is the buoyancy work. The last equal sign holds because the Boussinesq approximation $\rho '/\rho  =  - \delta T'/T$ is adopted. The physical meaning of Eq.~(\ref{OVequation}) in the overshoot region is obvious. The buoyancy work term is negative in the overshoot region and $\varepsilon _{\rm{turb}}$ is always positive. Therefore the r.h.s. of the equation, which is the local net turbulent kinetic energy generation rate, is always negative in overshoot region. Integrating Eq.~(\ref{OVequation}) in the whole overshoot region shows a negative $L_{\rm{K,bc}}$ or $F_{\rm{K,bc}}$, which is the input energy flux to maintain the convective overshoot. There are three turbulent variables in Eq.~(\ref{OVequation}): $F_{\rm{K}}$, $F_{\rm{C}}$, and $\varepsilon _{\rm{turb}}$; thus the equation is not closed. We need two extra equations to make it possible to solve Eq.~(\ref{OVequation}).

Statistical turbulent convection models \citep{xio89,zl12a} have shown that $ {-{\delta g{F_{\rm{C}}}}}/({{\rho {c_P}T}}) \approx \eta {\varepsilon _{\rm{turb}}}$ and $\eta$ in most of the overshoot region is basically a constant. We adopt this property and Eq.~ (\ref{OVequation}) becomes:
\begin{eqnarray} \label{OVequation2}
\frac{{\partial L_{\rm{K}}}}{{\partial {m_r}}} =   - (1+\eta) {\varepsilon _{\rm{turb}}},
\end{eqnarray}%

Analyses of turbulent convection models \citep{xio89,zl12a} have shown that the turbulent variables (e.g., $k$, $F_{\rm{K}}$, and $\varepsilon _{\rm{turb}}$) are basically exponentially decreasing in the overshoot region. Although those turbulent convection models may result in too simple a representation of $F_{\rm{K}}$, numerical simulations \citep{Freytag96} have also obtained exponential decreasing turbulent variables. Therefore we set $L_{\rm{K}}$ as an exponentially decreasing function as follows:
\begin{eqnarray} \label{TKLexp}
{L_{\rm{K}}} = {L_{{\rm{K,bc}}}} \exp (x) \ \ {\rm{for}}\ x \leq 0,
\end{eqnarray}%
where
\begin{eqnarray}
x = \frac{{r - R_{\rm{bc}}}}{{\theta {H_P}}},
\end{eqnarray}
${L_{{\rm{K,bc}}}}$ is the turbulent kinetic energy luminosity at the BCZ, $\theta$ is a parameter, and $\theta {H_P}$ is the e-folding length of ${L_{\rm{K}}}$ in the overshoot region (i.e., the scale height of ${L_{\rm{K}}}$). $\theta$ and ${L_{{\rm{K,bc}}}}$ need to be determined.

The parameter $\theta$ can be estimated as follows. \citet{JCD11} have shown that the length of the overshoot region with modification in $\nabla$ is $l_{\nabla} \approx 0.03R_{\odot}$. Theoretical analysis of the turbulent convection model has shown  that $l_{\nabla} \approx H_{\rm{K}}$ where $H_{\rm{K}} = | {\rm{d}}r/{\rm{d}} \ln k |$ is the turbulent kinetic energy scale height \citep{zl12b}. This property does not depend on model parameters and can be validated from other turbulent models \citep{Marik02} and numerical simulations \citep{Meakin10}, so that it can be taken to be a general property of convective overshoot. Equations (\ref{OVequation2}) and (\ref{TKLexp}) show that ${\varepsilon _{\rm{turb}}}$ should also be an exponentially decreasing function with the same e-folding length as ${L_{{\rm{K,bc}}}}$; thus ${L_{{\rm{K,bc}}}} \propto {\varepsilon _{\rm{turb}}}$. It is well known that turbulence theory shows ${\varepsilon _{\rm{turb}}} \approx k^{3/2}/l$; therefore ${L_{{\rm{K,bc}}}} \propto k^{3/2}$ when the characteristic length $l$ is a slowly varying function. Finally we can estimate $\theta$ as follows:
\begin{eqnarray} \label{thetavalue}
\theta  = \frac{1}{{{H_P}}}\frac{{dr}}{{d\ln \left| {{L_{\rm{K}}}} \right|}} = \frac{1}{{{H_P}}}\frac{2}{3}\frac{{dr}}{{d\ln k}} = \frac{2}{3}\frac{{{H_{\rm{K}}}}}{{{H_P}}} \approx 0.2.
\end{eqnarray}%

${L_{{\rm{K,bc}}}}$ is a key parameter in this model. The local convection theory (i.e., mixing-length theory) predicts ${L_{{\rm{K,bc}}}}=0$ because it ignores the turbulent transport of turbulent kinetic energy. However, as it has been discussed, the existence of overshoot requires a negative ${L_{{\rm{K,bc}}}}$ to support the energy of fluid moving in the overshoot region. At present, statistical turbulent convection models and numerical simulations show significant difference on ${L_{{\rm{K,bc}}}}$. In Xiong's \citeyearpar{xio81} and Li \& Yang's \citeyearpar{LY07} models, the typical value of the turbulent kinetic energy flux at the base of thick convective envelopes (for solar models or red-giant models) is of order $L_{\rm{K,bc}} = - ( 10^{-3}  - 10^{-2} ) L_{\rm{total}}$. In contrast, numerical simulations \citep[e.g.,][]{Singh95,Tian09,Hotta14,Kapyla17} show significant turbulent kinetic energy in a convective envelope and the $L_{\rm{K,bc}}/L_{\rm{total}}$ could be as significant as $\sim -40\%$ \citep[e.g.,][]{Singh95}. However, the gradient type approximations adopted to model the third-order correlations in those statistical turbulent convection models could be invalid near the BCZ \citep{Tian09}. On the other hand, these numerical simulations do not reproduce conditions, in particular the thermal timescale, at the base of the solar convection zone and hence likely exaggerate the kinetic energy flux. At present, it is difficult to determine ${L_{{\rm{K,bc}}}}$ from the hydrodynamic equations directly. In this case, we estimate ${L_{{\rm{K,bc}}}}$ by using an indirect method to calibrate the structure of the solar convective envelope \citep{TKFZ14}: for the given composition in the solar convection zone, we can adjust the values of ${L_{{\rm{K,bc}}}}$ and the parameter $\alpha_{\rm{MLT}}$ for the mixing-length theory to obtain a solar convective envelope which has correct ${R_{\rm{bc}}}$ and a density profile in the best agreement with helioseismic inferences. It is shown in \citet{TKFZ14} that, for the original AGSS09 composition, the required ratio ${L_{{\rm{K,bc}}}}/{L_{\odot}}$ is about $-(0.13  -  0.19)$. The required ${L_{{\rm{K,bc}}}}$ for AGSS09Ne should be a little weaker since the upward revised Ne abundance enhances the opacity near the BCZ. We have tested and found that ${L_{\rm{K,bc}}}=-0.13{L_{\odot}}$ is suitable for the AGSS09Ne composition. Now the turbulent kinetic energy flux below the BCZ has been determined.

It is also required to investigate the turbulent kinetic energy flux in the convection zone. Due to the shortcomings of the third-order correlation models in statistical turbulent convection models and the required huge amount of numerical calculation, it is difficult to determine $L_{\rm{K}}$ in the solar convection zone from the analyses or simulations of hydrodynamic equations. However, $L_{\rm{K}}$ should not significantly affect the structure of the convection zone for the following reason. In the deep convection zone with $T>10^5 \ {\rm K}$, convection is very efficient so that the temperature gradient is nearly adiabatic and insensitive to $L_{\rm{K}}$. In this region, $L_{\rm{K}}$ has little effect on the stellar structure because the possible variation of $F_{\rm{K}}$ should be exactly compensated by $F_{\rm{C}}$ to ensure a $F_{\rm{R}}$ determined by the nearly adiabatic temperature gradient. Therefore setting $L_{\rm{K}}$ in the deep convection zone ($T>10^5\,{\rm K}$) is essentially arbitrary. In the uppermost part of the convection zone, with $T<10^5\, {\rm K}$, convection is not sufficiently efficient to ensure an adiabatic stratification, so that the temperature gradient is determined by the balance between total flux, $F_{\rm{R}}$, $F_{\rm{C}}$, and $F_{\rm{K}}$, depending on $F_{\rm{K}}$. The stellar radius is sensitive to the temperature gradient in this thin envelope; thus the integrated effects of the temperature gradient should be calibrated such that the radius of solar models is consistent with observations. Therefore, variations in $F_{\rm{K}}$ should be compensated by a change of $F_{\rm{C}}$ to maintain a required integral of $F_{\rm{R}}$. For example, when we calculate $F_{\rm{C}}$ by using the mixing-length theory, a negative $F_{\rm{K}}$ in the uppermost part of the convection zone will lead to a larger $\alpha_{\rm{MLT}}$ parameter than the SSM. Because of the calibration on stellar radius, the integral of the temperature gradient in the uppermost part of the convection zone should also be insensitive to the variations on $F_{\rm{K}}$. For this reason, we could ignore $F_{\rm{K}}$ in the uppermost part of the convection zone and it should not significantly affect the stellar structure. However, it should be remembered that the modifications of $F_{\rm{K}}$ will affect the profile of the temperature gradient and the structure of the uppermost part of the convection zone. Based on this analysis there is an arbitrariness in the definition of $L_{\rm{K}}$ within the convection zone.

Because $L_{\rm{K}}$ satisfies a differential equation (i.e., Eq.~(\ref{OVequation})), $L_{\rm{K}}$ should be smooth, which means that $L_{\rm{K}}$ and ${\rm{d}}L_{\rm{K}}/{\rm{d}}r$ must be continuous at the BCZ. Based on the smooth nature, and the arbitrariness in the definition, of $L_{\rm{K}}$ within the convection zone, we use the following formula for $L_{\rm{K}}$ within the convection zone:
\begin{eqnarray} \label{TKLexpCZ}
{L_{\rm{K}}} = {L_{{\rm{K,bc}}}} \exp \left(\frac{x}{{\sqrt {{x^2} + 1} }}\right) f(\log T,5.7,6.2) \ \ {\rm{for}}\ x > 0,
\end{eqnarray}
where $f$ is a smooth decaying function for arbitrary independent variables $y$, $a$ and $b$ as
\begin{eqnarray}
f(y,a,b) = \left\{ {\begin{array}{*{20}{c}}
   {1,y > b}  \\
   {\frac{1}{2} + \frac{1}{2}\sin [(\frac{{y - a}}{{b - a}} - \frac{1}{2})\pi ],a \le y \le b}  \\
   {0,y < a}  \\
\end{array}} \right..
\end{eqnarray}
This formula ensures that $L_{\rm{K}}$ is smooth at the BCZ and that $L_{\rm{K}}$ will not increase exponentially without limit in the convection zone because $x/\sqrt{x^2+1}<1$. The decaying function $f$ ensures that $L_{\rm{K}}=0$ in the uppermost part of the convection zone so that the resulting $\alpha_{\rm{MLT}}$ is not affected by $L_{\rm{K}}$ and comparable with that in the SSM.

\begin{figure}
\centering
\includegraphics[scale=0.5]{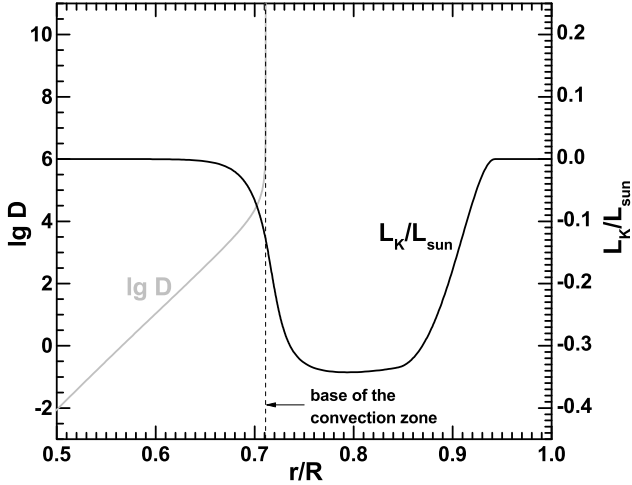}
\caption{The turbulent kinetic energy luminosity and the overshoot mixing diffusion coefficient profiles of solar Model TWA. We set ${L_{\rm{K,bc}}}=-0.13{L_{\odot}}$. The overshoot mixing diffusion coefficient is based on Eq.~\ref{ovmixing} with $C_{\rm {X}} = 5\times10^{-4}$. The location of the BCZ is shown by the vertical dashed line. }\label{figLkD}
\end{figure}

The adopted profiles of ${L_{\rm{K}}}$ in the convection zone and overshoot region are defined by Eq.~(\ref{TKLexp}) and Eq.~(\ref{TKLexpCZ}) with ${L_{\rm{K,bc}}}=-0.13{L_{\odot}}$, respectively. The profile of ${L_{\rm{K}}}/L_{\odot}$ of a typical solar model is shown in Fig.~\ref{figLkD}.

\subsubsection{The overshoot mixing} \label{Secovershootmix}

Another important effect of convective overshoot is the overshoot mixing. The overshoot mixing is usually assumed to be very efficient to keep the overshoot region always homogeneous and with the same composition as the adjacent convection zone, as was tested by \citet{Z12Li}. However, this assumption is questionable. It implies that a $0.37H_P$ overshoot region \citep{JCD11} below the BCZ results in ${\rm{A(Li)}}<-4$, which is much lower than observed \citep[e.g.,][shows ${\rm{A(Li)}}=1.05\pm0.1$]{AGSS09}. Another viewpoint is that the overshoot mixing can be assumed to be a diffusion process and the diffusion coefficient exponentially decreases in the overshoot region \citep[e.g.,][]{Herwig00,OVMZ13}. A model for convective overshoot mixing based on statistical turbulent convection model has been developed by \citet{OVMZ13}. It shows that the diffusion coefficient is related to the turbulent dissipation and the effect of buoyancy as:
\begin{eqnarray} \label{ovmixing}
	D_{\rm{OV}}=C_{\rm{X}} \frac {\varepsilon _{\rm{turb}}} {N^2_1} = - \frac{{{C_{\rm{X}}}g{L_{\rm K}}}}{{4\theta (1 + \eta)\pi {r^2}P{N^2_1}}},
\end{eqnarray}%
where $C_{\rm{X}}$ is a parameter much less than unity \citep{OVMZ13} and
\begin{equation} \label{ovmixingN2}
	N_1^2 =  - \frac{{\delta g}}{{{H_P}}}(\nabla  - {\nabla _{\rm ad}} - \chi \nabla_\mu) \; ,
\end{equation}
is the effective squared buoyancy frequency for overshoot mixing,
where
\begin{equation}
\label{nablamu}
\nabla_\mu = \sum\limits_i {{{\left(\frac{{\partial \ln T}}{{\partial {X_i}}}\right)}_{{X_j},\rho ,P}}\frac{{d{X_i}}}{{d\ln P}}} \; ;
\end{equation}
here $X_i$ is the abundance of the $i-$th species and $\chi$ is a positive parameter of order unity based on turbulent convection models \citep{OVMZ13}. If $\chi=1$, $N_1^2=N^2$ is exactly the squared Brunt-V\"{a}is\"{a}l\"{a} frequency. We will adopt Eq.~(\ref{ovmixing}) to calculate the overshoot mixing below the BCZ. The value of $\chi$ depends on turbulent convection models and is difficult to determine from first principles. However, in the solar overshoot region in the solar evolution models, $\nabla_\mu$ in the overshoot region is of order less than $0.01$, while ${\nabla _{\rm ad}}-\nabla$ is of order $0.1$. Therefore the contribution of the $\chi\nabla_\mu$ term to $N_1^2$ is not significant when $\chi$ is of order unity. In this case, we ignore the contribution of the $\nabla_\mu$ term, i.e., assuming $\chi=0$.  The effect of $C_{\rm{X}}$ and $\eta$ is determined only by their combination $C_{\rm{X}}/(1+\eta)$; thus $\eta$ is set to $0.1$ which is a typical result shown in statistical turbulent convection model \citep{zl12b}.

The typical profile of $D_{\rm{OV}}$ below the BCZ is shown in Fig.~\ref{figLkD}. From the BCZ downward to the solar center, $D_{\rm{OV}}$ quickly decreases near the BCZ because $N_1^2$ is zero at the BCZ and increases as $r$ decreases. After a short distance away from the BCZ, $D_{\rm{OV}}$ exponentially decreases because the dissipation rate ${\varepsilon _{\rm{turb}}}$ exponentially decreases and ${N_1^2}$ changes much more slower than ${\varepsilon _{\rm{turb}}}$. We set the parameter $C_{\rm{X}}=5 \times 10^{-4}$ in order to approximately reproduce the observation of solar Li abundance in solar models. When $\chi$ is set to be nonzero or $\eta$ is enlarged, $C_X$ should be slightly enlarged to compensate their variations.

\subsection{A helium-poor mass loss: the solar wind} \label{Secsolarwind}

The solar wind, which is not included in the SSMs, may result in an effect detectable by comparing solar models with helioseismic inferences. \citet{windmass} investigated mass-loss rates of solar-like stars and found that ${\rm{d}}M/{\rm{d}}t \propto t^{-2.00\pm0.52}$ and that the mass-loss rate should be roughly constant before $ \sim 0.1\,{\rm{Gyr}}$, corresponding to a saturation of the stellar X-ray flux; this implies that the solar wind may have been $10^{3}$ times more massive in the early stage. Based on the present solar wind mass-loss rate ${\rm{d}}M/{\rm{d}}t = -2 \times 10^{-14}M_{\odot} {\rm{yr}}^{-1}$ \citep{solarwindpresent}, the estimated typical value of the total mass loss is of the order of magnitude of $10^{-3}  -  10^{-2} M_{\odot}$, which is comparable with the mass of the solar convection zone. Observations have revealed an important property of the solar wind that the composition of solar wind is different from the photosphere, especially the helium abundance in the solar wind is only about a half of $Y_{\rm{S}}$ \citep[e.g.,][]{windcomp,Reames95,vS00,vS10}. This is so called the first ionization potential (FIP) effect, i.e., abundances of the elements with FIP less than 10eV (e.g., Mg, Si, and Fe) are enhanced in the corona with respect to photospheric values and high FIP elements (e.g., O, Ne, and He) have much smaller abundance enhancements or even abundance depletions in the corona \citep{fiprev}. It is commonly interpreted as caused by the ponderomotive force resulting from the propagation or reflection of magnetohydrodynamic waves in the chromosphere \citep{fiprev}. We assume that this helium-poor property is always satisfied in the evolution of the solar wind. The typical value of the solar-wind mass loss and its helium-poor property show that taking into account the solar wind could enhance $Y_{\rm{S}}$ at a level of $\sim0.01$, which is about three times the uncertainty of the helioseismically inferred $Y_{\rm{S}}$. Therefore the solar wind should be included in solar models.

We take into account the helium-poor mass loss caused by the solar wind in solar models. The abundances of the lost mass are set as:
\begin{eqnarray} \label{solarwindXYZ}
(X_{\rm{L}},Y_{\rm{L}},Z_{\rm{L}}) = \frac{({{X_{\rm{S}}}},\lambda_{Y}Y_{\rm{S}},\lambda_{Z}Z_{\rm{S}})}{{X_{\rm{S}}}+\lambda_{Y}{Y_{\rm{S}}}+\lambda_{Z}{Z_{\rm{S}}}},
\end{eqnarray}%
where $\lambda_{Y}$ is the relative efficiency of helium mass loss relative to hydrogen, $\lambda_{Z}$ is the relative efficiency of heavy-elements mass loss. $\lambda_{Y}=0.5$ is adopted based on observations. $\lambda_{Z}=1$ is adopted as default and the effects of varied $\lambda_{Z}$ will also be investigated. Variations of $\lambda_{Z}$ correspond to the possible FIP or reversed FIP effects on heavy elements escaping in the solar wind.

Since the mass-loss rate should be saturated before $ \sim 0.1\,{\rm{Gyr}}$ \citep{windmass} and the main effect of mass loss occurs in the MS stage, the adopted time-dependent mass loss starts from $0.1\,{\rm{Gyr}}$ and the mass-loss rate is calculated as:
\begin{eqnarray} \label{solarwinddMdt}
\frac{{\rm{d}}M} {{\rm{d}}t} = - C{t^\gamma }.
\end{eqnarray}%
Integrating the above equation shows a total mass loss as:
\begin{eqnarray} \label{solarwindML}
M_{\rm{L}} =  \frac{C}{{\gamma  + 1}}({t_1}^{\gamma  + 1} - {t_0}^{\gamma  + 1}),
\end{eqnarray}%
where $t_0 = 0.1\, {\rm{Gyr}}$ and $t_1 = 4.57 \,{\rm{Gyr}}$. The total mass loss $M_{\rm{L}}$ is a free parameter. $C$ and $\gamma$ are adjusted to satisfy Eq.~(\ref{solarwindML}) and to reproduce the present solar-wind mass-loss rate $ C {t_1^\gamma } = 2 \times 10^{-14} M_{\odot} {\rm{yr}}^{-1}$ \citep[e.g.,][]{solarwindpresent}.

\subsection{Inhomogeneous disk accretion in the solar PMS stage} \label{Secpmsaccr}

Because of angular momentum conservation, circumstellar disks are an unavoidable consequence during the stellar formation through the gravitational collapse. Much observational evidence has shown that accretion disks are commonly found to surround T-Tauri stars and the stars are accreting from the disks \citep{Baccr90,BSaccr91,skaccr90,Haccr16}. The relations of mass accretion rates of T-Tauri stars to stellar mass and age can be measured by using the excess hot continuum emission caused by accretion onto stars \citep{GHaccr98}. For the solar-mass stars, the typical accretion duration is of the order of magnitude of $1  -  10\,$Myr and the typical accretion rate is $10^{-8}  -  10^{-9} M_{\odot}/{\rm{yr}}$ \citep[e.g.,][]{Haccr98,Haccr16}. Those lead to a typical total accreted mass of about $10^{-2}  -  10^{-1} M_{\odot}$ for a solar-mass star. For the Sun, the disk should exist because the minimum-mass solar nebula, i.e., the lowest mass of the disk which formed planets in the solar system, can be estimated from the planetary composition as $0.01  -  0.07 M_{\odot}$ \citep{MMSNw77}. Therefore it is reasonable to assume that the Sun experienced disk accretion at an age less than some $10 \,{\rm{Myr}}$ in the PMS stage with a typical accretion rate $10^{-9}  -  10^{-8} M_{\odot}/{\rm{yr}}$. The adopted PMS accretion rate in solar models is a combination of the relations for accretion rate versus stellar mass and stellar age (both are given by \citet{Haccr16}):
\begin{eqnarray} \label{AccrRate}
	&& \log \left[\frac{{d(M/{M_{\odot}})}}{{d(t/{\rm{yr}})}}\right] = \\ \nonumber
	&&  - 1.32 - 1.07\log (t/{\rm{yr}}) + 2.1\log \left(\frac{{M/{M_{\odot}}}}{{0.7}} \right) \; ,
\end{eqnarray}%
with a dispersion $0.5$dex.

The solar PMS accretion with varied metallicity has been investigated \citep[e.g.,][]{ser11} and no satisfactory solar model has been found. They have suggested to investigate PMS accretion with varied helium abundance. We now mainly consider the variation on helium abundance of the accreted material in the solar PMS stage. The effects of the variation on metallicity will also be investigated. Possible justifications for the inhomogeneous accretion may be related to planet formation and details of the accretion process. The planet formation, including the condensation of heavy elements, may lead to varied composition of the accreted material if the timescales of the accretion and planet formation processes are comparable \citep{Melendez09, ser11}. On the other hand, the physical processes in the accretion process may also lead to varied composition of the accreted material. Since the mechanism of PMS accretion from protoplanetary disk is believed to be in the scenario of magnetosphere accretion \citep{Haccr94,Haccr16} driven by the magnetorotational instability \citep{Balbus91}, this mechanism requires coupling between the gas and the magnetic field. Since the temperature in the disk is low, the bulk of the disk is only weakly ionized. The coupling mainly occurs in the surface layers of the disk where the ionization fraction is enhanced by nonthermal ionization processes such as cosmic rays and X-rays \citep{Gammie96}. Because these ionization processes are affected by the FIP of each element, it is possible that the element abundances in ions in the coupling region are not the same of the element abundances in neutrals. The ions could be accelerated and separated from neutrals by the ponderomotive force due to magneto-hydrodynamic waves and this process is also affected by the FIPs of elements leading to FIP or inverse FIP effects \citep{fiprev}. If the separation is sufficient, the accreted material should be ion-dominated and it is possible that its composition is different from the bulk of the disk.

We take into account inhomogeneous PMS accretion in solar models. Since we mainly investigate the effects of a varied helium abundance of accreted material, we set the accreted helium abundance $Y_{\rm{acc}}$ as a free parameter. The metallicity of the accreted mass is set to $Z_{\rm{acc}}=0.015$ as default, but variations in $Z_{\rm{acc}}$ will also be investigated. In the fully convective phase in the early PMS stage, the accretion is in the form of free fall so that the composition of the accreted mass should be same as the protostar. Even if the composition of the accreted mass is different from the protostar, the protostar is also homogeneous before it develops a radiative core due to the complete convective mixing. Therefore we let the accretion start from $2\,{\rm{Myr}}$ when the Sun is developing its radiative core. In this case, the `initial' abundances in our solar models may be a little different from the primordial abundances because the possible inhomogeneous accretion in the fully convective phase may change the composition of the star. The accretion duration $\tau_{\rm{acc}}$ is a free parameter. A $0.5\,$dex dispersion for the accretion rate Eq.~(\ref{AccrRate}) will be considered.

\section{Results} \label{Secresult}

The properties of solar models with convective overshoot, the inhomogeneous mass loss caused by solar wind, and the inhomogeneous PMS accretion will be investigated in this section. At first, a solar model with only the convective overshoot will be discussed, then the properties of several typically improved solar models with all three extra physical processes will be described, finally we will investigate the effects of parameters of those extra physical processes. Because there are six free parameters (i.e., $\lambda_{Z}$ and $M_{\rm{L}}$ for solar wind mass loss, $Y_{\rm{acc}}$, $Z_{\rm{acc}}$, $\tau_{\rm{acc}}$ and the dispersion of the accretion rate for the inhomogeneous PMS accretion), the required number of solar models is tremendous if those parameters completely cover their ranges independently, i.e., $N^6$ where $N\sim10$ is the typical number of sampling points for each parameter. In order to reduce the amount of numerical calculations, we classify those parameters into two sets. The primary parameters ($M_{\rm{L}}$ and $Y_{\rm{acc}}$) are varied over the full ranges. The effects of variations on $\lambda_{Z}$, $Z_{\rm{acc}}$, $\tau_{\rm{acc}}$ and the strength of the accretion rate will be investigated in turn. It reduces the amount of solar models to be on the order of magnitude of $10^{3-4}$, which is sustainable.

The main factor of the evaluation of solar models is the sound-speed profile. The comparisons of sound-speed and density profiles between solar models and the helioseismic inferences are done in two ways. For some representative solar models (e.g., SSMs, Model OV09Ne, and some typical improved models, e.g., Model TWA, see below), we carried out inversions for the relative differences $\delta c_{\rm{s}} / c_{\rm{s}}$ and $\delta \rho/\rho$ in sound speed and density between the Sun and the model, with the technique of optimally localized averages \citep[e.g.,][]{Gough1991, Rabell1999}, and using the so-called `Best set' of observed frequencies, introduced by \citet{Basu1997}. For other solar models, since the number of them is huge, we compare the sound-speed or density profile between solar models and a reference solar sound-speed or density profile derived from the helioseismic inversion on Model TWA. The reference profile slightly depends on the solar model used as reference in the inversion, but the variations are small in the most part of solar interior except in a shallow envelope with $r>0.96R_{\odot}$ when different solar models are used as reference. Therefore the second way is a reasonable alternative which significantly reduces the time of numerical calculation. In the second way, the sound speed and density of models are compared with the reference sound-speed and density profiles only in the solar interior $r<0.96R_{\odot}$.

\subsection{Effects of the convective overshoot} \label{SecresultOV}

\begin{figure}
\centering
\includegraphics[scale=0.5]{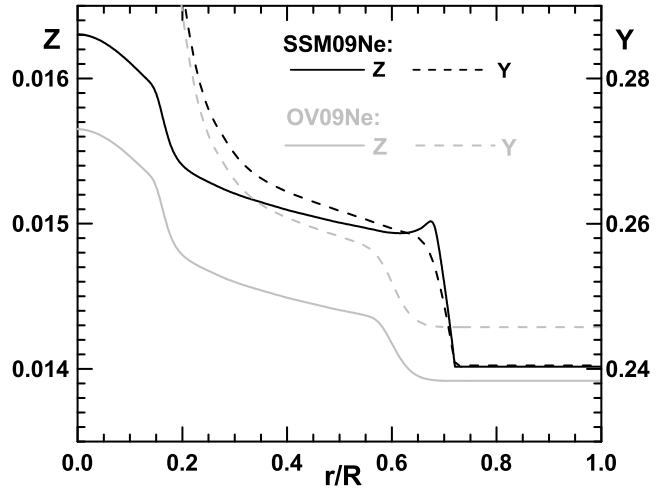}
\caption{Helium abundance (right ordinate) and metallicity (left ordinate) in the interior of Models OV09Ne and SSM09Ne. }.
\label{OV09Ne}
\end{figure}

The basic properties of Model OV09Ne are listed in Table~\ref{modelinfo} and shown in Fig.~\ref{std_c}. The only distinction between Model OV09Ne and the standard Model SSM09Ne is that the helioseismically based convective overshoot model introduced in Section \ref{Secovershoot} is applied in OV09Ne. The location of the BCZ $R_{\rm{bc}}$ of Model OV09Ne is 0.7155, which is significantly improved and close to the value of Model SSM98. The lithium abundance of Model OV09Ne is significant depleted to close to the observations. The surface helium abundance $Y_{\rm{S}}$ is also improved in Model OV09Ne. All those improvements are directly resulted from the convective overshoot: the negative turbulent kinetic energy moves the BCZ downward, the overshoot mixing contributes to the lithium depletion, and the mixing counters the helium settling near the BCZ thus leading to a higher $Y_{\rm{S}}$ as shown in Fig.~\ref{OV09Ne}.

It is no surprise that those improvements in the properties of the convection zone are found in Model OV09Ne, because the parameters ($L_{\rm{K,bc}}$, $\theta$ and $C_{\rm{X}}$) of the adopted overshoot model are actually based on the requirements for improvements on helioseismic properties and the lithium abundance. However, the helioseismically based overshoot model cannot solve the ``solar abundance problem" in solo, since it cannot improve the sound-speed profile in the solar radiative interior, i.e., the sound-speed deviations of Model OV09Ne in the region of $r<0.6R_{\odot}$ are more significant than for the standard Model SSM09Ne even though $R_{\rm{bc}}$ is significantly improved. The reason is that, since $(Z/X)_{\rm{S}}$ is calibrated, the overshoot mixing counters the settling of heavy elements and hence leading to a lower metallicity in the solar radiative interior as shown in Fig.~\ref{OV09Ne}. A lower metallicity results in a lower opacity thus making the sound-speed profile worse. On the other hand, because the lower metallicity leads to a lower $T_{\rm{C}}$ in Model OV09Ne, its $^8{\rm{B}}$ neutrino flux is lower than that of Model SSM09Ne and is close to the lower limit of the $1\sigma$ range of the $^8{\rm{B}}$ neutrino flux taking into account both the observational and theoretical uncertainties.

\subsection{Improved solar models} \label{SecresultTWA}

We have calculated about 900 solar models with $\lambda_{Z}=1$, $Z_{\rm{acc}}=0.015$ and different $\tau_{\rm{acc}}$, $M_{\rm{L}}$ and $Y_{\rm{acc}}$. We have tested and found that the helium-poor accretion could improve the helioseimic properties of solar model and vice versa. Therefore we mainly consider helium-poor accretion so that $Y_{\rm{acc}}$ varies in the range from $0.00$ to $0.26$ with a step $0.02$. $M_{\rm{L}}$ varies in the range from $0.00$ to $0.01$ with a step $0.001$. Six sample points of $\tau_{\rm{acc}}$ are $8$, $10$, $12$, $15$, $20$ and $30$Myr, since observations shows accretion occurs for stars with age in the order of magnitude of $1-10\,$Myr. For models with accretion duration $\tau_{\rm{acc}}<8$Myr, their properties cannot be well improved simultaneously. The total accreted masses are mainly determined by the accretion duration and slightly varying with total mass loss: $0.0462-0.0472M_{\odot}$ for $\tau_{\rm{acc}}=8$Myr, $0.0529-0.0540M_{\odot}$ for $\tau_{\rm{acc}}=10$Myr, $0.0582-0.0594M_{\odot}$ for $\tau_{\rm{acc}}=12$Myr, $0.0645-0.0658M_{\odot}$ for $\tau_{\rm{acc}}=15$Myr, $0.0723-0.0738M_{\odot}$ for $\tau_{\rm{acc}}=20$Myr, $0.0829-0.0846M_{\odot}$ for $\tau_{\rm{acc}}=30$Myr.

\begin{figure*}
\centering
\includegraphics[scale=0.5]{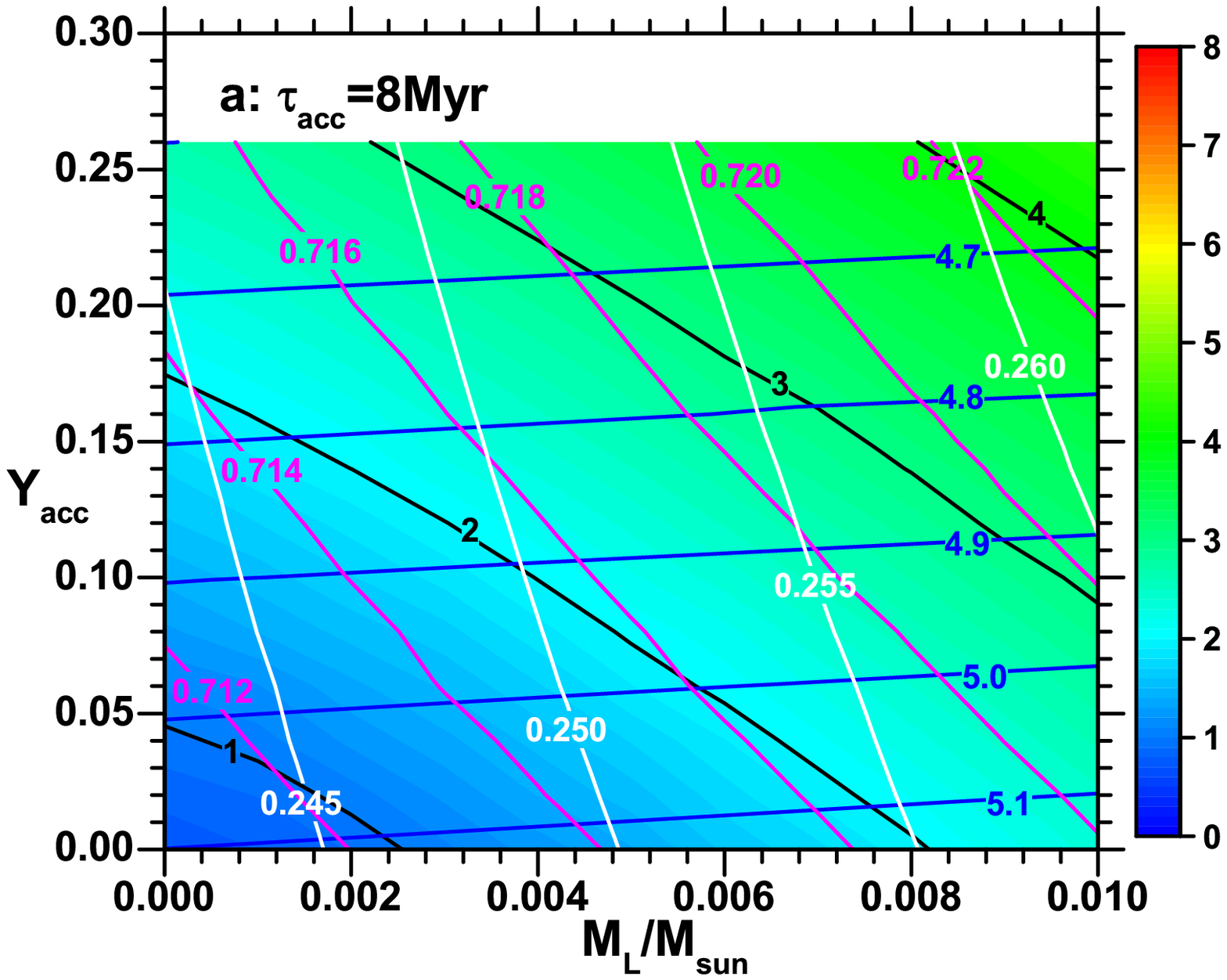}
\includegraphics[scale=0.5]{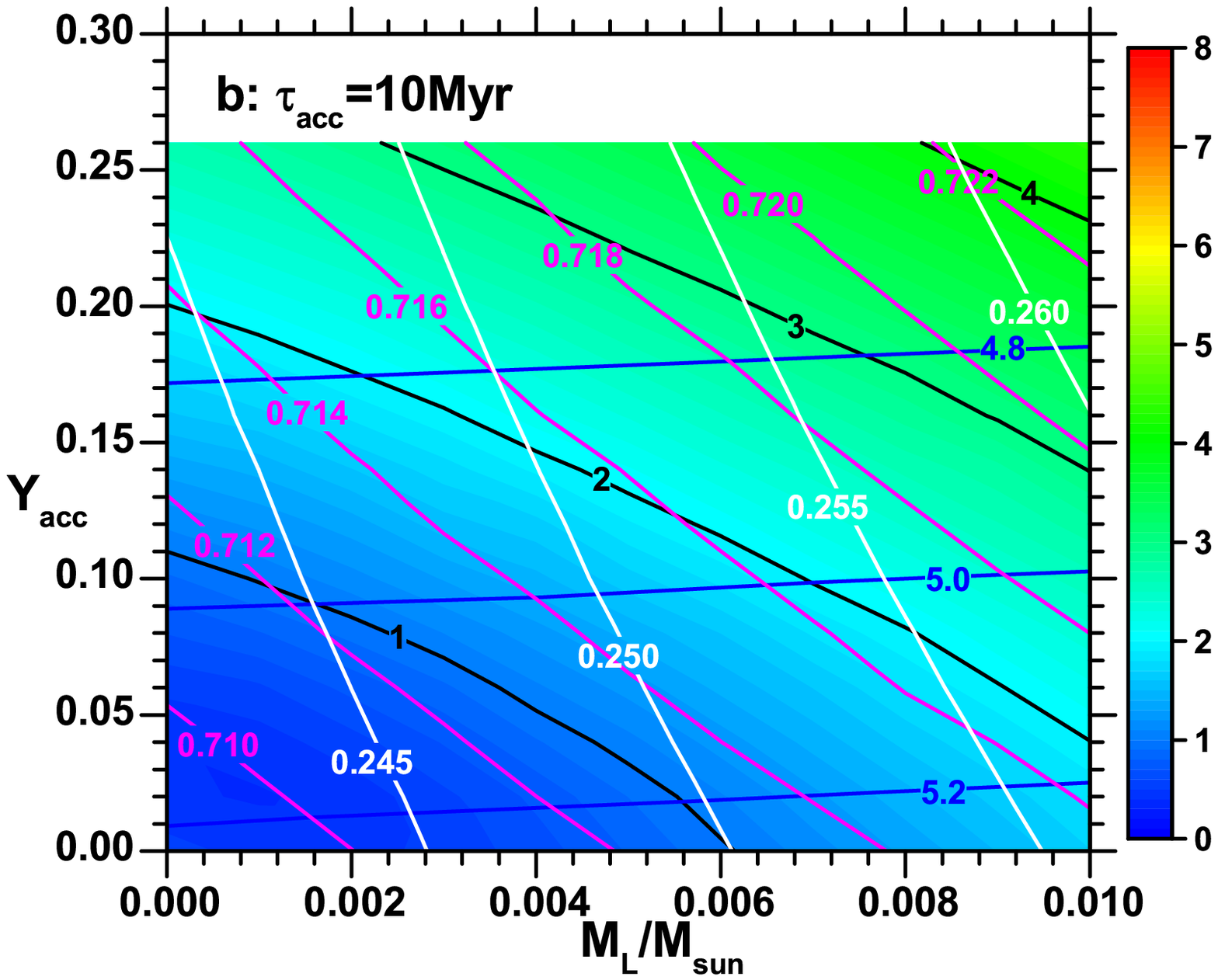}
\includegraphics[scale=0.5]{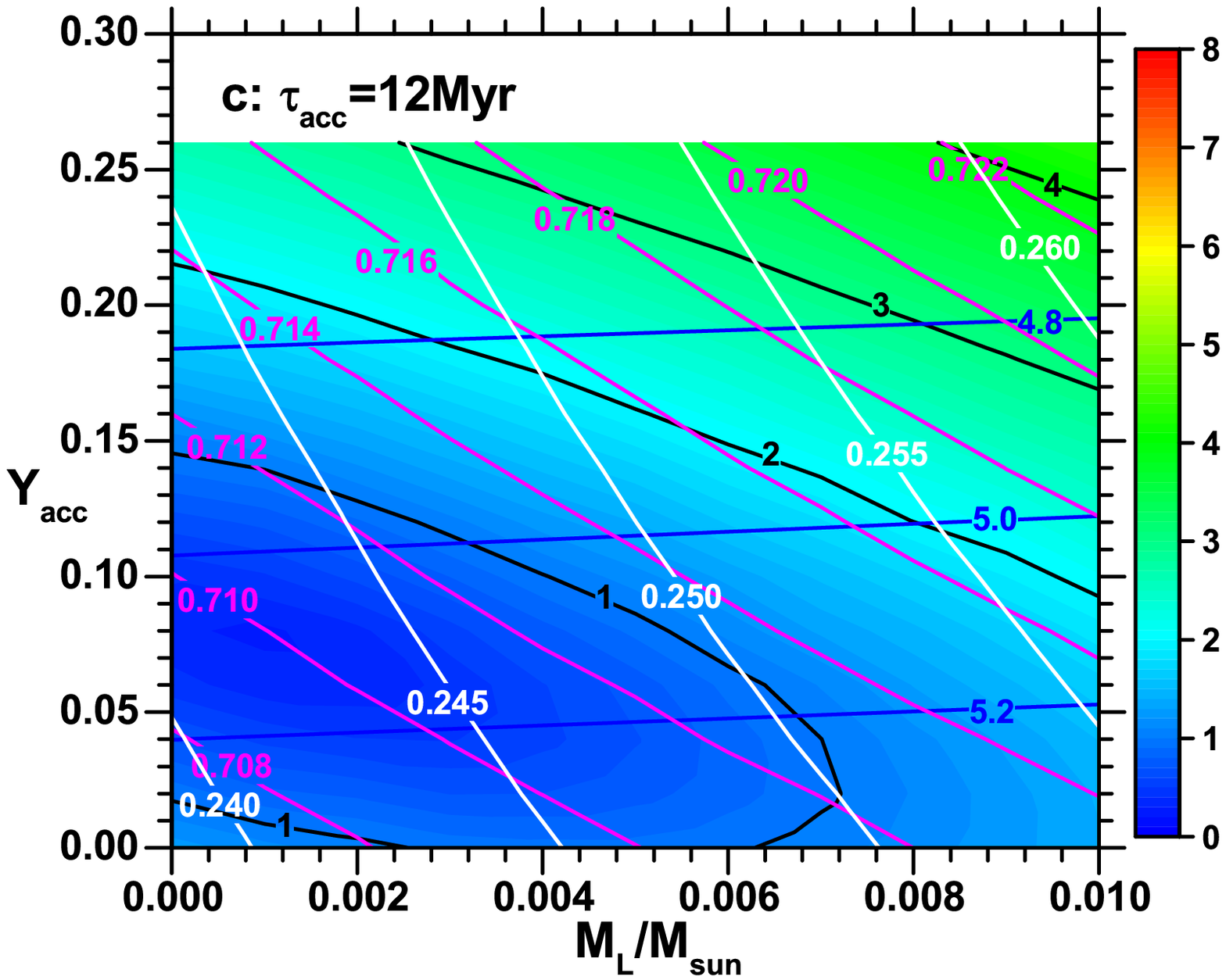}
\includegraphics[scale=0.5]{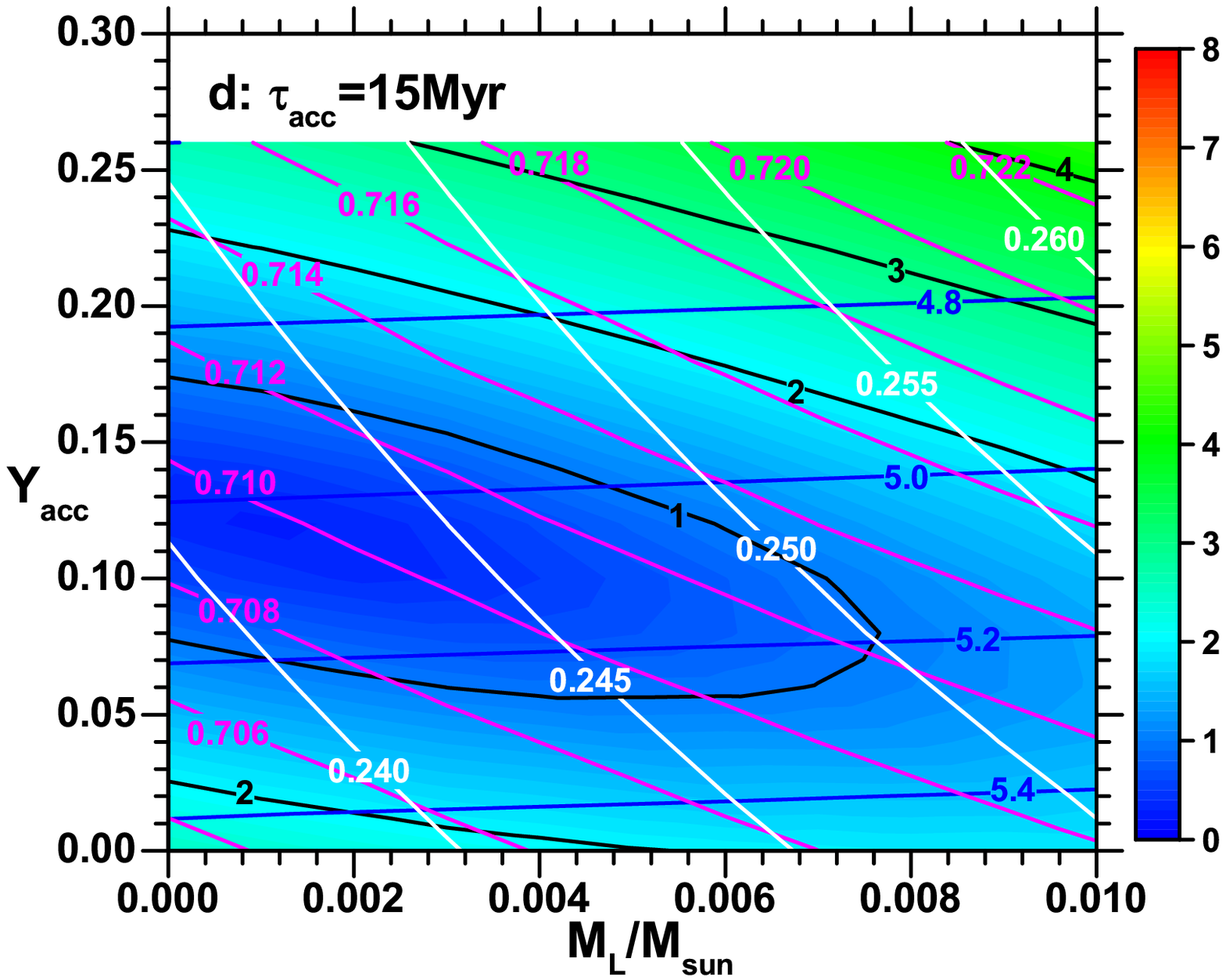}
\includegraphics[scale=0.5]{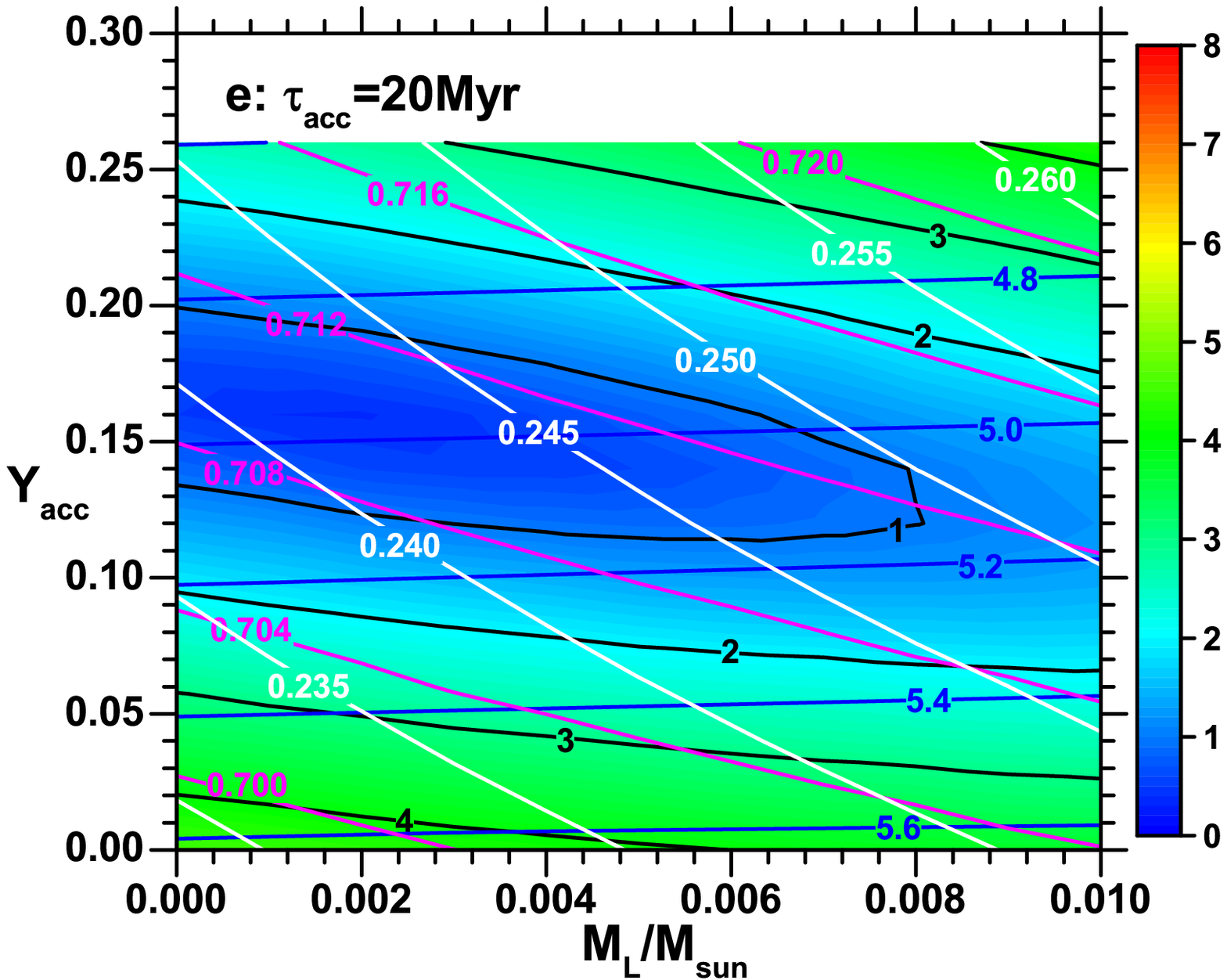}
\includegraphics[scale=0.5]{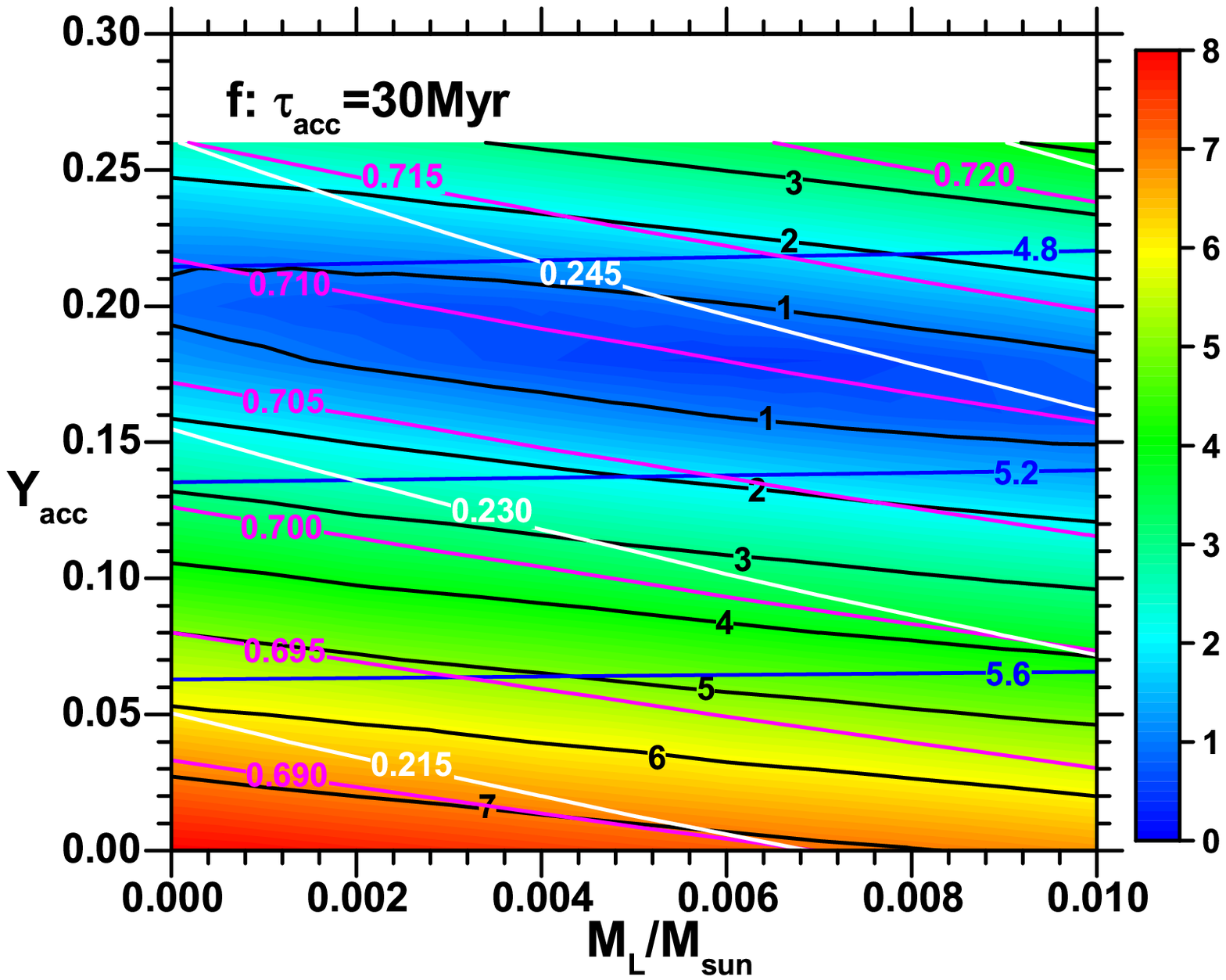}
	\caption{ Properties of solar models with different duration of accretion $\tau_{\rm{acc}}$, total mass loss $M_{\rm{L}}$ and helium abundance of accreted material $Y_{\rm{acc}}$. Colors and black contours represent r.m.s. deviations of squared sound speed $\sqrt{\langle(\Delta c_{\rm{s}}/c_{{\rm{s}},{\rm{ref}}})^2\rangle}$ in $r<0.96R_{\odot}$ multiplied by 1000, where $\Delta c_{\rm{s}}=c_{{\rm{s}},{\rm{ref}}}-c_{\rm{s}}$ and $c_{{\rm{s}},{\rm{ref}}}$ is the reference solar sound speed inferred from helioseismic inversion based on Model TWA. Purple contours represent the location of the BCZ $R_{\rm{bc}}/R_{\odot}$. Blue contours represent neutrino fluxes ${\rm{\Phi}} (^8{\rm{B}}) /10^6$. White contours represent $Y_{\rm{S}}$.
}\label{Type2}
\end{figure*}

Some main properties related to observations for those solar models (sound-speed deviations, $^8$B neutrino flux, location of BCZ, and surface helium abundance) are shown in Fig.~\ref{Type2}. The abscissas are the total mass loss and the ordinates are the helium abundance of the accreted material. The color and the black contours show the r.m.s. sound-speed deviations multiplying by 1000, the white contours show the surface helium abundance, the blue contours show the $^8$B neutrino flux multiplying by $10^{-6}$ and the purple contours show the location of BCZ $R_{\rm{bc}}/R_{\odot}$. In the following we discuss the relations between those model properties and the parameters ($Y_{\rm{acc}}$, $M_{\rm{L}}$ and $\tau_{\rm{acc}}$) and offer explanations for these relations.

The surface helium abundance is positively correlated with both $Y_{\rm{acc}}$ and $M_{\rm{L}}$. The former is obvious. The latter arises because the helium-poor mass loss concentrates helium in the CZ. $R_{\rm{bc}}$ is also positively correlated with both $Y_{\rm{acc}}$ and $M_{\rm{L}}$. The main reason could be that, at the BCZ, opacity is anti-correlated with $Y_{\rm{S}}$ because both hydrogen and helium are fully ionized and hydrogen is more efficient to contribute electron scattering opacity than helium. The r.m.s. sound-speed deviation is correlated with $R_{\rm{bc}}$ for $R_{\rm{bc}}>0.71R_{\odot}$ and anti-correlated with $R_{\rm{bc}}$ for $R_{\rm{bc}}<0.71R_{\odot}$. This is because the r.m.s. sound-speed deviation is mainly contributed by the sound-speed deviation in about $0.6<r/R<0.7$ below the BCZ and the deviation in that region is strongly affected by $R_{\rm{bc}}$. It is shown that, for all solar models, models with $R_{\rm{bc}}=0.710R_{\odot}$ show best sound-speed deviation. This is a little deeper than the helioseismic inference $R_{\rm{bc}}=0.713R_{\odot}$ \citep{JCD91,ba97}. The model $R_{\rm{bc}}$ should be improved if the temperature gradient modification caused by $F_{\rm{C}}$ overshoot is taken into account. When the negative turbulent heat flux in the overshoot is taken into account, it will enhance the temperature gradient below the BCZ \citep[e.g.,][]{xio01,zl12a}. Therefore, in order to keep a sound-speed profile, the location of the BCZ should move upward.

The $^8$B neutrino flux is anti-correlated with $Y_{\rm{acc}}$ and weakly correlated with $M_{\rm{L}}$. A lower $Y_{\rm{acc}}$ requires a lower $X_{\rm{0}}$ for the calibrations of solar model, which leads to a higher central temperature because of the lower central hydrogen abundance and the calibration of model luminosity. A higher central temperature leads to a higher $^8$B neutrino flux. The correlation with $M_{\rm L}$ arises because the helium-poor mass loss slightly depletes heavy elements in the CZ (i.e., see Eq.~\ref{solarwindXYZ}, $Z_{\rm{L}}<Z_{\rm{S}}$ for the case of $\lambda_{\rm{Y}}=0.5$ and $\lambda_{\rm{Z}}=1$), thus leading to a slightly higher $Z_{\rm{C}}$ because of the calibration of $(Z/X)_{\rm{S}}$, and hence to a slightly higher $^8$B neutrino flux.

Comparing each panel in Fig.~\ref{Type2}, it is found that, for models with a longer $\tau_{\rm{acc}}$, the effects of $Y_{\rm{acc}}$ are more significant. The main effect of the helium-poor PMS accretion is to lead to a helium-abundance gradient and a helium-poor envelope for the ZAMS model. For a model with a longer $\tau_{\rm{acc}}$, the helium-abundance gradient occurs in a more extended region owing to the continues retreat of the convective envelope, which also reduces the mass of the convective envelope and hence amplifies the effect of the helium-poor accretion, enhancing the reduction in the helium abundance in the envelope. Therefore a longer $\tau_{\rm{acc}}$ amplifies the effects of the helium-poor accretion.

\begin{table}
\centering
\caption{ Basic information of some improved solar models. }\label{Type2Best}
\begin{tabular}{lcccccc}
\hline\noalign{\smallskip}
                                     & Best08   & Best10   & TWA      & Best15  & Best20  & Best30 \\
\hline
$\alpha_{\rm{MLT}}$                  & 2.3560   & 2.3711   & 2.3708   & 2.3697  & 2.3707  & 2.3627  \\
$X_{\rm{0}}$                         & 0.7087   & 0.7082   & 0.7096   & 0.7114  & 0.7130  & 0.7134  \\
$Z_{\rm{0}}$                         & 0.01469  & 0.01471  & 0.01472  & 0.01472 & 0.01474 & 0.01488 \\
$X_{\rm{C}}$                         & 0.3485   & 0.3472   & 0.3485   & 0.3503  & 0.3519  & 0.3522  \\
$Z_{\rm{C}}$                         & 0.01569  & 0.01571  & 0.01572  & 0.01572 & 0.01574 & 0.01589 \\
$\log T_{\rm{C}}$                    & 7.1924   & 7.1928   & 7.1925   & 7.1922  & 7.1919  & 7.1919  \\
$\log  \rho_{\rm{C}}$                & 2.1859   & 2.1870   & 2.1857   & 2.1839  & 2.1823  & 2.1813  \\
$\tau_{\rm{acc}}/{\rm{Myr}}$         & 8        & 10       & 12       & 15      & 20      & 30      \\
$M_{\rm{acc}}/M_{\odot}$             & 0.0464   & 0.0532   & 0.0585   & 0.0649  & 0.0729  & 0.0843  \\
$Y_{\rm{acc}}$                       & 0.000    & 0.023    & 0.070    & 0.117   & 0.155   & 0.179   \\
$M_{\rm{L}}/M_{\odot}$               & 0.0016   & 0.0025   & 0.0028   & 0.0031  & 0.0038  & 0.0084  \\
$\gamma$                             & -1.663   & -1.827   & -1.868   & -1.904  & -1.975  & -2.242  \\
$Y_{\rm{S}}$                         & 0.2448   & 0.2450   & 0.2450   & 0.2450  & 0.2448  & 0.2456  \\
$(Z/X)_{\rm{S}}$                     & 0.0188   & 0.0188   & 0.0188   & 0.0188  & 0.0188  & 0.0188  \\
$ R_{\rm{bc}}/R_{\odot}$             & 0.7118   & 0.7110   & 0.7110   & 0.7110  & 0.7110  & 0.7115  \\
A(Li)                                & 0.77     & 0.77     & 0.82     & 0.90    & 1.03    & 1.34    \\
\hline
neutrino fluxes  & & & & & & \\
in $({\rm{cm}}^{-2}{\rm{s}}^{-1})$ & & & & & & \\
\hline
$pp$ $(10^{10})$                     & 5.97     & 5.97     & 5.98     & 5.98    & 5.98    & 5.98 \\
$pep$ $(10^{8})$                     & 1.47     & 1.47     & 1.47     & 1.47    & 1.47    & 1.47 \\
$hep$ $(10^{3})$                     & 8.14     & 8.12     & 8.13     & 8.14    & 8.15    & 8.14 \\
$^7$Be $(10^{9})$                    & 4.83     & 4.87     & 4.84     & 4.80    & 4.77    & 4.77 \\
$^8$B $(10^{6})$                     & 5.10     & 5.18     & 5.13     & 5.05    & 4.99    & 4.99 \\
$^{13}$N $(10^{8})$                  & 2.18     & 2.20     & 2.19     & 2.17    & 2.16    & 2.18 \\
$^{15}$O $(10^{8})$                  & 1.62     & 1.65     & 1.63     & 1.61    & 1.60    & 1.61 \\
$^{17}$F $(10^{6})$                  & 3.57     & 3.62     & 3.59     & 3.54    & 3.50    & 3.54 \\
\hline
\end{tabular}
	\tablecomments{See notes for Table~\ref{tab:basmod}.}
\end{table}

\begin{figure}
\centering
\includegraphics[scale=0.5]{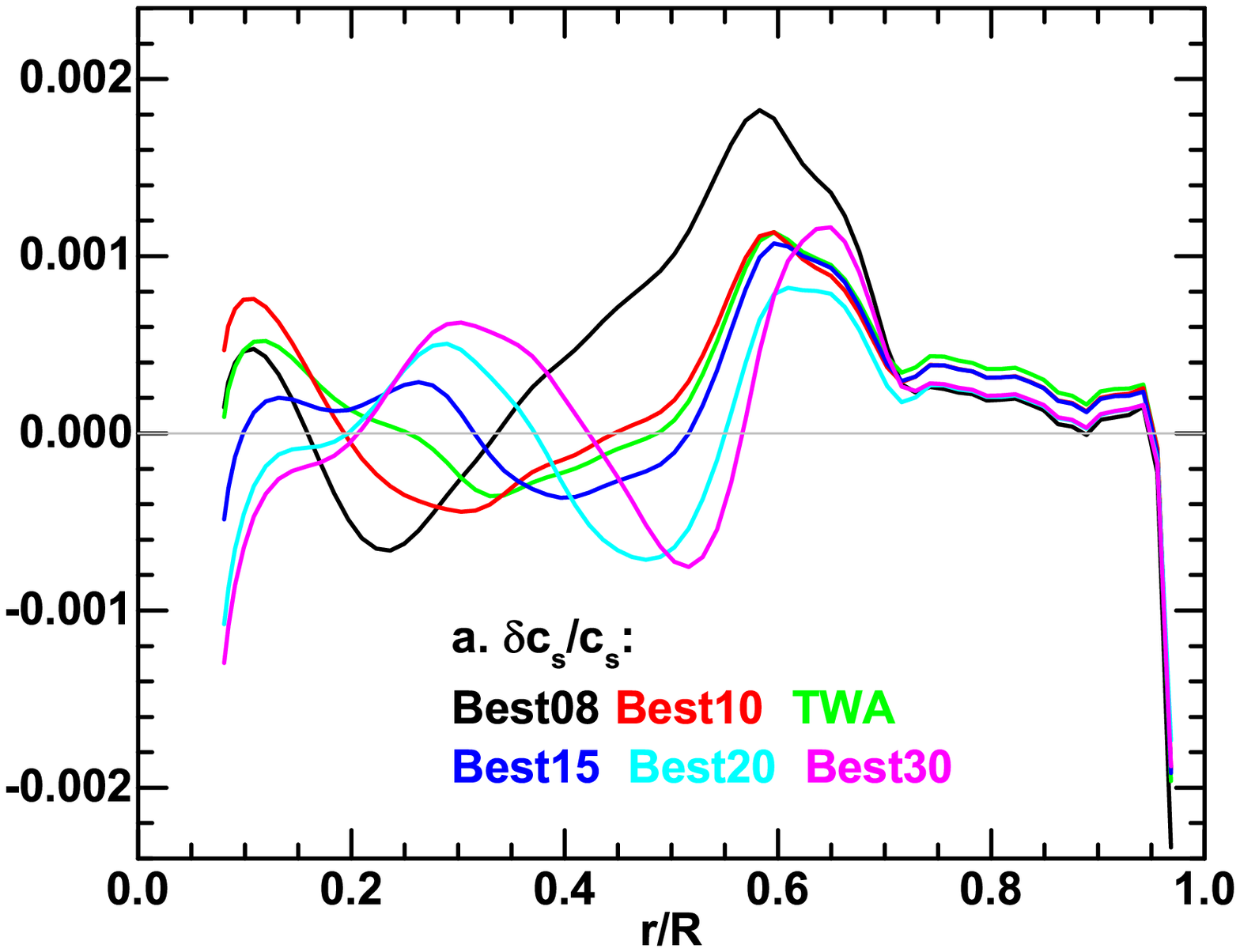}
\includegraphics[scale=0.5]{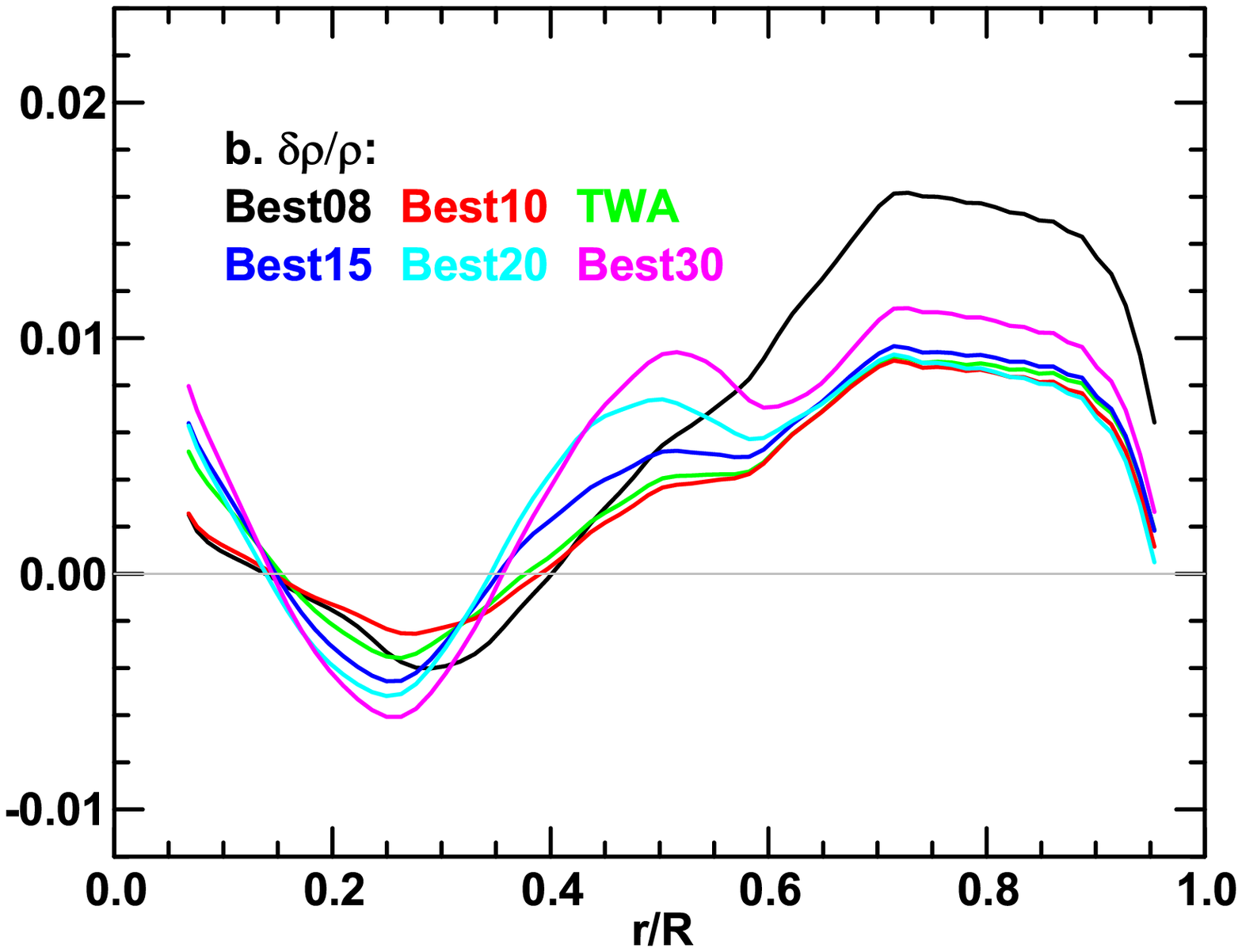}
\caption{Sound-speed (a) and density (b) deviations from helioseismic inversions (similar to Fig.\ref{std_c}) of some improved solar models.
}\label{Type3}
\end{figure}

Since the solar sound-speed profile strongly constrains solar models, we are mainly concerned with the sound-speed deviations. For those models with helium-poor accretion, the sound-speed profiles can be significantly improved. For the solar models with improved sound-speed profiles with r.m.s. deviation less than 0.1\%, their $^8$B neutrino fluxes are consistent with observation in the 1-$\sigma$ observational and theoretical uncertainty. Helioseismic inference on surface helium abundance suggested that $Y_{\rm{S}}=0.245-0.252$ \citep{ba04}. For each $\tau_{\rm{acc}}$, solar models with the best sound-speed profiles and in the required range of $Y_{\rm{S}}$ have been obtained by adjusting the model parameters $M_{\rm{L}}$ and $Y_{\rm{acc}}$ along the contours of $Y_{\rm{S}}=0.245$ in Fig.~\ref{Type2}. The information on the best model for each $\tau_{\rm{acc}}$ is shown in Table~\ref{Type2Best} and their sound-speed and density deviations are shown in Fig.~\ref{Type3}.The sound-speed and density profiles are significantly improved. The sound-speed deviations of those models with $\tau_{\rm{acc}}\geq10$Myr are basically less than 0.1\%. We therefore take Model TWA (with $\tau_{\rm{acc}}=12$Myr) as an example to investigate the reason for the improvements in the sound-speed profile.

\subsubsection{Evolution of the helium-abundance profile} \label{SecresultTWAHe}

\begin{figure}
\centering
\includegraphics[scale=0.5]{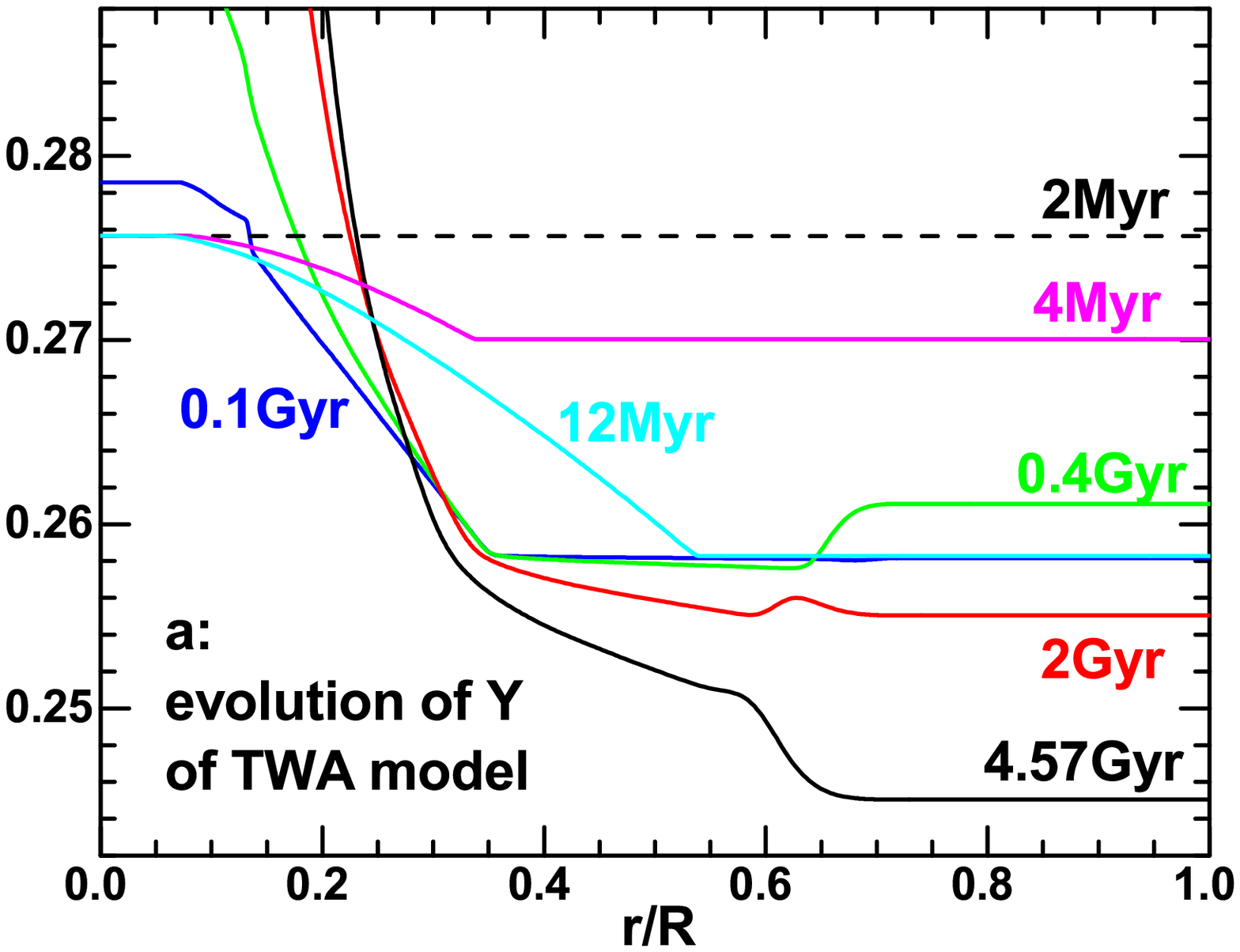}
\includegraphics[scale=0.5]{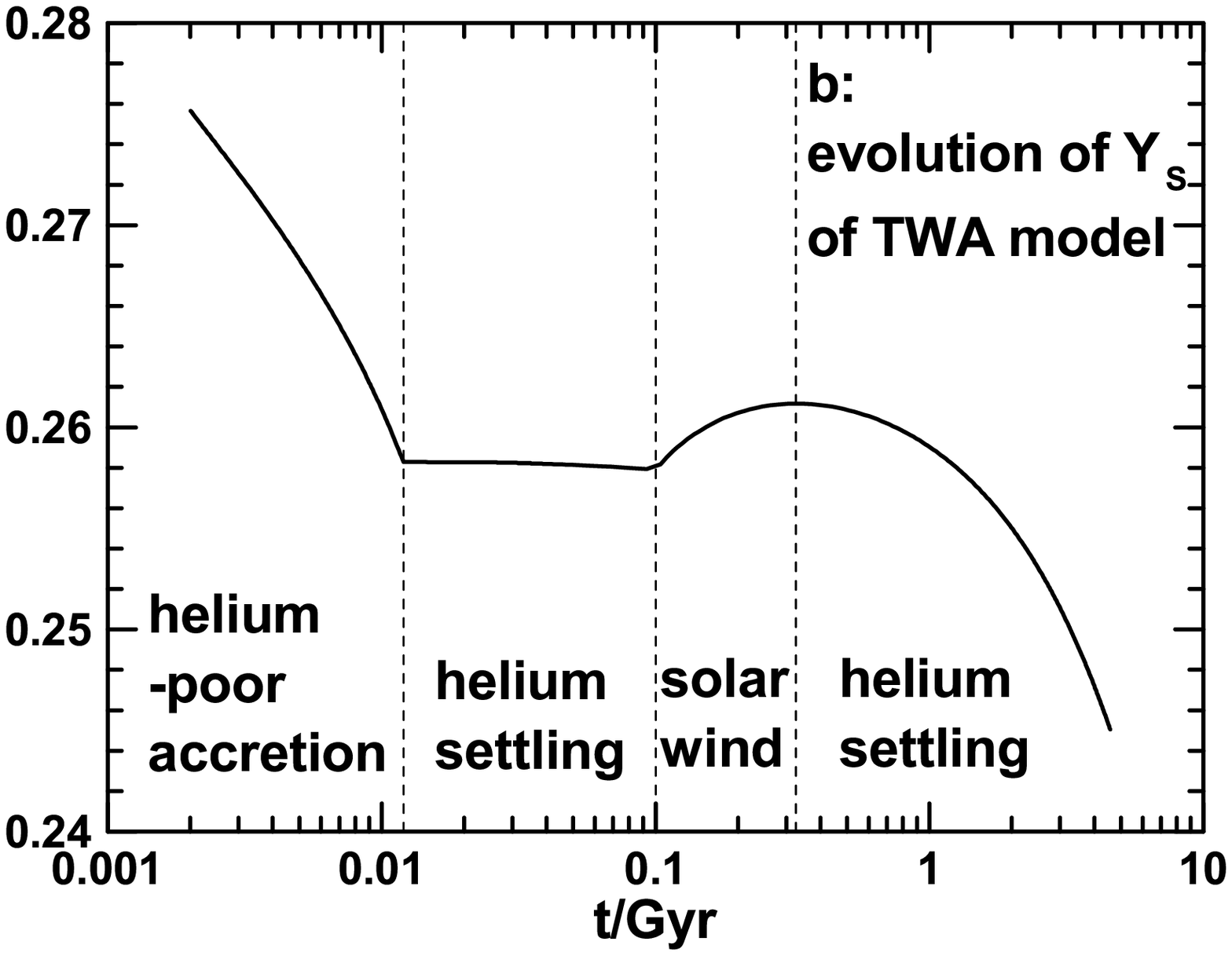}
	\caption{Evolution of helium abundance: a. helium profiles in Model TWA at various ages, b. surface helium abundance $Y_{\rm{S}}$ of Model TWA as a function of age. In panel b, the stages of the $Y_{\rm{S}}$ evolution are split by the vertical lines at age $12\,$Myr, $0.1\,$Gyr, and $0.3258\,$Gyr. The mechanism dominating the evolution of $Y_{\rm{S}}$ in each phase is indicated. }\label{Yevol}
\end{figure}

The difference between Model TWA and Model OV09Ne is the inclusion of inhomogeneous accretion and mass loss. The main effect of both processes is on the evolution of abundance profiles. Therefore it is natural to start the investigation of the properties of Model TWA from its evolution of abundance profiles. The evolution of the helium-abundance profile in the interior of solar Model TWA is shown in Fig.~\ref{Yevol}a. The Sun is fully convective before 2\,Myr and the helium abundance is homogeneous as shown with the dashed line. From 2\,Myr, the Sun is developing its radiative core and the convective envelope is retreating. The helium-poor accretion and the retreating convective envelope result in a negative helium-abundance gradient $dY/dr<0$ in the interior, shown as the purple ($4\,$Myr) and cyan ($12\,$Myr) lines. After the helium-poor accretion stops (at $12\,$Myr) and before the solar wind becomes active (at $0.1\,$Gyr), the effect of molecular diffusion is not significant owing to its long characteristic timescale; thus the mantle, which we define roughly as the radiative region with $0.3<r/R_{\odot}<0.7$, and the convective envelope stay homogeneous, shown as the blue line for the age of $0.1\,$Gyr. From $0.1\,$Gyr, the helium-poor mass loss caused by the solar wind is active; thus helium is concentrated in the convective envelope, and the helium abundance in the convective envelope is higher than that in the mantle, shown as the green line for the age of $0.4\,$Gyr. After about $0.33\,$Gyr, the mass loss decays and the helium settling caused by the molecular diffusion dominates the helium-abundance profile in the solar mantle and envelope, shown as red and black lines.

It is shown that there is a helium-abundance bump during the evolution of the helium-abundance profile (e.g., at about $r=0.6R_\odot$ for the 2Gyr case). It is caused by the combined effects of mass loss, molecular diffusion and convective overshoot mixing. Helium settling caused by molecular diffusion forces the region with $dY/dr>0$ in the helium profile, which is caused by the helium-poor mass loss, to move downward. Therefore the strength of overshoot mixing, which decays with depth, in that region becomes weaker and weaker and cannot completely remove the positive gradient on the helium profile. Although the helium settling can reduce the helium gradient, the effect is slow because its timescale is too long. Therefore the positive gradient in the helium profile remains on a timescale of $\sim 1$Gyr. On the other hand, the molecular diffusion causes a negative gradient in the helium profile below the convective overshoot region. Those lead to a helium-abundance bump at about $0.6R_\odot$. The helium-abundance bump leads to a region with $\nabla_{\mu} <0$, where $\nabla_{\mu}$ is defined in Eq.~(\ref{nablamu}), which results in thermohaline instability. Thermohaline mixing is not taking into account in the TWA solar model. Its effect will be discussed later.

The evolution of the helium abundance $Y_{\rm{S}}$ at the surface of Model TWA is shown in Fig.~\ref{Yevol}b. At the helium-poor accretion stage, $Y_{\rm{S}}$ is decreasing due to the accumulation of the helium-poor material in the convective envelope. After the end of the accretion and before the solar wind gets active, $Y_{\rm{S}}$ is slightly reduced by helium settling. From $0.1\,$Gyr, the helium-poor mass loss concentrates helium in the convective envelope so that $Y_{\rm{S}}$ is increasing. However, the solar wind significantly decreases with age. At about $0.33\,$Gyr, $Y_{\rm{S}}$ reaches its maximum and then decreases due to helium settling.

\subsubsection{Sound-speed profile} \label{SecresultTWAC}

\begin{figure}
\centering
\includegraphics[scale=0.5]{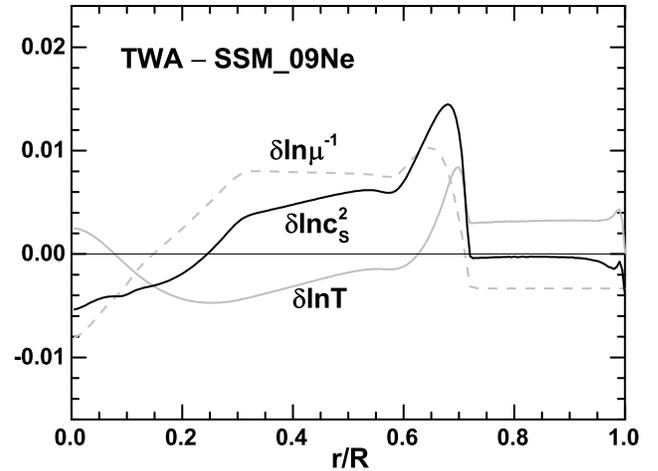}
\caption{Differences in squared sound speed, temperature, and mean molecular weight between solar models: TWA vs. SSM09Ne. }\label{C2vs}
\end{figure}
Assuming as an approximation an ideal gas it follows from Eq.~(\ref{soundspeed}) that the sound speed depends on temperature and abundances as:
\begin{eqnarray} \label{soundspeedapp}
{c^2_{\rm{s}}} \approx {\Gamma _1} \frac{k_{\rm{B}}}{m_{\rm{u}}} \frac{T}{\mu },
\end{eqnarray}%
where $k_{\rm{B}}$ is Boltzmann's constant, ${m_{\rm{u}}}$ is the atomic mass unit, and $\mu$ is the mean molecular weight. The differences of sound speed, temperature, and mean molecular weight between Model TWA and Model SSM09Ne are shown in Fig.~\ref{C2vs}, which helps to understand how the sound speed of Model TWA is improved.

In the convection zone, $Y_{\rm{S}}$ of Model TWA is 0.245, higher than for Model SSM09Ne. This leads to a higher mean molecular weight, shown as the dashed line. Accordingly, the temperature in the convection zone of Model TWA is also higher than that of Model SSM09Ne because the sound speed in the convection zone is well defined by the hydrostatic equation and the polytropic relation \citep[e.g.,][]{JCD86}; thus Model TWA requires higher temperature to compensate its higher mean molecular weight. The direct reason for the higher temperature in the convection zone of Model TWA is as follows. The higher mean molecular weight leads to a higher density to maintain the pressure; thus the pressure scale height $H_P$ is smaller than for Model SSM09Ne. This leads to a higher $ | d \ln T/dr | $ in Model TWA because $ | d \ln T/dr | = \nabla / H_P$.

In the solar radiative mantle the mean molecular weight in Model TWA is lower than in Model SSM09Ne because the PMS helium-poor accretion leaves a helium-poor solar mantle. Below the BCZ, the bumps of the temperature difference $\delta \ln T$ around $0.6<r/R_\odot<0.7$ are caused by the negative turbulent kinetic energy flux enhancing the temperature gradient. Going downward, the differences in the temperature decreases. The lower mean molecular weight and the higher temperature in Model TWA result in a higher sound speed in the solar radiative mantle than in Model SSM09Ne.

In the solar core, the main contribution to $\delta \ln c^2_{\rm{S}}$ is $\delta \ln \mu$. The helium abundance in the core in Model TWA is higher than Model SSM09Ne due to the helium-poor accretion leaving a negative helium-abundance gradient. Thus the mean molecular weight is higher than in Model SSM09Ne, resulting in a lower sound speed.

The overall effect is that the sound speed in the mantle of Model TWA is higher than in Model SSM09Ne and the sound speed in the core of Model TWA is lower than in Model SSM09Ne. This almost exactly compensates for the differences of sound speed between Model SSM09Ne and helioseismic inferences. As shown in Fig.~\ref{Type3}a, the maximum deviation of the sound speed in Model TWA is only about $0.1$\% in $r<0.95R_\odot$. Comparing the temperature and mean molecular weight modifications with the sound-speed modification, it can be found that the modification of mean molecular weight in the solar radiative interior is the main factor for the improvements in the model sound-speed profile and the temperature modification is significant only in the overshoot region.

Once the density profile is given, the mass profile can be determined by integrating the density profile, and then the pressure profile can be determined by integrating the hydrostatic equation. Since $\Gamma_1$ is almost constant in most of the solar interior, according to Eq.~(\ref{soundspeed}) the sound-speed profile is also determined, and hence the sound-speed profile is related to the density profile. Therefore the density profile of Model TWA is also in good agreement with helioseismic inferences at the level of less than 1\%, as shown in Fig.~\ref{Type3}b.

\begin{figure}
\centering
\includegraphics[scale=0.5]{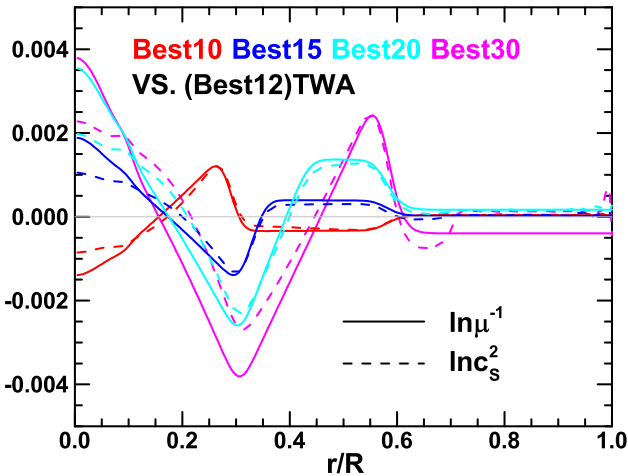}
\caption{ Differences in squared sound speed and mean molecular weight between improved solar models (Best10, Best15, Best20 and Best30) and Model TWA, in the sense (Model Bestxx) -- (Model TWA).
}\label{T3Vs}
\end{figure}

\begin{figure}
\centering
\includegraphics[scale=0.5]{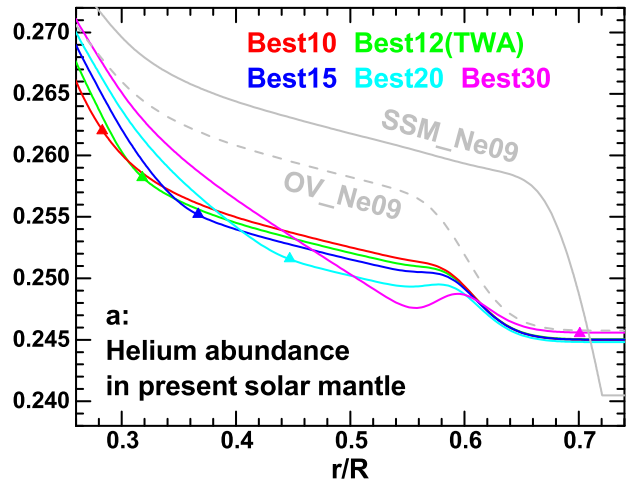}
\includegraphics[scale=0.5]{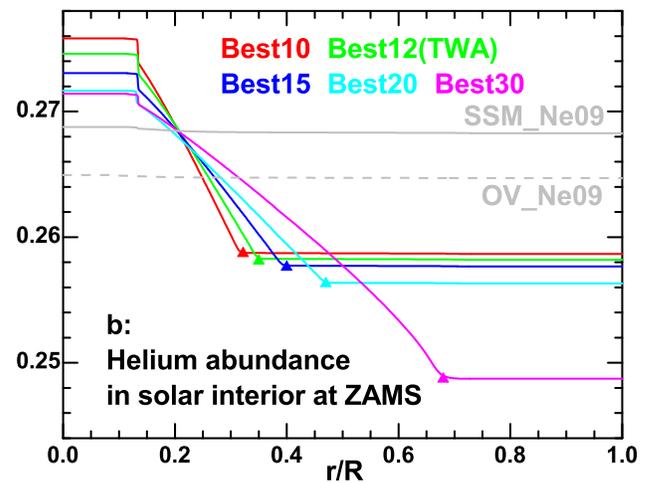}
	\caption{ Helium-abundance profiles of improved solar models: a: for models at the present solar age, b. for models at ZAMS. The triangles denote the locations corresponding to the locations in mass of the BCZ when the PMS accretion stops for each model.
}\label{T3Yprofile}
\end{figure}

Differences in sound speed and mean molecular weight between other improved models with different $\tau_{\rm{acc}}$ (Best10, Best15, Best20 and Best30) and Model TWA are shown in Fig.~\ref{T3Vs}. It is found that the main factor contributing to the sound-speed differences is the mean molecular weight, which implies that the main factor contributing to the improvements on their sound-speed profiles is also the mean molecular weight, as for Model TWA. Since the differences in mean molecular weight represent the differences in the model helium-abundance profiles, the result that the sound-speed profiles of those improved models are consistent with helioseismic inferences could help to investigate the solar helium-abundance profile.

The helium-abundance profiles of improved solar models and Models OV09Ne and SSM09Ne at the present solar age are shown in Fig.~\ref{T3Yprofile}a. The improved models have steeper abundance profiles in the solar core and a helium-reduced solar mantle. Both distinctions result from the inhomogeneous PMS accretion. Since the accretion stops before ZAMS and other mechanisms to affect element abundances (i.e., nuclear reactions and diffusion) have little effect before ZAMS, the effects of the inhomogeneous PMS accretion will be clear when we investigate the abundance profile of ZAMS models. The helium-abundance profiles of the improved solar models (Best10, TWA, Best15, Best20 and Best30), and Models OV09Ne and SSM09Ne are shown in Fig.~\ref{T3Yprofile}b. At the ZAMS, Models OV09Ne and SSM09Ne have nearly homogeneous helium-abundance profiles, and each improved model has a helium-poor envelope and a helium-abundance gradient region between the convective core and the envelope, resulting from the helium-poor accretion and the retreat of the convective envelope. Because the accretion rates of those models are the same, the helium-abundance gradient is determined by $Y_{\rm{acc}}$, such that a lower $Y_{\rm{acc}}$ leads to a steeper gradient. The boundary of the helium-abundance gradient for each ZAMS model is the location of the BCZ when the accretion stops, denoted by triangles in Fig.~\ref{T3Yprofile}a. The helium-abundance gradient regions of the improved models at ZAMS in Fig.~\ref{T3Yprofile}b correspond to their steeper abundance profiles in the solar core in Fig.~\ref{T3Yprofile}a and their helium-poor envelopes at ZAMS in Fig.~\ref{T3Yprofile}b correspond to the helium-poor mantles in Fig.~\ref{T3Yprofile}a, separated by the triangles.

The differences in the helium-abundance profile among the improved models are not significant (less than 0.003 in most part of the solar mantle). A characteristic of those models is that their helium abundance in most of the mantle (for $0.3<r/R_{\odot}<0.65$) is about 0.01 less than for Model SSM09Ne. Since those models are selected from many models with varied parameters, it implies that a solar model with this characteristic could have an improved sound-speed profile.

\subsubsection{Neutrino fluxes} \label{SecresultTWAFi}

Beside helioseismology, observations of the solar neutrino fluxes place constraints on the properties of the solar core. As shown in Table~\ref{modelinfo}, when the uncertainties of observations and models are taken into account, the $pp$ chain neutrino fluxes of Models SSM09Ne, OV09Ne and the improved models are all in acceptable ranges compared with observations. Since $^7{\rm{Be}}$ and $^8{\rm{B}}$ neutrino fluxes more strongly depend on temperature and the luminosity of solar models is calibrated, higher $T_{\rm{C}}$ leads to higher $^7{\rm{Be}}$ and $^8{\rm{B}}$ neutrino fluxes and lower $pp$ and $pep$ neutrinos fluxes \citep[see, e.g.,][]{Bah96,ser11}, which is consistent with the relation between $T_{\rm{C}}$ and neutrino fluxes shown in Tables \ref{modelinfo} and \ref{Type2Best}.

\subsubsection{Lithium abundance} \label{SecresultTWALi}

\begin{figure}
\centering
\includegraphics[scale=0.5]{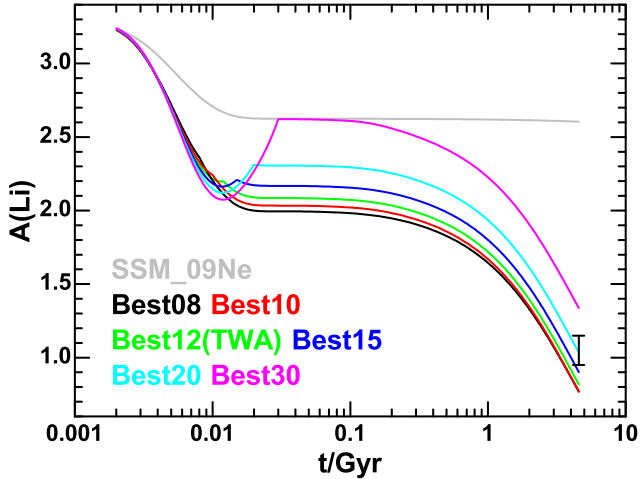}
\caption{The evolution of Li abundances of the SSM and improved solar models. The black error bar at solar age is the solar $\rm{A(Li)}$ taken for \citet{AGSS09}. }\label{ALiEvol}
\end{figure}

The observed solar Li abundance provides another constraint on the evolution and structure of solar models. The solar Li depletion is thought to be related to mixing below the solar convection zone \citep[e.g.,][]{Schlattl99,xio02,xio09,Z12Li}. The Li abundances of the improved models show significant depletion compared with Model SSM09Ne and are close to the observation as shown in Table~\ref{modelinfo}. This is because we set an appropriate value for the overshoot mixing parameter $C_{\rm{X}}$. The evolution of the Li abundance of the SSM and improved models is shown in Fig.~\ref{ALiEvol}. There are two main differences between improved models and the SSM: improved models show more Li depletion in the PMS stage (i.e., at an age less than about 0.01\,Gyr), and it shows Li depletion in the MS stage but the SSM does not. Both effects are due to convective overshoot: the former is because the negative turbulent kinetic energy flux makes BCZ deeper \citep{TKFZ14} and the latter is because of the overshoot mixing.

Figure~\ref{ALiEvol} shows that the Li abundance is increasing at the later stage of the PMS accretion. Lithium is not depleted in the accreted material, and the convective envelope is retreating so that the mass of the envelope is decreasing. Therefore the contribution to the envelope of the accreted material with no Li depletion is increasing and results in a Li-enrichment in the convective envelope. A model with larger $\tau_{\rm{acc}}$ shows more Li-enrichment in the envelope because its mass of the envelope at the later stage of accretion is less due to the retreat of the convective envelope; thus this Li-enrichment mechanism is more efficient.

\subsection{Effects of parameters of accretion and mass loss} \label{SecresultPara}

\subsubsection{The strength of the accretion rate} \label{SecresultDisp}

We have calculated about 2000 solar models to investigate the effects of the strength of the accretion rate. This is varied by multiplying the accretion rate given in Eq.~(\ref{AccrRate}) by a factor $10^{\eta}$, corresponding to the 0.5dex dispersion of the formula; ${\eta}$ varies in the range from $-0.5$ to $0.5$ with a step $0.1$. $Y_{\rm{acc}}$ varies in the range from $0.00$ to $0.26$ with a step $0.02$. Seven sample points of $M_{\rm{L}}$ are considered: $0$, $0.004$, $0.007$, $0.010$, $0.015$, $0.020$ and $0.030$ solar mass. The default values $\lambda_{Z}=1$ and $Z_{\rm{acc}}=0.015$ are adopted. Two cases of $\tau_{\rm{acc}}$, i.e., 12Myr and 20Myr, have been investigated.

We introduce a `lacking helium mass' defined as follows:
\begin{eqnarray} \label{MYlack}
m_{Y,\rm{lack}}=\frac{1-X_0-Z_0-Y_{\rm{acc}}}{X_0+Z_0+Y_{\rm{acc}}} m_{\rm{acc}}(t),
\end{eqnarray}%
which represents the lack of the mass of helium at age $t$ during the helium-poor accretion comparing with a homogeneous accretion. $m_{\rm{acc}}(t)$ is the accreted mass at age $t$. The total accreted mass ${M_{\rm{acc}}}=m_{\rm{acc}}(\tau_{\rm{acc}})$ satisfies,
with $\tau _{\rm{acc}}$ and $t$ in Myr,
\begin{eqnarray} \label{Macc2}
&&\frac{{{M_{\rm{acc}}}}}{{{M_{\odot}}}} = 1 + \frac{{{M_{\rm{L}}}}}{{{M_{\odot}}}} - \\ \nonumber
&&\left[\left(1 + \frac{{{M_{\rm{L}}}}}{{{M_{\odot}}}}\right)^{1.1} - 0.607 \times {10^\eta }\left(\frac{1}{{{\tau _{\rm{acc}}}}^{0.07}} - \frac{1}{{{2^{0.07}}}}\right)\right]^{ - \frac{1}{{1.1}}},
\end{eqnarray}%
which is derived from the adopted accretion rate Eq.~(\ref{AccrRate}), the presence of the solar wind mass loss, and final mass equaling to the solar mass. The total `lacking helium mass' is defined by $M_{Y,\rm{lack}}=m_{Y,\rm{lack}}(t=\tau_{\rm{acc}})$. Because $\dot{M} \propto t^{-1.07}$, we obtain
\begin{eqnarray} \label{macct}
{m_{\rm{acc}}} = {M_{\rm{acc}}}\frac{{{t^{ - 0.07}} - {2^{ - 0.07}}}}{{{\tau _{\rm{acc}}}^{ - 0.07} - {2^{ - 0.07}}}}
\end{eqnarray}%
and
\begin{eqnarray} \label{mylackt}
M_{Y,\rm{lack}}^{-1}\frac{{d{m_{Y,\rm{lack}}}}}{{dt}} =
M_{\rm{acc}}^{-1}\frac{{d{m_{\rm{acc}}}}}{{dt}} = \frac{{ - 0.07{t^{ - 1.07}}}}{{{\tau _{\rm{acc}}}^{ - 0.07} - {2^{ - 0.07}}}} \; .
\end{eqnarray}%

\begin{figure*}
\centering
\includegraphics[scale=0.5]{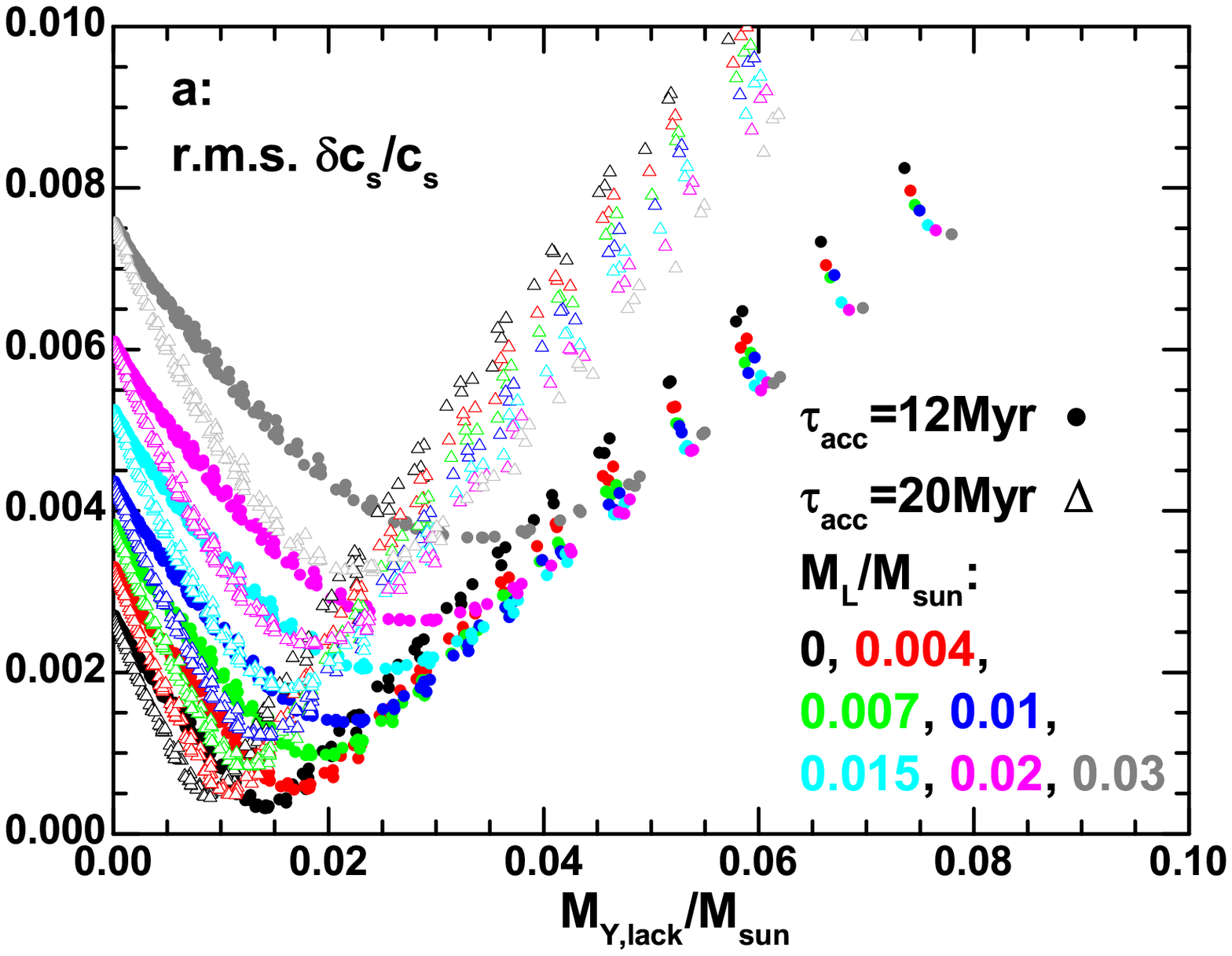}
\includegraphics[scale=0.5]{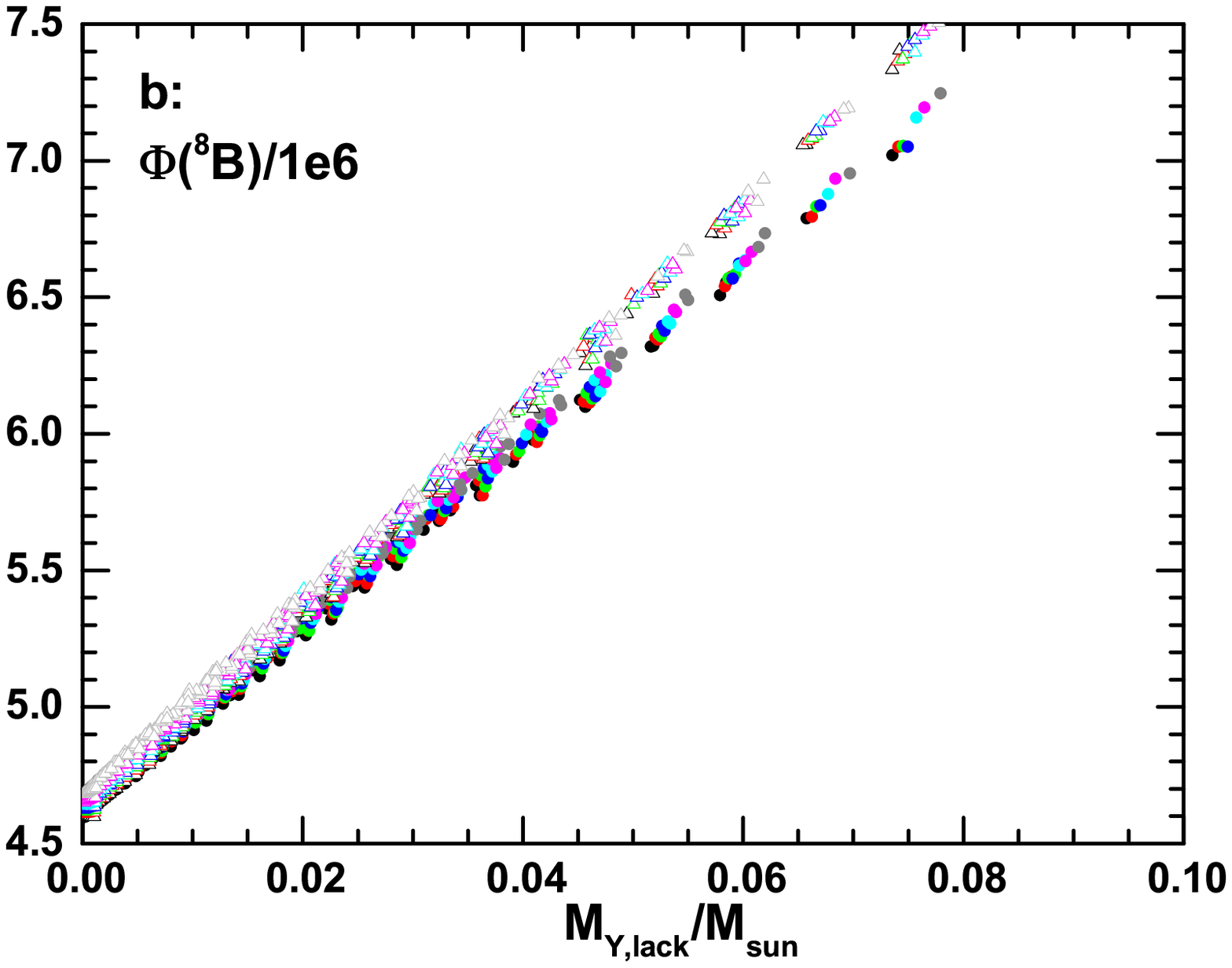}
\includegraphics[scale=0.5]{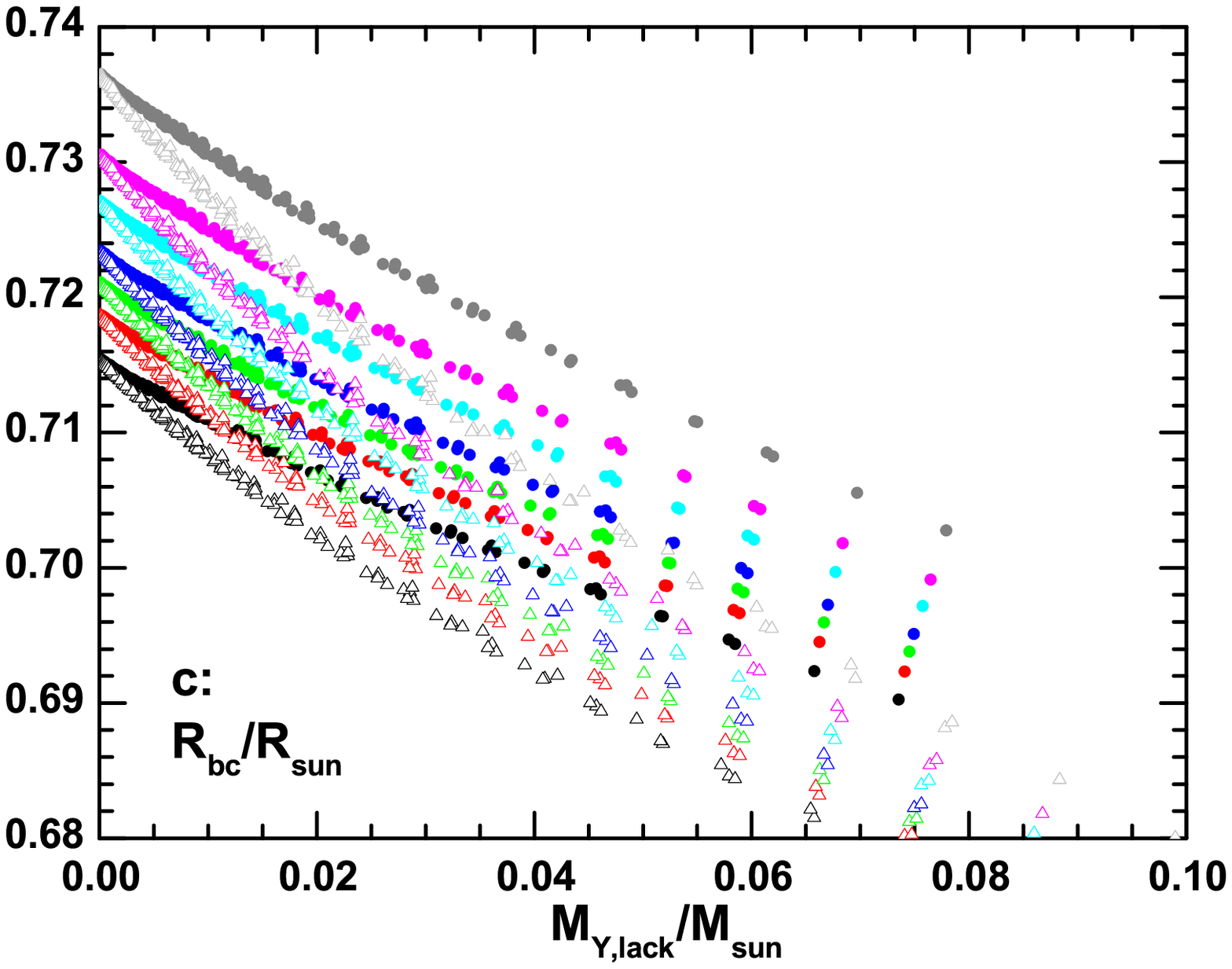}
\includegraphics[scale=0.5]{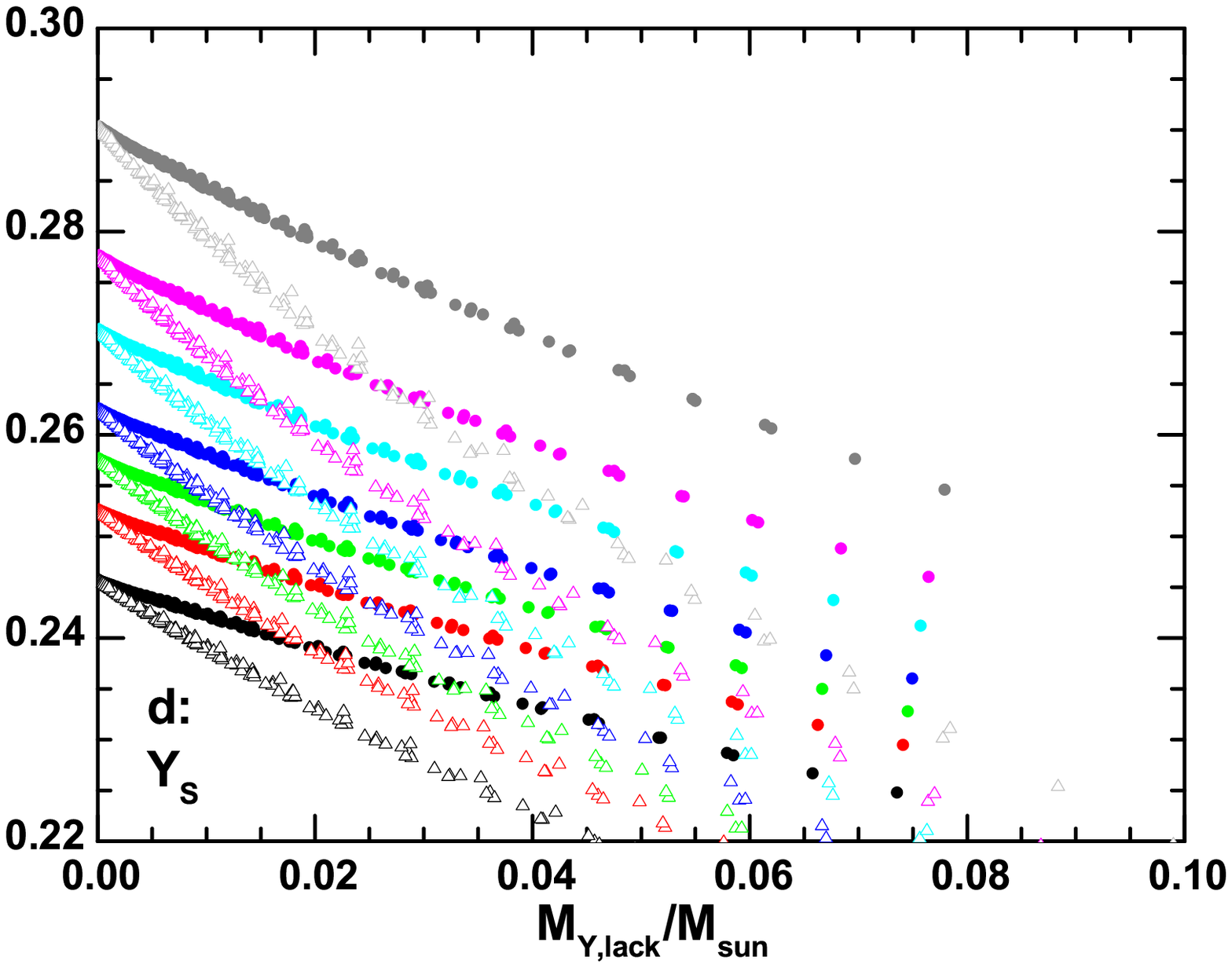}
\caption{ Properties of solar models with different strength of accretion rate, $Y_{\rm{acc}}$ and mass loss:
a. rms deviations of sound speed $\sqrt{\langle(\delta c_{\rm{s}}/c_{\rm{s,ref}})^2\rangle}$ from the solar structure inferred using Model TWA as reference, b. the $^8$B neutrino flux $\Phi(\rm{^8B})$, c. the location of the BCZ $R_{\rm{bc}}$, d. the surface helium abundance $Y_{\rm{S}}$. Filled circles are for models with $\tau_{\rm{acc}}=12\,$Myr and empty triangles are for models with $\tau_{\rm{acc}}=20\,$Myr. The total mass loss $M_{\rm L}$ is indicated by the color of the symbols. The factor $\eta$ multiplying the accretion rate and $Y_{\rm acc}$ are not shown for each models. They are represented in the abscissas by the total `lacking helium mass' $M_{Y,\rm{lack}}$ determined by $\eta$ and $Y_{\rm acc}$ through Eqs~(\ref{MYlack}) -- (\ref{mylackt}) with $t=\tau_{\rm acc}$.
}\label{Type1}
\end{figure*}

For those solar models with given $\tau_{\rm{acc}}$ and $M_{\rm{L}}$, there are two free parameters: ${\eta}$, $Y_{\rm{acc}}$ . Therefore the properties of each of those models should be determined by the values of the two parameters. However, as shown by the symbols in each color in Fig.~\ref{Type1}, if the total `lacking helium mass' is adopted as the independent variable, the properties of solar models show little dispersion. Therefore the effect of ${\eta}$ and $Y_{\rm{acc}}$ can be combined in $M_{Y,\rm{lack}}$. This can be explained as follows. For a ZAMS stellar model, nuclear fusion dominates the energy release; thus the gravitational and thermal energy release which is the only time-dependent term in stellar structure equations is negligible. Therefore the structure of the ZAMS stellar model is mainly determined by its interior abundance profiles and it basically cannot remember the details of its PMS stage. For a given $\tau_{\rm{acc}}$, Eq.~({\ref{mylackt}}) shows that the evolution of the lack of helium mass $d{m_{Y,\rm{lack}}}/dt$ is determined only by $M_{Y,\rm{lack}}$.
Because the accreted material is mixed with the material in the convective envelope, the variation $\Delta Y_{\rm{S}}$ of the helium abundance in the convective envelope during a time step $\Delta t$ is determined by the adding of `lacking helium mass' in the time step such that $\Delta Y_{\rm{S}}/\Delta t \approx -(d{m_{Y,\rm{lack}}}/dt)/m_{\rm{CZ}}$. Therefore the evolution of $Y_{\rm{S}}$ in the accretion stage is determined mainly by $M_{Y,\rm{lack}}$ since $d{m_{Y,\rm{lack}}}/dt$ is determined only by $M_{Y,\rm{lack}}$ and $m_{\rm{CZ}}$ is insensitive to the accretion. Since the convective envelope retreats in the PMS stage,
the variation of helium abundance in the convective envelope will determine the helium abundance profile in the radiative region below the convective envelope. Therefore the structure of the ZAMS models are mainly determined by $M_{Y,\rm{lack}}$ for a given $\tau_{\rm{acc}}$.

An exceptional model property for which the effects of ${\eta}$ and $Y_{\rm{acc}}$ cannot be combined into $M_{Y,\rm{lack}}$ is the surface lithium abundance of solar models. This is because lithium in the convective envelope is significantly depleted in the PMS stage due to the high temperature at the BCZ. Since the accretion refreshes lithium in the convective envelope, the evolution of lithium abundance is directly affected by the strength of accretion rate described by ${\eta}$ but not by $M_{Y,\rm{lack}}$.

\begin{table}
\centering
\caption{ Basic information of some solar models with different $\eta$ comparing with Model TWA. }\label{TWA_eta_comp}
\begin{tabular}{lcccc}
\hline\noalign{\smallskip}
                                        & E1       & TWA      & E2      & E3   \\
\hline\noalign{\smallskip}
$\alpha_{\rm{MLT}}$                     & 2.3651   & 2.3708   & 2.3739  & 2.3692  \\
$X_{\rm{0}}$                            & 0.7105   & 0.7096   & 0.7088  & 0.7080  \\
$Z_{\rm{0}}$                            & 0.01472  & 0.01472  & 0.01471 & 0.01466 \\
$X_{\rm{C}}$                            & 0.3493   & 0.3485   & 0.3482  & 0.3487  \\
$Z_{\rm{C}}$                            & 0.01572  & 0.01572  & 0.01571 & 0.01567 \\
$\log T_{\rm{C}}$                       & 7.1924   & 7.1925   & 7.1926  & 7.1924  \\
$\log  \rho_{\rm{C}}$                   & 2.1849   & 2.1857   & 2.1861  & 2.1857  \\
$\tau_{\rm{acc}}/{\rm{Myr}}$            & 12       & 12       & 12      & 12      \\
$\eta$                                  & -0.17    & 0.00     & 0.20    & 0.50    \\
$M_{\rm{acc}}/M_{\odot}$                & 0.0401   & 0.0585   & 0.0897  & 0.1635  \\
$Y_{\rm{acc}}$                          & 0.000    & 0.070    & 0.132   & 0.193   \\
$M_{Y,\rm{lack}}/M_{\odot}$             & 0.01519  & 0.01515  & 0.01512 & 0.01507 \\
$Y_{\rm{S}}$                            & 0.2453   & 0.2450   & 0.2449  & 0.2451  \\
$(Z/X)_{\rm{S}}$                        & 0.0188   & 0.0188   & 0.0188  & 0.0188  \\
$ R_{\rm{bc}}/R_{\odot}$                & 0.7114   & 0.7110   & 0.7108  & 0.7111  \\
A(Li)                                   & 0.79     & 0.82     & 0.87    & 1.03    \\
\hline
neutrino fluxes  & & & &   \\
in $({\rm{cm}}^{-2}{\rm{s}}^{-1})$ & & & &   \\
\hline
$pp$ $(10^{10})$                        & 5.98     & 5.98     & 5.97    & 5.98 \\
$pep$ $(10^{8})$                        & 1.47     & 1.47     & 1.47    & 1.47 \\
$hep$ $(10^{3})$                        & 8.13     & 8.13     & 8.12    & 8.13 \\
$^7$Be $(10^{9})$                       & 4.83     & 4.84     & 4.85    & 4.83 \\
$^8$B $(10^{6})$                        & 5.10     & 5.13     & 5.14    & 5.11 \\
$^{13}$N $(10^{8})$                     & 2.18     & 2.19     & 2.19    & 2.18 \\
$^{15}$O $(10^{8})$                     & 1.62     & 1.63     & 1.63    & 1.62 \\
$^{17}$F $(10^{6})$                     & 3.57     & 3.59     & 3.59    & 3.56 \\
\hline
\end{tabular}
\label{tab:etamod}
\tablecomments{The accretion rate has been multiplied by a factor $10^\eta$ (cf.\ Eq.~\ref{Macc2}). See notes for Table~\ref{tab:basmod}.}
\end{table}

\begin{figure}
\centering
\includegraphics[scale=0.5]{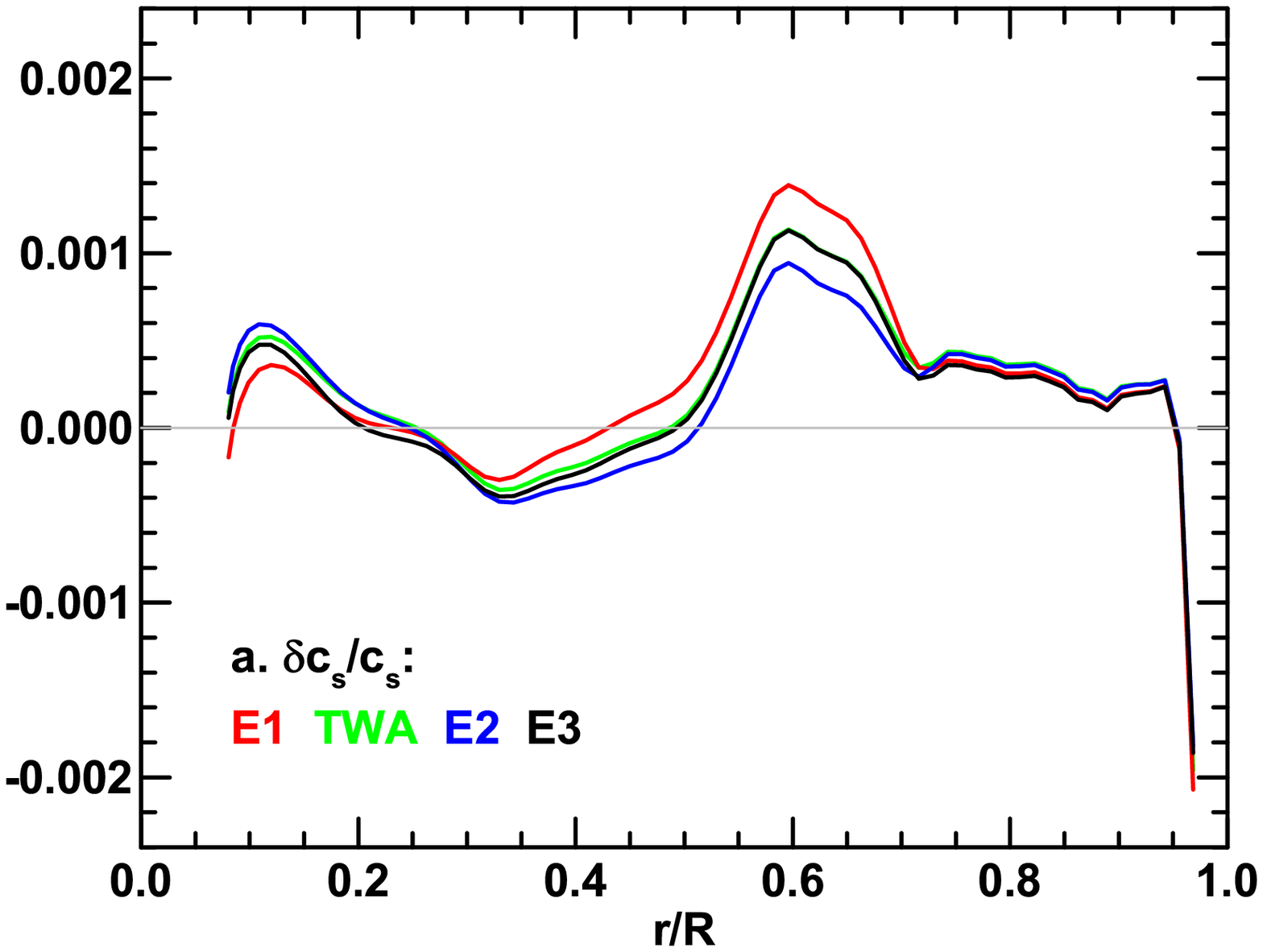}
\includegraphics[scale=0.5]{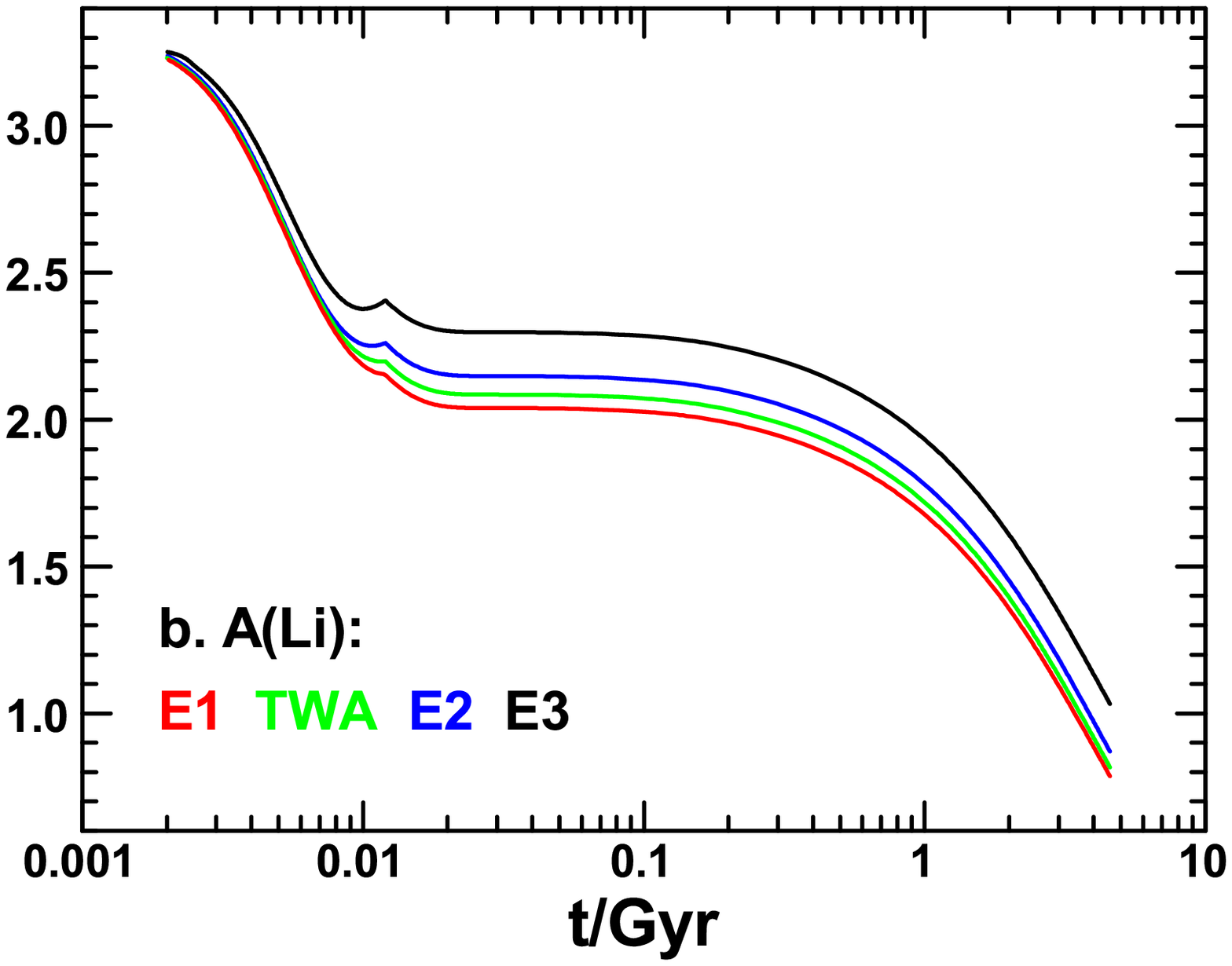}
\caption{Sound-speed deviations from helioseismic inversions (similar to Fig.\ref{std_c}a) (a) and lithium-abundance evolution (b) for some solar models with different accretion rates characterized by $\eta$ (see Table~\ref{tab:etamod}) comparing with Model TWA.
}\label{TWA_eta}
\end{figure}

For example, three solar models with the same $M_{Y,\rm{lack}}$, $M_{\rm{L}}$, and $\tau_{\rm{acc}}$ as Model TWA but different ${\eta}$ and $Y_{\rm{acc}}$ are compared with Model TWA. The information on those models is listed in Table~\ref{TWA_eta_comp} and their sound-speed deviations and lithium-abundance evolution are shown in Fig.~\ref{TWA_eta}. Their sound-speed deviations are very close to that of Model TWA and the differences in the deviations between those models and Model TWA are less than 0.03\%, implying that their interior structures are very close to that of Model TWA. On the other hand, lithium abundances of those models show significant differences. For a model with stronger accretion rate (a larger $\eta$), the effect of the accretion of material with no Li depletion refreshing the lithium abundance is more significant.

Since the parameters ${\eta}$ and $Y_{\rm{acc}}$ affect most of the model properties only by their combination function $M_{Y,\rm{lack}}$, an enhancement of the strength of the accretion rate can be compensated by an enhancement of the accreted helium abundance, keeping a fixed $M_{Y,\rm{lack}}$. Therefore, we can eliminate a free parameter and then reduce the amount of required solar models by setting ${\eta}=0$ without loss of generality. It should be mentioned that setting a fixed ${\eta}$ may reduce the range of $M_{Y,\rm{lack}}$ because of $Y_{\rm{acc}} \geq 0$.

\subsubsection{The metallicity of the accreted material} \label{SecresultZacc}

In order to investigate the effects of varied $Z_{\rm{acc}}$, we have calculated about 1800 solar models with ${\eta}=0$, $\lambda_{Z}=1$, $\tau_{\rm{acc}}=12$Myr and different $Z_{\rm{acc}}$, $M_{\rm{L}}$ and $Y_{\rm{acc}}$. $Y_{\rm{acc}}$ varies in the range from $0.00$ to $0.26$ with a step $0.02$. $M_{\rm{L}}$ varies in the range from $0.00$ to $0.01$ with a step $0.001$. $Z_{\rm{acc}}$ varies in the range from $0.005$ to $0.025$ with a step $0.002$. The cases of $Z_{\rm{acc}}=0.010$ and $Z_{\rm{acc}}=0.020$ are also calculated.

\begin{figure*}
\centering
\includegraphics[scale=0.5]{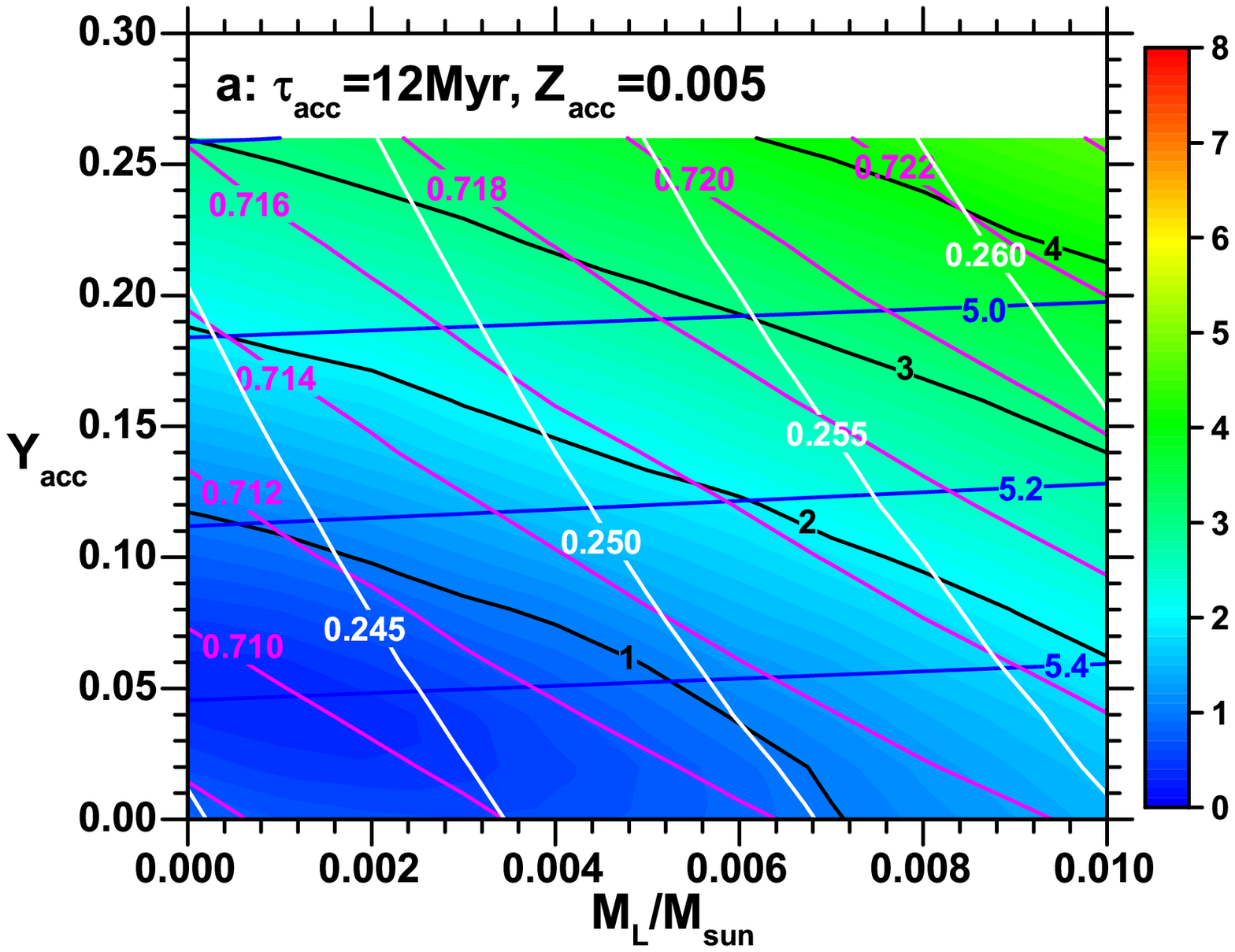}
\includegraphics[scale=0.5]{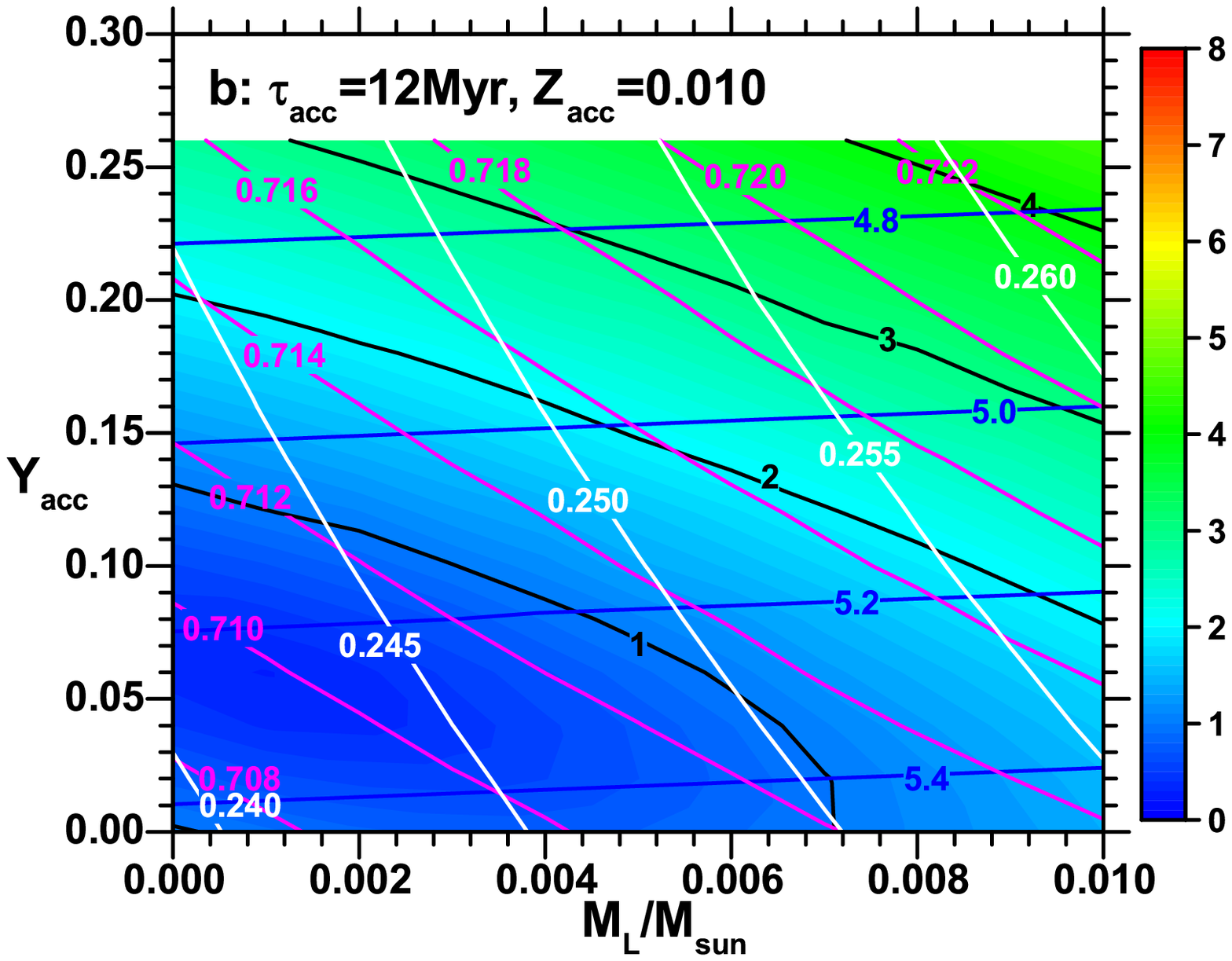}
\includegraphics[scale=0.5]{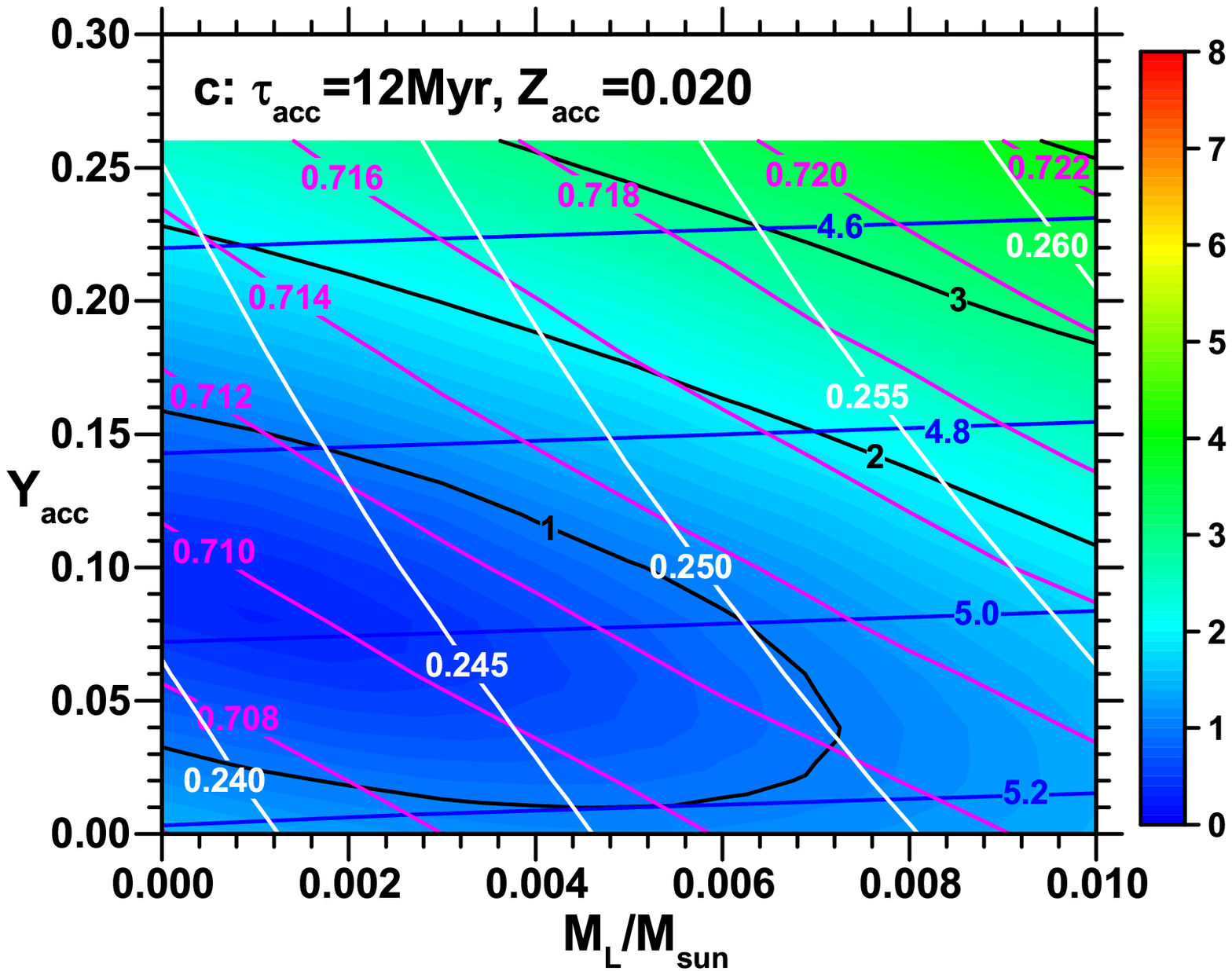}
\includegraphics[scale=0.5]{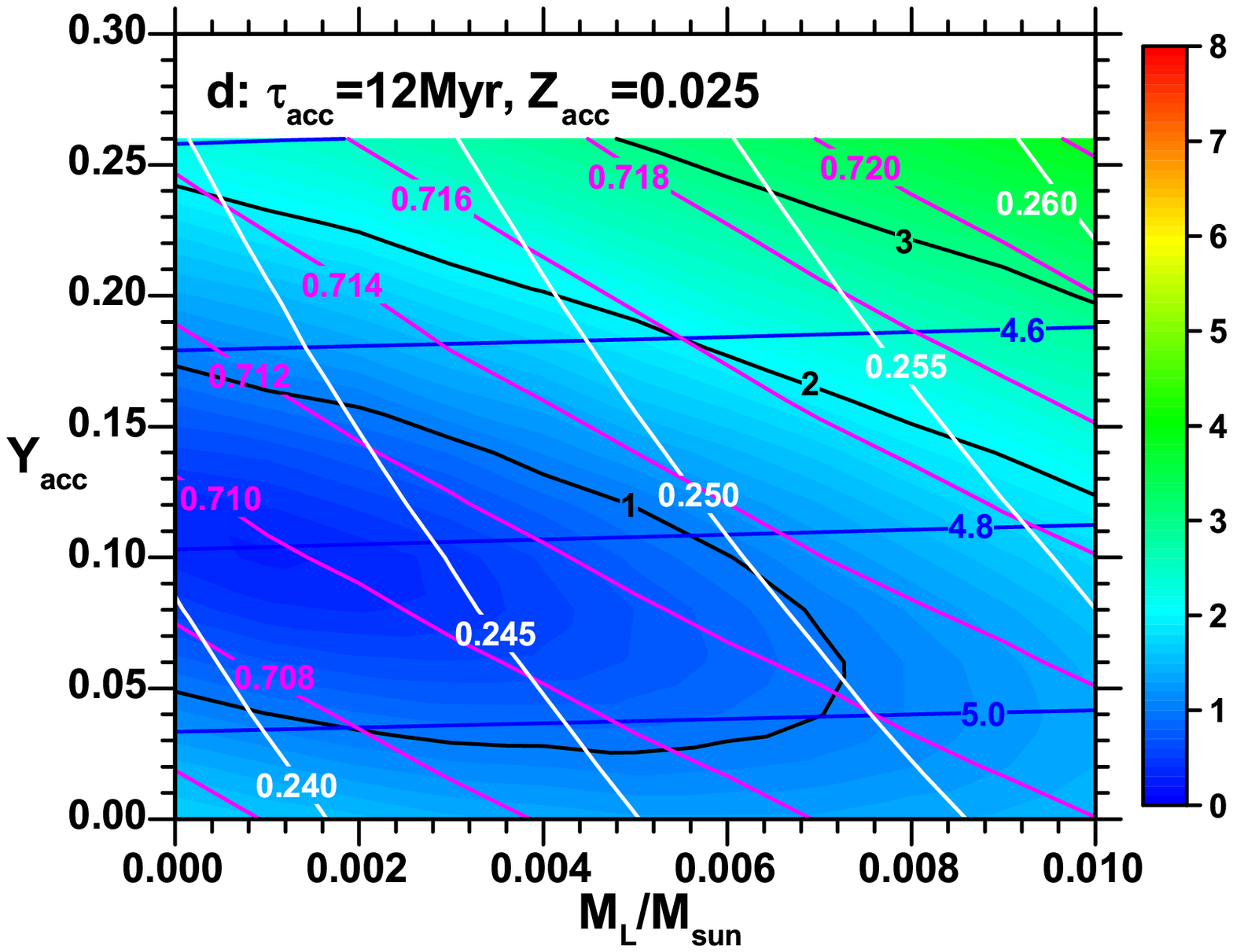}
	\caption{ Similar to Fig.~\ref{Type2}c, but for solar models with different heavy-element abundance $Z_{\rm{acc}}$ of the accreted material.
}\label{Type4}
\end{figure*}

Properties (sound-speed deviations, $^8$B neutrino flux, location of BCZ, and surface helium abundance) of those solar models with $Z_{\rm{acc}}=0.005$, $Z_{\rm{acc}}=0.010$, $Z_{\rm{acc}}=0.020$ and $Z_{\rm{acc}}=0.025$ are shown in Fig.~\ref{Type4}. The case of $Z_{\rm{acc}}=0.015$ is shown in Fig.~\ref{Type2}c. Solar models with other $Z_{\rm{acc}}$ are not shown. The properties of those solar models are continuously varied from $Z_{\rm{acc}}=0.005$ to $Z_{\rm{acc}}=0.025$. Comparing Fig.~\ref{Type4} with Fig.~\ref{Type2}c, it is found that there are two main effects of varied $Z_{\rm{acc}}$. One is that a higher $Z_{\rm{acc}}$ significantly reduces the $^8$B neutrino flux of solar models. The second is that the required degree of helium depletion in PMS accretion for improving the helioseismic quantities of solar models is reduced when a higher $Z_{\rm{acc}}$ is used. Now, we investigate the reasons for these effects.

\begin{table*}
\centering
\caption{ Basic information of some solar models with different $\lambda_{Z}$ or $Z_{\rm{acc}}$ comparing with Model TWA. }\label{TWA_z_comp}
\begin{tabular}{lccccccccc}
\hline\noalign{\smallskip}
                                      & TWA      & ZA05    & ZA10    & ZA20    & ZA25    & ZM06    & ZM08    & ZM12    & ZM14   \\
\hline\noalign{\smallskip}
$\alpha_{\rm{MLT}}$                   & 2.3708   & 2.3711  & 2.3730  & 2.3715  & 2.3761  & 2.3714  & 2.3692  & 2.3723  & 2.3722 \\
$X_{\rm{0}}$                          & 0.7096   & 0.7061  & 0.7077  & 0.7115  & 0.7130  & 0.7120  & 0.7108  & 0.7090  & 0.7085 \\
$Z_{\rm{0}}$                          & 0.01472  & 0.01563 & 0.01518 & 0.01425 & 0.01379 & 0.01423 & 0.01454 & 0.01483 & 0.01492\\
$X_{\rm{C}}$                          & 0.3485   & 0.3452  & 0.3467  & 0.3505  & 0.3519  & 0.3508  & 0.3497  & 0.3481  & 0.3476 \\
$Z_{\rm{C}}$                          & 0.01572  & 0.01659 & 0.01616 & 0.01526 & 0.01480 & 0.01520 & 0.01552 & 0.01584 & 0.01593\\
$\log T_{\rm{C}}$                     & 7.1925   & 7.1939  & 7.1933  & 7.1918  & 7.1911  & 7.1917  & 7.1921  & 7.1927  & 7.1929 \\
$\log  \rho_{\rm{C}}$                 & 2.1857   & 2.1852  & 2.1855  & 2.1858  & 2.1864  & 2.1860  & 2.1855  & 2.1855  & 2.1856 \\
$z_{\rm{acc}}$                        & 0.015    & 0.005   & 0.010   & 0.020   & 0.025   & 0.015   & 0.015   & 0.015   & 0.015  \\
$Y_{\rm{acc}}$                        & 0.070    & 0.043   & 0.055   & 0.085   & 0.092   & 0.050   & 0.067   & 0.073   & 0.075  \\
$M_{\rm{acc}}/M_{\odot}$              & 0.0585   & 0.0585  & 0.0585  & 0.0585  & 0.0585  & 0.0588  & 0.0586  & 0.0585  & 0.0584 \\
$M_{\rm{L}}/M_{\odot}$                & 0.0028   & 0.0027  & 0.0027  & 0.0028  & 0.0028  & 0.0050  & 0.0036  & 0.0023  & 0.0020 \\
$\gamma$                              & -1.868   & -1.855  & -1.855  & -1.868  & -1.868  & -2.069  & -1.956  & -1.797  & -1.746 \\
$\lambda_{Z}$                         & 1.0      & 1.0     & 1.0     & 1.0     & 1.0     & 0.6     & 0.8     & 1.2     & 1.4    \\
$Y_{\rm{S}}$                          & 0.2450   & 0.2451  & 0.2450  & 0.2449  & 0.2447  & 0.2450  & 0.2452  & 0.2450  & 0.2450 \\
$(Z/X)_{\rm{S}}$                      & 0.0188   & 0.0188  & 0.0188  & 0.0188  & 0.0188  & 0.0188  & 0.0188  & 0.0188  & 0.0188 \\
$ R_{\rm{bc}}/R_{\odot}$              & 0.7110   & 0.7110  & 0.7109  & 0.7109  & 0.7107  & 0.7109  & 0.7111  & 0.7109  & 0.7109 \\
A(Li)                                 & 0.82     & 0.77    & 0.78    & 0.84    & 0.86    & 0.93    & 0.86    & 0.78    & 0.77   \\
\hline
neutrino fluxes $^{\rm{g}}$ & & & & & & \\
in $({\rm{cm}}^{-2}{\rm{s}}^{-1})$ & & & & & & \\
\hline
$pp$ $(10^{10})$                      & 5.98     & 5.96    & 5.97    & 5.98    & 5.99    & 5.98    & 5.98    & 5.97    & 5.97   \\
$pep$ $(10^{8})$                      & 1.47     & 1.46    & 1.47    & 1.48    & 1.48    & 1.48    & 1.48    & 1.47    & 1.47   \\
$hep$ $(10^{3})$                      & 8.13     & 8.07    & 8.10    & 8.16    & 8.19    & 8.18    & 8.15    & 8.11    & 8.10   \\
$^7$Be $(10^{9})$                     & 4.84     & 4.95    & 4.90    & 4.78    & 4.73    & 4.76    & 4.80    & 4.86    & 4.88   \\
$^8$B $(10^{6})$                      & 5.13     & 5.42    & 5.28    & 4.97    & 4.84    & 4.94    & 5.04    & 5.17    & 5.21   \\
$^{13}$N $(10^{8})$                   & 2.19     & 2.38    & 2.29    & 2.09    & 2.00    & 2.07    & 2.14    & 2.22    & 2.24   \\
$^{15}$O $(10^{8})$                   & 1.63     & 1.81    & 1.72    & 1.54    & 1.46    & 1.53    & 1.59    & 1.66    & 1.68   \\
$^{17}$F $(10^{6})$                   & 3.59     & 3.99    & 3.79    & 3.39    & 3.20    & 3.35    & 3.45    & 3.64    & 3.69   \\
\hline
\end{tabular}
\label{tab:lammod}
	\tablecomments{Here $\lambda_Z$ characterizes the relative efficiency of heavy-element mass loss (cf.\ Eq.~\ref{solarwindXYZ}) and $Z_{\rm acc}$ is the heavy-element abundance of the accreted material during PMS evolution. See notes for Table~\ref{tab:basmod}.}
\end{table*}

\begin{figure}
\centering
\includegraphics[scale=0.5]{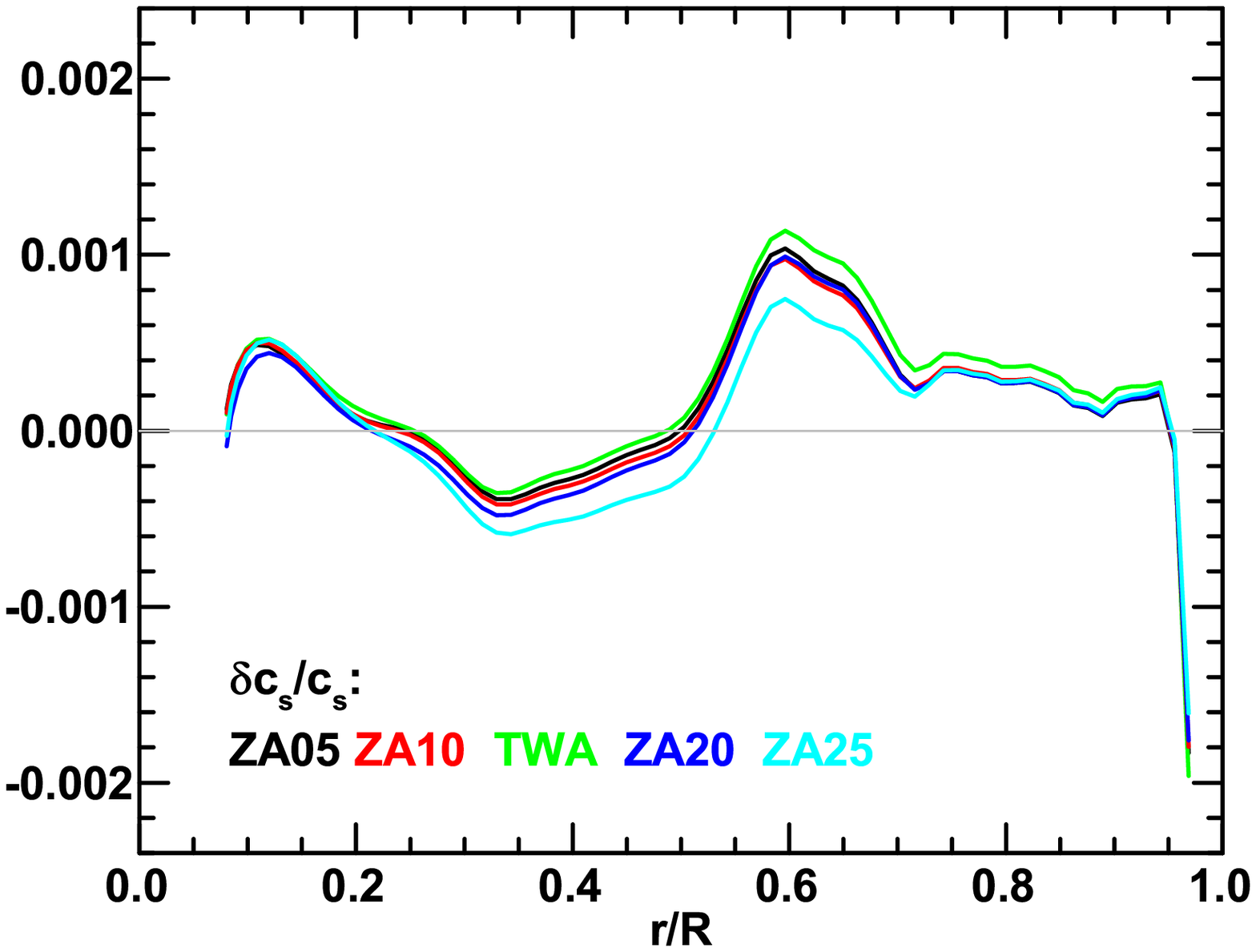}
\caption{Sound-speed deviations from helioseismic inversions (similar to Fig.\ref{std_c}a) for some solar models (see Table~\ref{tab:lammod}) with different $Z_{\rm{acc}}$ comparing with Model TWA.
}\label{Type4vsTWA}
\end{figure}

\begin{figure}
\centering
\includegraphics[scale=0.5]{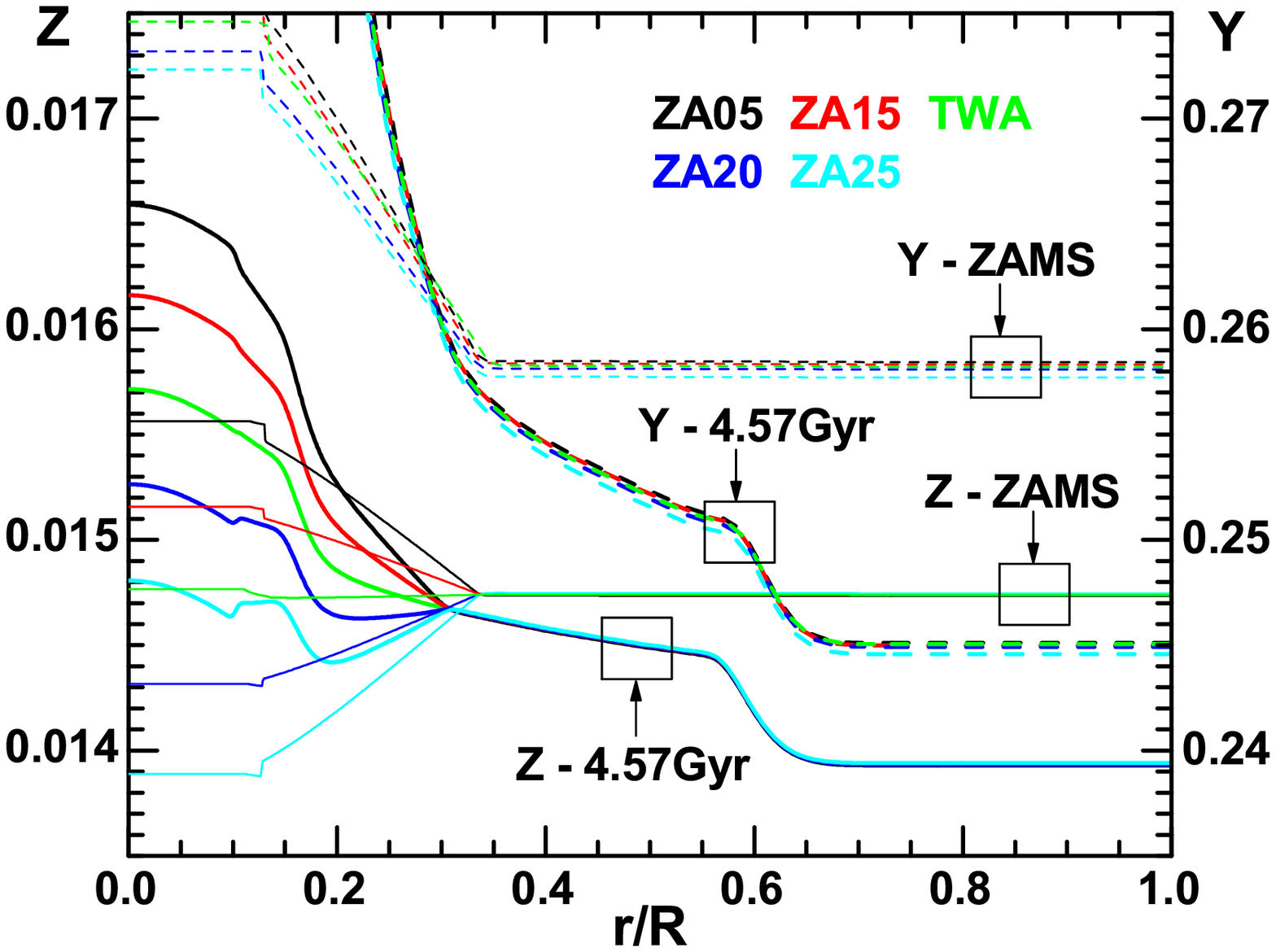}
\caption{ Helium-abundance and metallicity profile of some solar models (see Table~\ref{tab:lammod}) with different $Z_{\rm{acc}}$ comparing with Model TWA.
}\label{Type4vsTWAYZ}
\end{figure}

Solar models (ZA05, ZA10, ZA20, and ZA25) similar to Model TWA but with different $Z_{\rm{acc}}$ are compared with Model TWA to investigate the effects of varied $Z_{\rm{acc}}$. Information on those models is shown in Table~\ref{TWA_z_comp} and the sound-speed deviations are shown in Fig.~\ref{Type4vsTWA}. The ZA models are the selected best models with the minimum r.m.s. sound-speed deviation and $Y_{\rm{S}}$ in the helioseismically inferred range. The main differences between ZA models and Model TWA are the abundance profiles shown in Fig.~\ref{Type4vsTWAYZ}. Because $(Z/X)_{\rm{S}}$ of the models is calibrated to 0.0188 and the efficient mixing in the convective envelope, a reduction of the initial metallicity $Z_0$ is required to compensate an enhanced $Z_{\rm{acc}}$, as shown by $Z_0$ in Table~\ref{TWA_z_comp} and the metallicity in the core of the ZAMS models in Fig.~\ref{Type4vsTWAYZ}. Between the convective core boundary and the mass layer which is the BCZ at age $t=\tau_{\rm{acc}}$, there is a range with gradient of metallicity and the gradient is determined by $Z_{\rm{acc}}$ such that $Z_{\rm{acc}}>0.015$ leads to $dZ/dr>0$ and vice versa. Although the metallicity in the core will increase during the evolution because of molecular diffusion and CNO cycle nuclear reactions, the differences of the core metallicity among those ZAMS models remain in the models with present solar age. Therefore $Z_{\rm{C}}$ of those solar models are anti-correlated with $Z_{\rm{acc}}$. A higher $Z_{\rm{C}}$ leads to a higher temperature gradient and then a higher temperature in the core. Therefore it leads to higher $^7$Be and $^8$B neutrino fluxes. A consequence of the higher $T_{\rm{C}}$ is a lower $X_{\rm{C}}$ and then a lower $X_{\rm{0}}$ and a higher $Y_{\rm{0}}$. Therefore the required $Y_{\rm{acc}}$ is lower for improving the sound-speed profile. In a word, a lower $Z_{\rm{acc}}$ leads to a higher $Z_{\rm{C}}$; thus the $^8$B neutrino flux is higher and the required $Y_{\rm{acc}}$ is lower. This explains the properties shown in Fig.~\ref{Type4}.

We recall that the interior structure of a main-sequence stellar model is completely determined by the abundance profiles. On the other hand, for a solar model, there are four conditions at the stellar surface, i.e., luminosity, radius, temperature and density. The number of conditions equals the number of stellar structure equations. In this case, the interior structure of a solar model can be integrated from the surface to the center when its abundance profiles are given. This means that the envelope structure of a solar model is completely determined by the abundance profiles in the envelope even if the abundance profiles in the core are unknown. Thus, if the sound-speed profile of a solar envelope model is constrained by the helioseismic inferences, the helium-abundance profile and the metallicity profile in the envelope should be in one-to-one correspondence, such that fixing one of the profiles essentially determines the other. For the ZA solar models, the abundance profiles in $r>0.3R$ are basically identical to that of Model TWA because they have the same abundances at the surface (i.e., $Y_{\rm{S}}\approx0.245$ and $(Z/X)_{\rm{S}}=0.0188$). As shown in Fig.~\ref{Type4vsTWA}, the sound-speed profiles of ZA models and Model TWA are in very good agreement with the helioseismic inferences, implying that the $Y$ and $Z$ profiles of Model TWA are suitable for each other, given the helioseismic constraint on sound speed. It should be pointed out here that the correspondence between the $Y$ and $Z$ profiles is affected by the details input physics involved in the structure equations of the envelope, e.g., opacity, EOS and the model of convection overshoot which affects the profiles of $L_{\rm{K}}$ and $L_{\rm{C}}$.

As shown in Fig.~\ref{Type4vsTWA}, the differences between sound-speed deviations of those solar models are quite small even in the core where the abundance profiles of those models are quite different. The possible reason is the anti-correlation between $X_{\rm{C}}$ and $T_{\rm{C}}$ leading to a compensation between temperature and $\mu$ so that the variation of sound-speed profile is slight.

\subsubsection{The loss rate of metallicity in the solar wind} \label{SecresultZwind}

About 600 solar models with ${\eta}=0$, $Z_{\rm{acc}}=0.015$, $\tau_{\rm{acc}}=12$Myr and different $\lambda_{Z}$, $M_{\rm{L}}$ and $Y_{\rm{acc}}$ have been calculated to investigate the effects of varied $\lambda_{Z}$, which represents the relative escape speed of heavy elements in the solar wind. $Y_{\rm{acc}}$ varies in the range from $0.00$ to $0.26$ with a step $0.02$. $M_{\rm{L}}$ varies in the range from $0.00$ to $0.01$ with a step $0.001$. $\lambda_{Z}$ varies in the range from $0.6$ to $1.4$ with a step $0.2$.

\begin{figure*}
\centering
\includegraphics[scale=0.5]{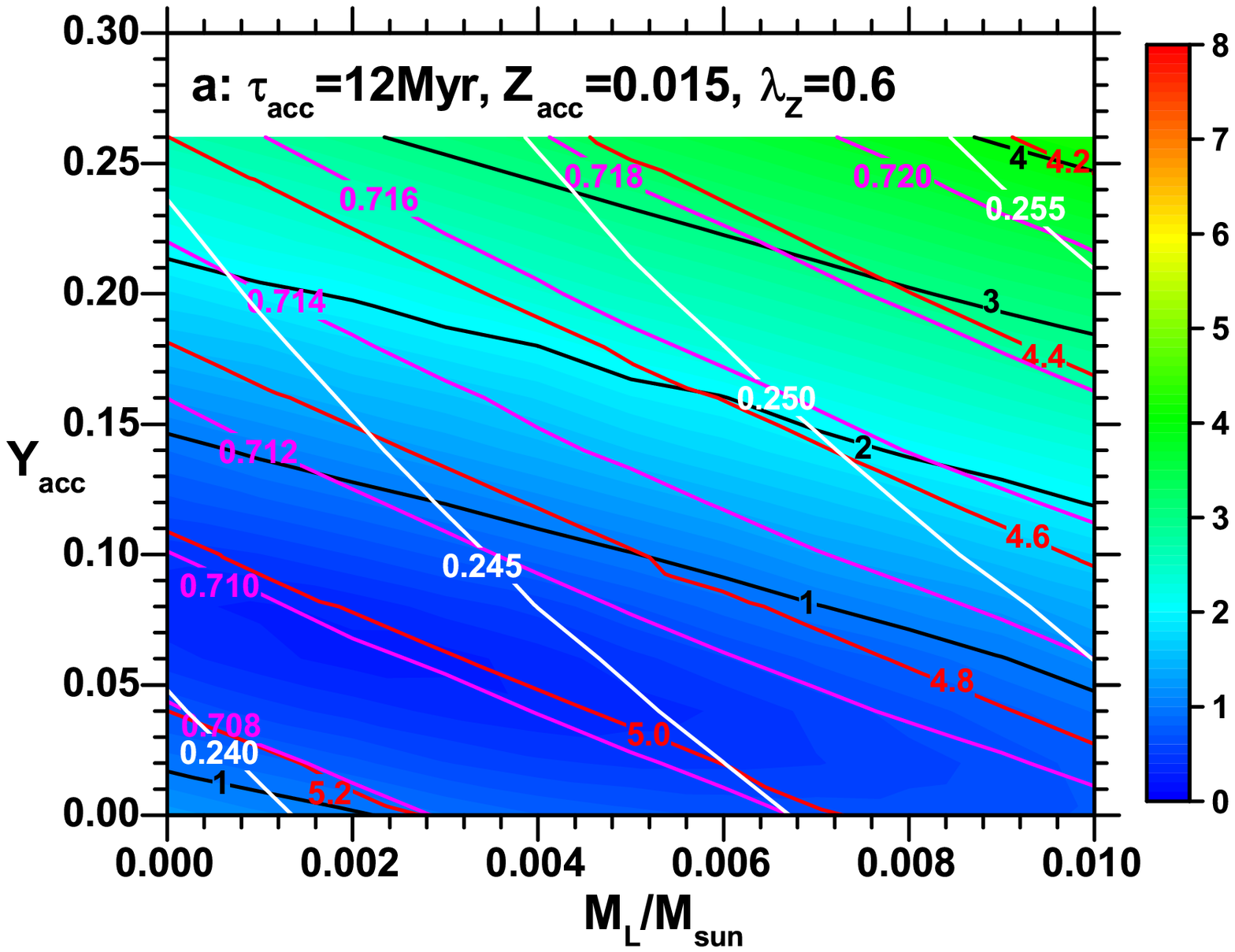}
\includegraphics[scale=0.5]{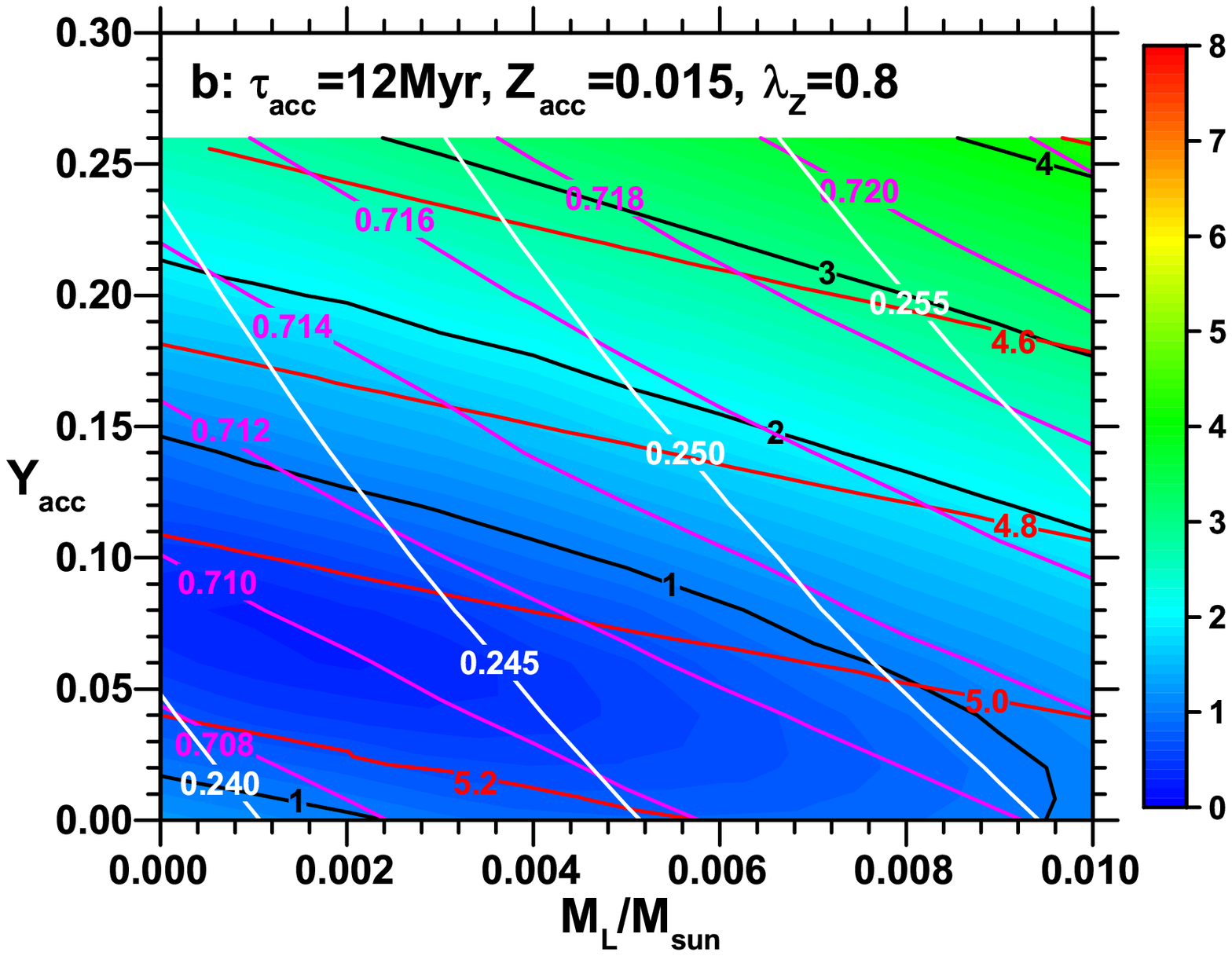}
\includegraphics[scale=0.5]{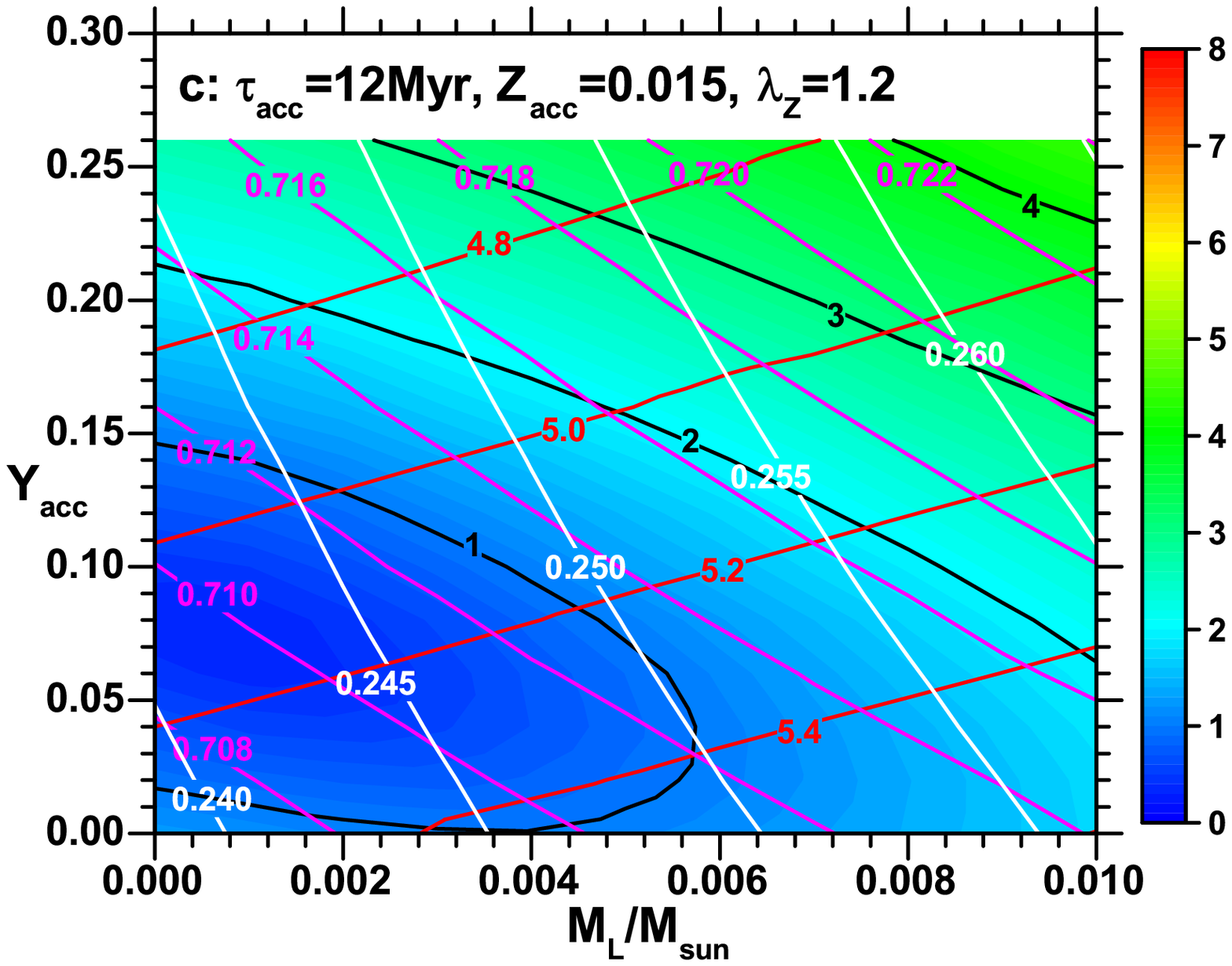}
\includegraphics[scale=0.5]{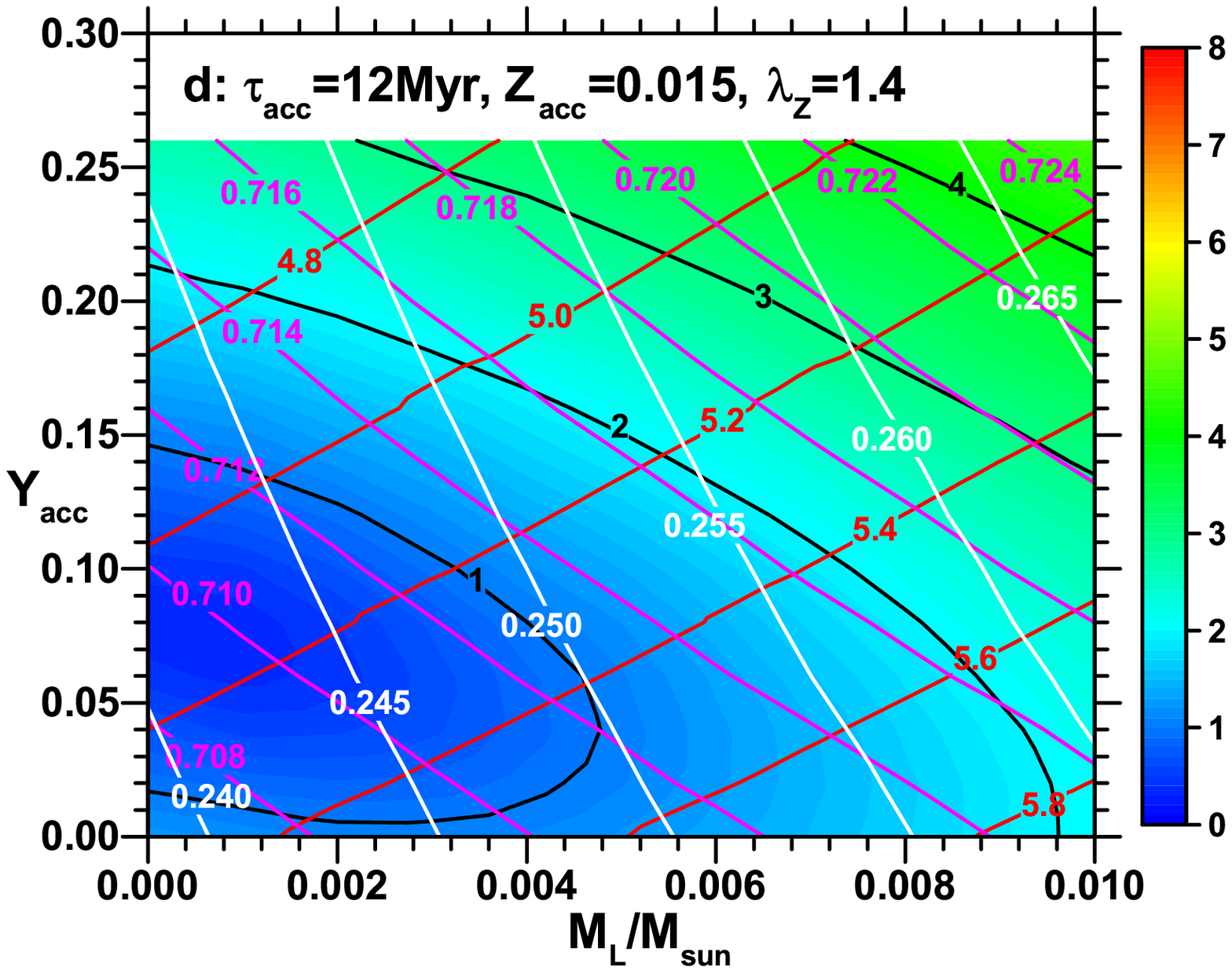}
	\caption{ Similar to Fig.~\ref{Type2}c, but for solar models with different relative efficiency $\lambda_{Z}$ of heavy-element mass loss (cf.\ Eq.~\ref{solarwindXYZ}). The $^8$B neutrino flux of solar models is now shown as the red contours.
}\label{Type5}
\end{figure*}

Sound-speed deviations, $^8$B neutrino fluxes, locations of BCZ, and surface helium abundances of those solar models with $\lambda_{Z}=0.6$, $\lambda_{Z}=0.8$, $\lambda_{Z}=1.2$ and $\lambda_{Z}=1.4$ are shown in Fig.~\ref{Type5}. The case of $\lambda_{Z}=1.0$ is shown in Fig.~\ref{Type2}c. There are two main effects of varied $\lambda_{Z}$ as shown in Fig.~\ref{Type5} and Fig.~\ref{Type2}c: $^8$B neutrino fluxes of solar models are anti-correlated with $M_{\rm{L}}$ for $\lambda_{Z}<1$ and positively correlated with $M_{\rm{L}}$ for $\lambda_{Z}>1$; and the required mass-loss for improving helioseismic quantities of a solar model is anti-correlated with $\lambda_{Z}$. We now investigate the reasons for these relations.

\begin{figure}
\centering
\includegraphics[scale=0.5]{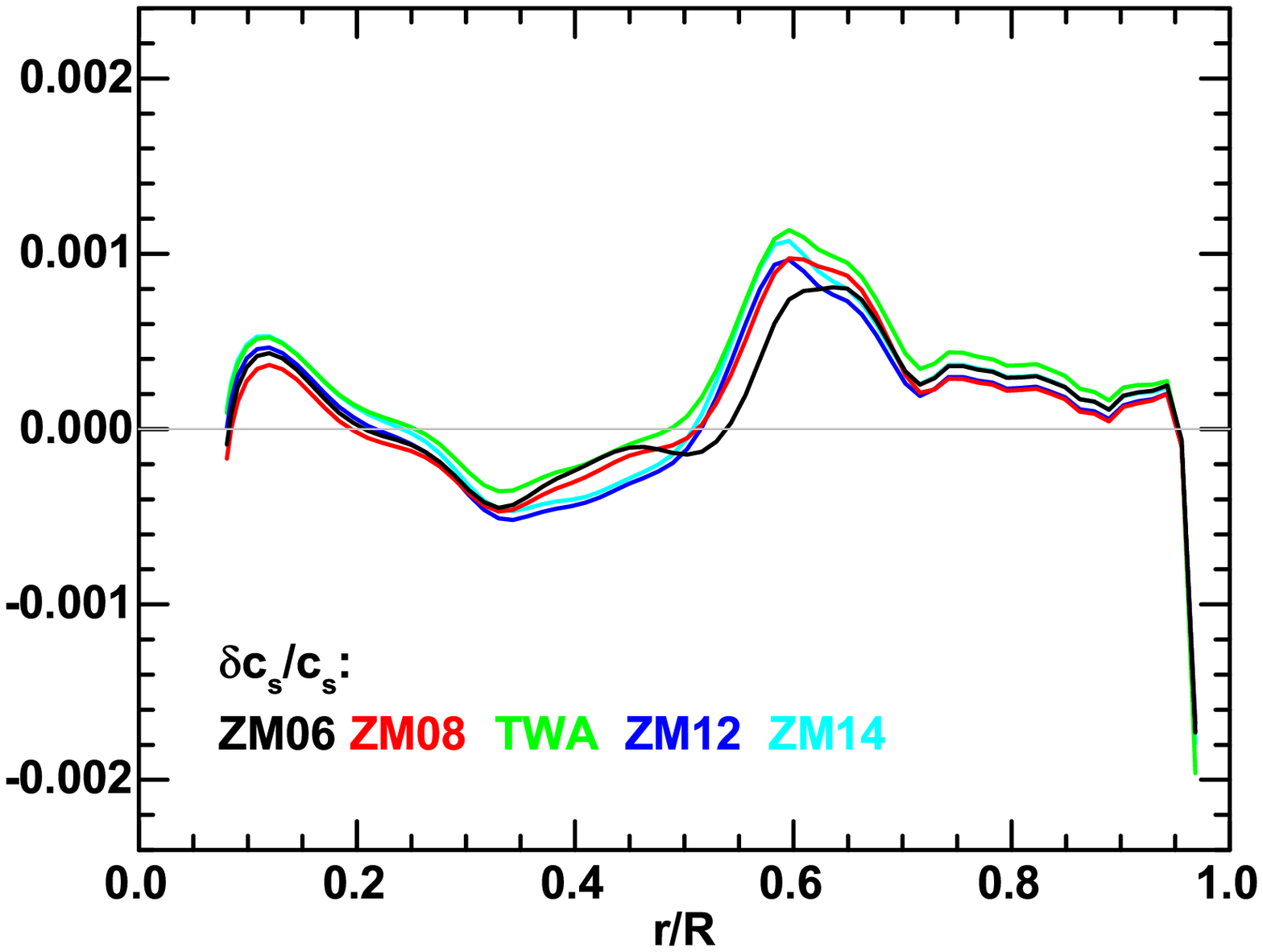}
\caption{Sound-speed deviations from helioseismic inversions (similar to Fig.\ref{std_c}a) for some solar models (see Table~\ref{tab:lammod}) with different $\lambda_{Z}$ comparing with Model TWA.
}\label{Type5vsTWA}
\end{figure}

\begin{figure}
\centering
\includegraphics[scale=0.5]{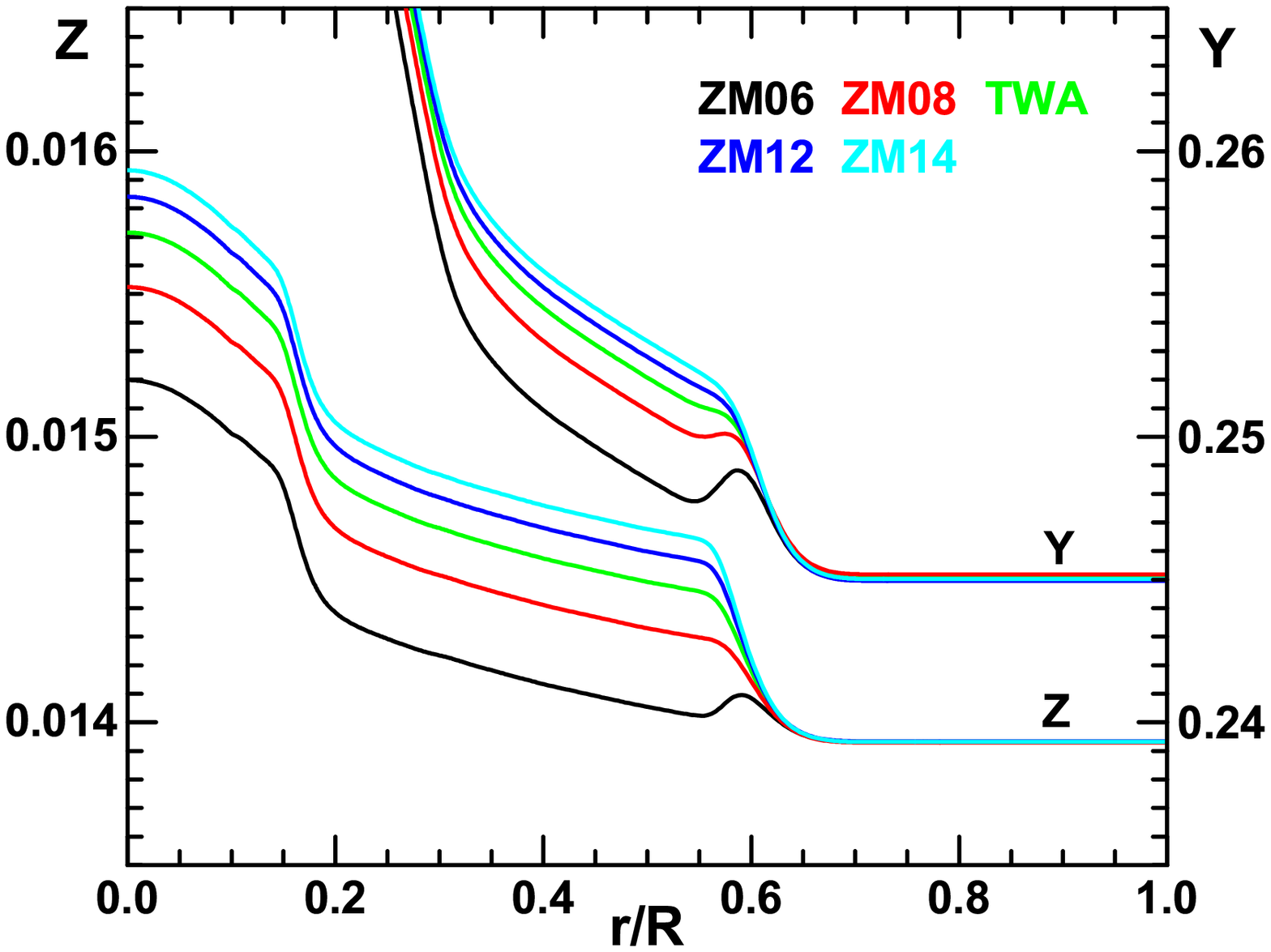}
	\caption{ Helium-abundance and metallicity profile of some solar models (see Table~\ref{tab:lammod}) with different $\lambda_{Z}$ comparing with Model TWA.
}\label{Type5YZ}
\end{figure}

Solar models (ZM04, ZM08, ZM12, and ZM14) similar to Model TWA but with different $\lambda_{Z}$ are compared with Model TWA to investigate the effects of varied $\lambda_{Z}$. The information of those models is shown in Table~\ref{TWA_z_comp} and the sound-speed deviations are shown in Fig.~\ref{Type5vsTWA}. The ZM models are the selected best models with the minimum r.m.s. sound-speed deviation and $Y_{\rm{S}}$ in the helioseismically inferred range. As discussed above, the structure of a solar model is determined by its abundance profiles. Therefore it is straightforward to compare the ZM solar models by comparing their abundance profiles, which are shown in Fig.~\ref{Type5YZ}.  Unlike the case of the ZA models, the metallicity profiles of the ZM models are significantly different from Model TWA such that a lower $\lambda_{Z}$ leads to a lower metallicity profile in the solar interior with $r<0.6R$. For $\lambda_{Z}<\lambda_{Z,cr}$, where $\lambda_{Z,cr}=({X_{\rm{S}}} + {\lambda _Y}{Y_{\rm{S}}})/(1 - {Z_{\rm{S}}}) \approx 0.9$ for $\lambda_{Y}=0.5$, we obtain $Z_{\rm{L}}<Z_{\rm{S}}$; thus the mass loss concentrates the heavy elements in the convective envelope, just as $\lambda_{Y}=0.5$ concentrates helium in the convective envelope. If $\lambda_{Z}>\lambda_{Z,cr}$, $Z_{\rm{L}}>Z_{\rm{S}}$; thus the mass-loss depletes heavy elements in the convective envelope. The effect of varied $\lambda_{Z}$ on the metallicity profile of solar model is similar to a modification of the strength of molecular diffusion near the BCZ, i.e., $\lambda_{Z}>\lambda_{Z,cr}$ is similar to an enhancement on diffusion and vice versa. Because $(Z/X)_{\rm{S}}$ is calibrated, the effect of $\lambda_{Z}$ on the metallicity profile is represented by the part of the solar interior with $r<0.6R$, as shown in Fig.~\ref{Type5vsTWA}. The final effect of $\lambda_{Z}$ on the metallicity profile is that a lower $\lambda_{Z}$ leads to a lower metallicity profile in the solar interior with $r<0.6R$ and vice versa. Therefore the $Z_{\rm{C}}$ is correlated (for $\lambda_{Z}>\lambda_{Z,cr}$) or anti-correlated (for $\lambda_{Z}<\lambda_{Z,cr}$) with the total mass-loss. Because of the correlation between $Z_{\rm{C}}$ and the $^7$Be and $^8$B neutrino fluxes, the correlation between the $^8$B neutrino flux and $M_{\rm{L}}$ shown in Fig.~\ref{Type5} is now explained.

Figure~\ref{Type5vsTWA} shows that the sound-speed profiles of the ZM models are in very good agreement with helioseismic inferences, implying that the helium-abundance profile is suitable for the metallicity profile for each ZA model. As shown in Fig.~\ref{Type5YZ}, a lower metallicity profile requires a lower helium-abundance profile to reproduce the helioseismically inferred sound-speed profile. This is because the a lower metallicity leads to lower temperature gradient and then lower temperature in the solar mantle; thus based on Eq.~(\ref{soundspeedapp}) a lower helium abundance in the mantle is required to retain the sound speed. Since $Y_{\rm{S}}$ of all ZM models is $0.245$, the helium-abundance profile in the mantle is determined by the total mass loss caused by the solar wind which concentrates helium in the convective envelope. Therefore, for a lower metallicity profile, a massive mass loss is required to obtain a lower helium-abundance profile in order to retain the sound-speed profile. This explains that the required mass loss for improving helioseismic quantities of a solar model is anti-correlated with $\lambda_{Z}$ as shown in Fig.~\ref{Type5}.

\section{Discussion} \label{Secdisc}

The standard solar model (SSM) based on AGSS09 composition is inconsistent with helioseismic inferences (of sound-speed and density profiles, helium abundance in the solar convection zone, and the location of the base of the solar convection zone) and observations of the solar Li abundance. This difficulty still stands even with the recent upward revised Ne abundance. We have investigated the possible mechanisms to improve the solar model with three extra physical processes, i.e., convective overshoot which leads to the turbulent kinetic energy flux $F_{\rm{K}}$ and the overshoot mixing, inhomogeneous mass loss caused by the solar wind, and PMS accretion with inhomogeneous materials. The turbulent kinetic energy flux is necessary to resolve the problem of the structure of the solar convective envelope. The convective overshoot mixing is required for the solar Li depletion. The mass loss caused by the solar wind shows a deficiency in helium \citep[e.g.,][]{windcomp} and the total mass loss is about $(10^{-3}  -  10^{-2})M_{\odot}$ \citep[e.g.,][]{windmass}. It is not difficult to estimate that the mass loss could increase $Y_{\rm{S}}$ by a level of $\sim 0.01$, which is larger than its uncertainty inferred by helioseismology. The motivation for PMS accretion with inhomogeneous materials is to adjust the abundance profiles in the solar ZAMS models and thereby affect the sound speed of solar models. The PMS accretion is commonly found in T Tauri stars which may resemble the Sun at its early stage. It is believed that the mechanism of PMS accretion in T Tauri stars is magnetospheric accretion from the disks around stars. For this scenario, we propose that the accreted materials may be inhomogeneous because the non-thermal ionization processes in the surface of the protoplanetary disk could be affected by the different first ionization potentials of elements.

The convective overshoot below the solar convection zone is described as a simple model based on an exponentially decreasing $F_{\rm{K}}$ and a given value $F_{\rm{K,bc}}$ of the turbulent kinetic energy flux at the BCZ. The parameters in the overshoot model are derived from helioseismic inferences and solar lithium abundance. The mass loss caused by the solar wind is modeled by using a decreasing power-law function of the stellar age. The composition of the lost material is assumed to be helium-poor, as revealed by observations. Effects of varied metallicities of the lost material are also investigated. The mass accretion rate of PMS accretion is based on observations and its dispersion is also taken into account. The duration of the PMS accretion and the composition of the accreted material are free parameters for the solar models. We have analyzed the solar evolutionary models with those extra physical processes and have found that, if the PMS accretion is helium-poor, there are solar models consistent with helioseismic inferences and the observation of the solar neutrino fluxes. A typical improved solar model is Model TWA shown in Table~\ref{modelinfo} and Figs~\ref{std_c}.

In this section, we carry out discussions on the following issues: implications of solar Li and Be abundances on the structure of the solar interior, the possible mechanisms to improve the sound-speed profile in solar models, and the possible thermohaline mixing in the solar interior.

\subsection{Fresh insight on the solar Li and Be abundance} \label{SecdiscLiBe}

It is useful to consider the constraints on the structure of the solar interior provided by the solar Li and Be abundances. $^7$Li and $^9$Be are fragile elements due to their proton capture reactions. The typical temperatures for those proton capture reactions are $\log T\approx 6.4$ and $\log T\approx 6.5$ for $^7$Li and $^9$Be, respectively, very close to (or a little higher than) the temperature of the BCZ in stars with mass close to the solar mass. Observations of the Li abundance of low-mass stars ($0.8<M/M_{\odot}<1.2$) in open clusters have shown more Li depletion than the predictions by the standard stellar models. The discrepancy is generally thought to be caused by some extra mixing missing in the standard stellar models, e.g., overshoot \citep{SBS76,Schlattl99,xio02,xio09,Z12Li,Baraffe17}, rotational mixing \citep{Pin90,CVZ92}, internal-wave mixing \citep{Montalban94} etc.. Therefore the abundances of $^7$Li and $^9$Be are tools to probe the mixing below the convective envelope in the interior of those stars.

\begin{figure}
\centering
\includegraphics[scale=0.5]{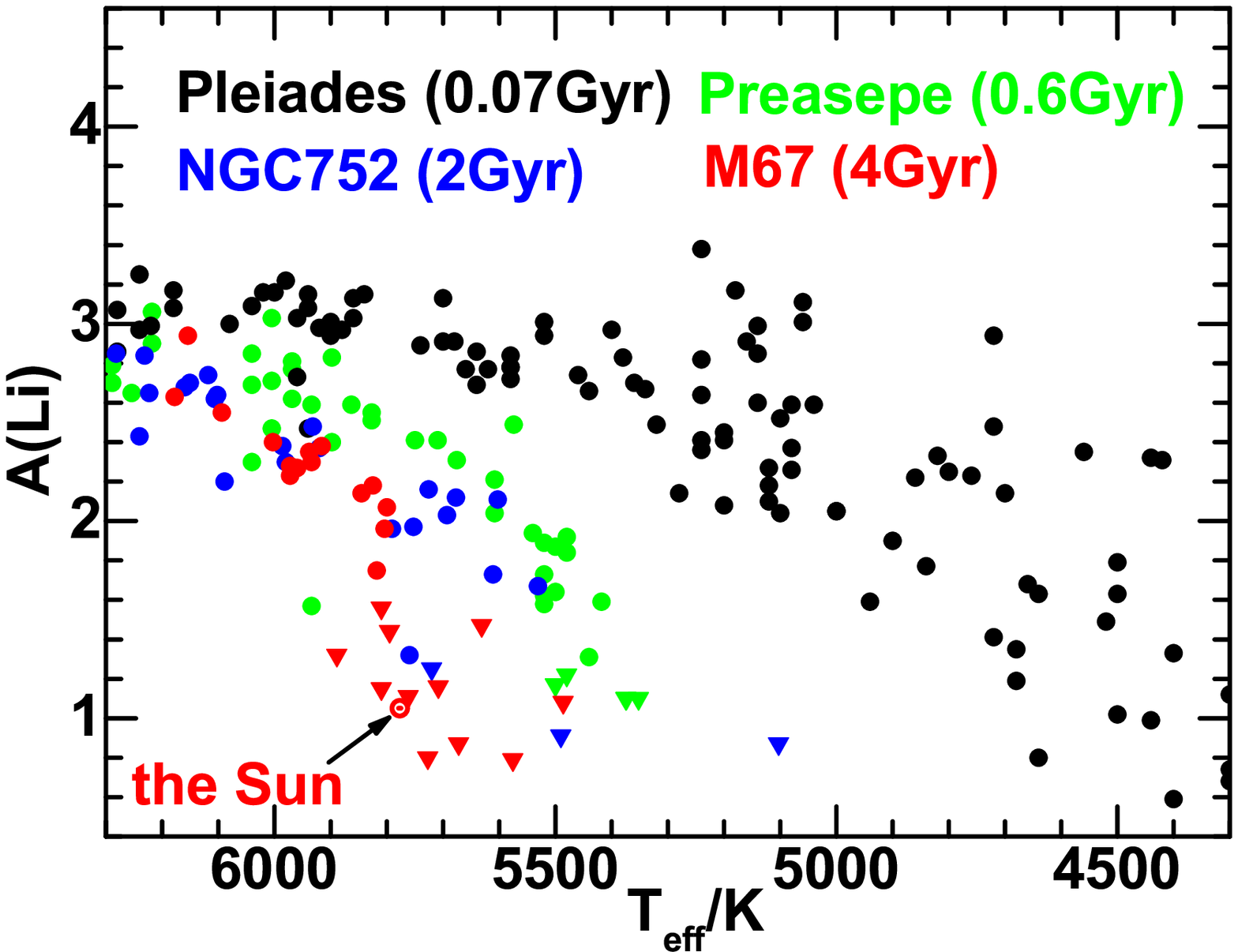}
	\caption{Observations of Li abundances of low-mass stars in open clusters: Pleiades \citep{PleiadesLi} with age $0.07\,{\rm{Gyr}}$ \citep{PPAge}, Praesepe \citep{PraesepeLi} with age $0.6\,{\rm{Gyr}}$ \citep{PPAge}, NGC752 \citep{PraesepeLi,NGC752Li} with age $2\,{\rm{Gyr}}$ \citep{NGC752Age} and M67 \citep{M67Li} with age $4\,{\rm{Gyr}}$ \citep{M67Age} in the range of effective temperature $4300 {\, \rm K} \leq T_{\rm{eff}} \leq 6300{\, \rm K} $ are shown in different colors. The solar Li abundance is shown as the red $\odot$ symbol. }\label{Licluster}
\end{figure}

The solar surface Li abundance is significantly depleted by about 2 dex relative to the meteoritic abundance \citep[e.g.][]{AGSS09}. Figure \ref{Licluster} shows the Li abundances of the Sun and some open cluster stars with metallicity close to the Sun. The Li abundances of the solar-age open cluster M67 stars with effective temperatures similar to the Sun has shown the same Li depletion level as the Sun, indicating that the solar Li depletion should not be unique, which is also supported by investigations of Li abundances of solar twins \citep[see, e.g.,][]{King97,Melendez07,LTK10,Castro11}. Since the effective temperature changes little for solar-mass stars during the main-sequence stage, the Li abundances of low-mass stars in open clusters with different ages shown in Fig.~\ref{Licluster} validate that Li is gradually depleted during the main-sequence stage. This evidence strongly indicates that the Sun gradually experienced Li depletion during its main-sequence stage. The Li abundances of young open clusters (e.g., Pleiades and Praesepe) indicate that the solar-mass stars lithium depletion is about $0.5 - 1.0$ dex in the PMS stage. Therefore the solar Li depletion from ZAMS to the present age is about $1.0 - 1.5$ dex, or equivalently, the e-folding timescale of Li depletion is about $\tau_{\rm{Li},\odot} \sim 1\,\rm{Gyr}$ in the solar main-sequence (MS) stage.

In the MS stage, the structure of the stellar envelope is basically in a quasi-static state. The variation of the ratio of $R_{\rm{bc}}$ to the stellar radius $R$ is small. The variation in the envelope of the relation between $m_r/M$ and $r/R$, is also small. Therefore we can analyze the issue of solar Li depletion in MS stage in the quasi-static case as follows.

The Li depletion timescale $\tau_{\rm{Li},\odot}$ is determined by the competition between burning and mixing. At a radius $r$ below the BCZ, there is an e-folding time of Li burning:
\begin{eqnarray} \label{LiBurningTime}
\tau_{\rm{n}}(r)=\frac{1}{\rho Y_1 {\cal R}(T)},
\end{eqnarray}%
where ${\cal R}(T)$ is the rate of the $^7$Li proton capture reaction, and $Y_1 \sim 1$ is the hydrogen abundance in mol$/$g. There is also a characteristic timescale of the mixing below the BCZ, in the region between $r$ and $R_{\rm{bc}}$:
\begin{eqnarray} \label{LiMixingTime}
\tau_{\rm{mix}}(r)=\frac{(R_{\rm{bc}}-r)^2}{D_{\rm{mix}}},
\end{eqnarray}%
where $D_{\rm{mix}}$ is the typical mixing diffusion coefficient between $r$ and $R_{\rm{bc}}$. Since the former decreases and the latter increases toward the solar center, there is a location $r_{*}$ where $\tau_{\rm{mix}}(r_{*})=\tau_{\rm{n}}(r_{*})$. For $r>r_{*}$, the mixing can efficiently mix materials during Li burning; thus the mixing could transport material going deeper to a higher-temperature region, so that $\tau_{\rm{Li},\odot} \leq \tau_{\rm{n}}(r)$. For $r<r_{*}$, the mixing cannot catch the Li burning; thus the rate of Li depletion in the envelope is lower than the Li burning rate at $r$, i.e., $\tau_{\rm{Li},\odot} \geq \tau_{\rm{n}}(r)$. The combination shows that $\tau_{\rm{Li},\odot}=\tau_{\rm{n}}(r_{*})=\tau_{\rm{mix}}(r_{*})$. Since $\tau_{\rm{Li},\odot} \sim 1\,\rm{Gyr}$, according to ${\cal R}(T)$ given by \citet{Nucl} and \citet{SFII}, the corresponding temperature for $\tau_{\rm{n}} \sim 1\,\rm{Gyr}$ is $\log T \approx 6.4$ at $r_{*} \approx 0.68R_{\odot}$. In particular, we find that $\tau_{\rm{mix}}(0.68R_{\odot}) \sim 1\,\rm{Gyr}$. Since the age of the Sun is much larger than $1\,\rm{Gyr}$, the solar envelope with $r>0.68R_{\odot}$ should be efficiently mixed and nearly homogeneous. The molecular diffusion has a characteristic timescale significantly longer than the solar age so that it cannot compete with the mixing.

A similar analysis can also be applied to the solar Be abundance. Because the solar Be abundance shows almost no depletion \citep[e.g.][]{AGSS09} and the temperature resulting in a characteristic timescale of $^9$Be burning as the solar age is about $\log T \approx 6.5$ at $r=0.6R_{\odot}$, we can conclude that $\tau_{\rm{mix}}(0.6R_{\odot})$ is not less than the solar age. Therefore any kind of mixing below the base of the solar convection zone cannot show considerable effect below $0.6R_{\odot}$.

\subsection{On improving the sound speed of solar models} \label{SecdiscSpeed}

The main problem of the SSMs with low-$Z$ composition (i.e., AGSS09 and AGSS09Ne) is that the sound-speed profile in the SSM is not consistent with the helioseismic inferences. It is shown in Figure \ref{std_c}a that, in the solar mantle, the sound speed in the solar models should be increased in order to be consistent with helioseismic inferences. It is shown in Eq.~(\ref{soundspeedapp}) that there are only two ways to increase $c_{\rm{s}}$: to increase temperature $T$ and to decrease the mean molecular weight $\mu$. The sound speed in the solar convection zone can be obtained from integrating the hydrostatic equation with the polytropic relation and is well defined \citep[e.g.,][]{JCD86}. The composition in the solar convection zone is also well defined from the spectral analyses and the helioseismic determination of helium abundance. Therefore the temperature in most of the solar convection zone is also well defined. To increase $T$ in the region $0.3<r/R_\odot<0.7$ is equivalent to increase the temperature gradient $ \nabla = {\rm{dln}} T / {\rm{dln}} P $ in that region. The mean molecular weight $\mu$ is determined by the composition as $\mu^{-1}=2-1.25Y-1.5Z$ (for the fully ionized case) where $Y$ is helium abundance and $Z$ is metallicity. Because the contribution of metallicity $Z$ to $\mu$ is only $ \sim 1\%$ in the Sun and $Z$ should not change too much, $\mu$ is mainly determined by the helium abundance $Y$. Therefore, in the solar mantle, to decrease the helium abundance could improve the sound speed. Now, we discuss the possible mechanisms for improving the sound-speed profile of the SSM.

Specifically, increasing the temperature gradient below the base of the solar convection zone can be achieved via increasing the opacity, enhancing the molecular diffusion, and taking into account extra energy inward transport mechanisms (e.g, inward energy fluxes caused by convective overshoot: convective heat flux \citep{zl12a,zdxc12}, turbulent kinetic energy \citep{TKFZ14} and internal gravity waves \citep{Arnett10}). The modifications of the opacity and the molecular diffusion have been extensively investigated, as introduced in Section \ref{SecIntro}. The effects of turbulent kinetic energy caused by convective overshoot on the solar model has been investigated in Section \ref{SecresultOV}. With the overshoot parameters inferred by helioseismology, the solar Model OV09Ne does not show a satisfactory sound-speed profile. This means that the overshoot model with helioseismically inferred parameters is not sufficient to solve the problem. Since the effects of negative $L_{\rm{K}}$ is equivalent to an enhancement of opacity \citep{TKFZ14}, the required modifications of the opacity for improving solar models \citep[e.g.,][]{ser09,JCD10} indicate that it is possible to use the overshoot model to solve the ``solar abundance problem" in solo if a longer e-folding length of $L_{\rm{K}}$ (i.e., a larger $\theta$) is adopted. It can be estimated from Fig.~13 in \citet{JCD10} that the required value of $\theta$ is about 5$-$7$H_{\rm{P}}$, significantly larger than the helioseismic suggestion.
However, a 5$-$7$H_{\rm{P}}$ e-folding length of $L_{\rm{K}}$ will show significant mixing in the solar mantle and leads to significant $^9$Be depletion at the solar surface which conflicts with observation.
Dissipations of other possible negative kinetic energy fluxes below the BCZ (e.g., convective heat flux and internal gravity waves) should also lead to mixing. Therefore taking them into account still cannot solve the ``solar abundance problem" in solo.

Except for increasing the temperature gradient $\nabla$, the only way to improve the sound speed of the solar model is to reduce the mean molecular weight $\mu$, or specifically, to reduce the helium abundance in the solar mantle. An example of the effects of changed helium abundance in that region in literature is the solar models with PMS accretion shown by \citet{ser11}: for the metal-rich early-accretion scenario, the initial hydrogen abundance becomes higher than the SSM due to the luminosity calibration and therefore the helium abundance is lower than the SSM; thus solar models with metal-rich early accretion have improved sound-speed profiles and $R_{\rm bc}$ compared to the SSM. However, we have to notice that, in the AGSS09 or AGSS09Ne SSMs, $Y_{\rm{S}}$ are already less than the helioseismic inferences. Therefore, in order to improve the sound speed and $Y_{\rm{S}}$ of SSMs simultaneously, extra mechanisms must be added to increase in $Y$ in the convective envelope and decrease $Y$ in the solar mantle. There are three possible mechanisms to achieve that: a mixing below the BCZ, a helium-poor mass accretion in the PMS stage, and a helium-poor mass loss in the MS stage. All three mechanisms have been investigated in this paper. The mixing below the BCZ competes with the molecular diffusion and thus is helpful to achieve the required changes of the $Y$ profile. However, the strength of the mixing is restricted by the observed light element depletion. As shown by the OV09Ne solar model, such a mixing is not enough to restore the sound-speed profile. The helium-poor PMS accretion is the main factor to improve the sound-speed profiles of the improved models listed in Tables \ref{Type2Best}, \ref{TWA_eta_comp} and \ref{TWA_z_comp}, because it leads to helium-poor envelopes in solar models. The effect of the helium-poor mass loss in the MS stage is to concentrate helium in the convective envelope which helps to solve the problem of low $Y_{\rm{S}}$ in SSM. As shown in this work, the combined effects of those processes could lead to suitable helium-abundance profiles to reproduce the helioseismically inferred sound-speed profile and $Y_{\rm{S}}$ simultaneously.

An interesting issue is the effects on solar models of the mass loss caused by the solar wind. As pointed out, varying $\lambda_{Z}$ works like a modification of molecular diffusion of heavy elements near the BCZ, as does $\lambda_{Y}$ on helium. Therefore the interior helium abundance and metallicity profiles can be adjusted by the variations of $\lambda_{Z}$ and $\lambda_{Y}$. Since the sound-speed profile could be significantly improved if the helium-abundance profile is suitable for the metallicity profile such as for Model TWA and the ZM models, one may wonder whether it is possible to obtain a solar model as good as Model TWA by only using the helioseismically based overshoot model and the mass loss with suitable values of $\lambda_{Z}$ and $\lambda_{Y}$. In this case, the helium-poor PMS accretion is not necessary. In order to investigate that, we have calculated some solar models with $0 \leq \lambda_{Z} \leq 2$ with a step 0.5, $0 \leq \lambda_{Y} \leq 2$ with a step 0.2 and $0\leq M_{\rm{L}}/M_{\odot}\leq0.01$ with a step 0.001, and without PMS accretion. However, no satisfactory model has been found.

Since the inhomogeneous PMS accretions is absent in the test, the helium-abundance profiles of ZAMS models in the test are homogeneous, which is the only difference between the models in the test and the improved models with all three extra processes. As mentioned above, the structure of a solar envelope model is completely determined by its $Y$ and $Z$ profile; thus the $Y$ and $Z$ profile should be in one-to-one correspondence in order to keep its sound-speed profile consistent with helioseismic inferences. Because the relative escape speed of heavy elements in the test varies over a large range, leading to a large range of $Z$ profiles, so does the relative escape speed of helium. No satisfactory solar model being found indicates the unsuitability of solar models with a homogeneous helium-abundance profile at ZAMS. Therefore it supports the necessity of the helium-poor PMS accretion and that an inhomogeneous ZAMS model with helium-poor envelope is necessary to reproduce a solar model with a sound-speed profile consistent with helioseismic inferences.

Since the relative escape speed of each heavy element assumes the same value $\lambda_{Z}$ in this paper, more comprehensive effects of the inhomogeneous mass loss on solar models could be investigated if the relative escape speed of each heavy element is treated independently and the varied heavy element abundances are considered in the interpolation of opacity in the solar interior.

\subsection{On the effects of thermohaline mixing} \label{SecdiscTHM}

A main factor in improving the sound speed in the TWA solar models is that the helium abundance in the mantle is lower than the SSM. As shown in Fig.~\ref{Yevol}a, there is a layer of positive helium-abundance gradient below the convective envelope during the evolution of Model TWA, e.g., at an age between 0.4$-$2Gyr. This positive helium-abundance gradient, which results from the helium-poor mass loss at the early MS stage, is required to appropriately compensate the helium settling so that the sound-speed profile and $Y_{\rm{S}}$ can be consistent with helioseismic inferences simultaneously. However, the positive helium-abundance gradient leads to thermohaline mixing since $\nabla_{\mu}<0$ in this layer. Thermohaline mixing is not included in our solar models. It could reduce or remove the positive helium-abundance gradient and then affect the helium-abundance profile of the solar model at the present age. If thermohaline mixing is too efficient to form a positive helium-abundance gradient layer, the TWA solar models would no longer reproduce a sound-speed profile and $Y_{\rm{S}}$ consistent with helioseismic inferences simultaneously.

A widely used formula for the diffusion coefficient for thermohaline mixing is \citep{U72,Kph80}:
\begin{eqnarray} \label{thermohaline1}
{D_{\rm{th}}} = {C_{\rm{th}}} \frac{\lambda}{\rho c_P} \frac{{ - {\nabla _\mu }}}{{{\nabla _{\rm{ad}}} - {\nabla }}}\ \ \rm{for} {\ } {\nabla _\mu }<0,
\end{eqnarray}%
where ${C_{\rm{th}}} = 8{(\pi \alpha_{\rm{th}} )^2}/3$ and $\alpha_{\rm{th}}$ is the aspect ratio of length to width of the fingers. By using the formula above, a value of ${C_{\rm{th}}} \approx 1000$ is required to explain abundance observations of RGB stars \citep{CZ07,CanL10,CL10}. We have tried to calculate a solar model with the same parameters of Model TWA and including thermohaline mixing with the diffusion coefficient defined by the above equation, setting ${C_{\rm{th}}} = 1000$. However, the resulting solar model does not show similar improvements as does Model TWA.

\begin{figure}
\centering
\includegraphics[scale=0.5]{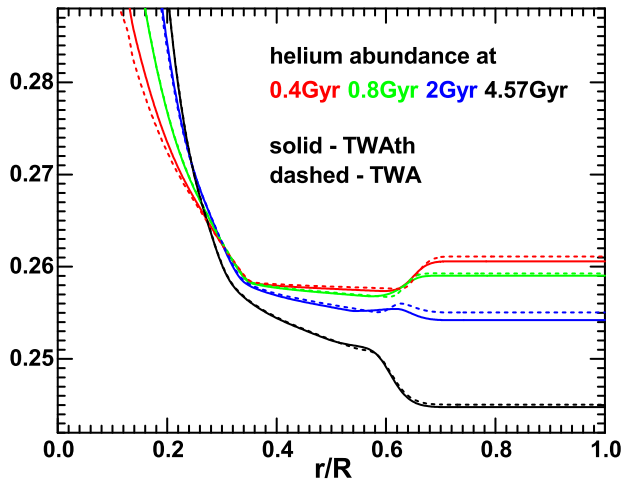}
	\caption{Evolution of helium abundance of Model TWAth, including thermohaline mixing, compared with Model TWA. }\label{TWAthYevol}
\end{figure}

On the other hand, numerical simulations have shown much lower diffusion coefficient for thermohaline mixing \citep{Den10,Trax11,BGS13}. \citet{Trax11} suggested a formula for the diffusion coefficient for thermohaline mixing based on their simulations (see Eq.~(24) in their paper). At the conditions of the base of the solar convection zone, it shows a typical value of diffusion coefficient for thermohaline mixing ${D_{\rm{th}}} \sim 100 \, {\rm cm^2 s^{-1}}$. We have calculated a solar model, TWAth, with the same parameters as Model TWA and a diffusion coefficient of thermohaline mixing ${D_{\rm{th}}} = 100\, {\rm cm^2s^{-1}}$. The evolution of the helium-abundance profile is shown in Fig.~\ref{TWAthYevol}. Thermohaline mixing slightly extended the $\nabla_{\mu}<0$ region and makes that be milder but it cannot completely erase the $\nabla_{\mu}<0$ region in the timescale of $\sim 1$Gyr. The seismic properties of Model TWAth are quite similar to Model TWA: the differences of the sound-speed profile are less than $0.03\%$, the differences of the density profile are less than $0.1\%$, $Y_{\rm{S}}=0.2448$ and $R_{\rm{bc}}=0.7109 R_\odot$ for Model TWAth. The lithium abundance of Model TWAth is ${\rm{A(Li)}}=0.76$, a little lower than Model TWA due to the thermohaline mixing enhancing the Li depletion in the early MS stage. $^7$Be and $^8$B neutrino fluxes of Model TWAth are $4.84 \times 10^9$ and $5.12 \times 10^6 \, {\rm cm^{-2} s^{-1}}$, respectively, a little less than Model TWA possibly due to the thermohaline mixing slightly offsetting the heavy-element settling and leading to a lower $Z_{\rm{C}}=0.01570$. It can be found that taking into account the diffusion coefficient of thermohaline mixing based on numerical simulations should not change the main results of the improved solar models.

\section{Summary and conclusions} \label{Secconc}

In this paper, we have focused on the ``solar abundance problem" that solar models with revised low-$Z$ composition are not consistent with helioseismic inferences. We have proposed to take into account three extra physical processes missed in SSM (i.e., convective overshoot which leads to turbulent kinetic energy flux and mixing, the helium-poor mass loss caused by the solar wind, and an inhomogeneous PMS accretion from the protoplanetary disk). Convective overshoot is predicted by many stellar convection models and is shown by many numerical simulations of stellar convection. The helium-poor property of the solar-wind mass loss is confirmed by observations. And PMS accretion is a common property of low-mass stars revealed by abundant observations of T-Tauri stars. Therefore those extra processes are supported by theory or observations. A possible physical justification for the crucial assumption of the material of the PMS accretion being inhomogeneous is that the magnetospherical accretion could be ion-dominated; thus the effect of the different first ionization potential of each element could lead to a different between its ion abundance and the neutral. Since \citet{ser11} have tested PMS accretion with variations in metallicity and have not found an overall satisfactory solar model, they have also suggested to investigate PMS accretion with variations in helium abundance.

With the recent upward revised Ne abundance \citep{Young18}, the standard solar model shows a little improvements comparing with the original AGSS09 SSM, but the ``solar abundance problem" remains. An inherent reason of the ``solar abundance problem" is the contradiction of the structure of solar convective envelope revealed in \citet{TKFZ14}. In order to eliminate the contradiction, we have taken into account the convective overshoot in solar model. The overshoot leads to negative turbulent kinetic energy, which result in a deeper $R_{\rm{bc}}$ so that it eliminates the contradiction, and overshoot mixing, which leads to significant lithium depletion. The main parameters in the overshoot model are derived from helioseimic inferences and the required solar lithium depletion. Although the resulting solar model shows good properties of the convective envelope (i.e., $R_{\rm{bc}}$ and $Y_{\rm{S}}$), the sound-speed profile in solar interior with $r<0.6R$ is worse than that of the SSM, indicating that the helioseismically based overshoot model is not sufficient to solve the ``solar abundance problem". Extra mechanisms must be taken into account to improve the sound-speed profile in solar interior.

We have then calculated solar models with the helioseismically based overshoot model and varied parameters of accretion and mass loss, i.e., duration of accretion, abundances of accreted material, and abundances of the solar wind. We found that significantly and overall improved solar models exist when the PMS accretion is helium-poor, such as Model TWA which is a typical improved solar model. Their sound-speed profiles, density profiles, surface helium abundances $Y_{\rm{S}}$ and $(Z/X)_{\rm{S}}$, lithium abundances, and neutrino fluxes are all in good agreements with helioseismic inferences or observations. Since some properties of the convective envelope (e.g., $R_{\rm{bc}}$ and lithium abundances) of the investigated models can be improved by the helioseismically based overshoot model, the main property of concern of the models is their sound-speed deviations in the solar radiative interior. The key to make the sound-speed profiles of the improved models consistent with helioseimic inferences is the evolution of the helium-abundance profile. For example, the helium abundance in the TWA solar model, which results from the helium-poor PMS accretion, is lower than that of the SSM in the solar mantle, thus leading to higher sound speed due to its lower $\mu$. The helium-poor mass loss (i.e., the solar wind) concentrates helium in the convective envelope; thus $Y_{\rm{S}}$ in the TWA models is higher than $Y_{\rm{S}}$ in the SSM.

The best set of parameters of accretion and mass loss cannot be determined because there are many solar models with different parameters (e.g., models in Tables \ref{Type2Best}, \ref{TWA_eta_comp} and \ref{TWA_z_comp}) showing similar improvements as Model TWA. A common property of those models is that they have a helium-poor PMS accretion with \textsl{lacking helium masses} (cf. Eq.~\ref{MYlack}) about $(1\% \sim 2\%)M_{\odot}$, indicating an inhomogeneous ZAMS solar interior in which the helium abundance in the envelope is lower than that in the core. The necessity of the helium-poor PMS accretion is indicated in the test that the sound-speed profile cannot be significantly improved by only use the helioseismic based overshoot model and varied abundances of both $Y$ and $Z$ in solar wind.

The analysis on improving the sound-speed profile shows that their are only two ways to improve: enhance the opacity or reduce the helium abundance in the solar mantle. Different from the \textsl{ad hoc} enhancements of opacity, neon abundance or molecular diffusion, which correspond to the former, this study shows that the latter is also a possible solution to the ``solar abundance problem".

The comparison of the solar Li abundance with the Li abundances of open cluster stars has indicated that there is a mixing process (very likely caused by convective overshoot) in the thin layer $0.68R_\odot<r<R_{\rm{bc}}$ and the characteristic timescale of the mixing in that region is about $\sim1\,$Gyr. Since the timescale is shorter than the present solar age, the layer should be well mixed so that the solar envelope with $r>0.68R_\odot$ should be almost homogeneous. The $\sim1\,$Gyr timescale of the mixing indicates that the overshoot mixing is a weak mixing process, which is consistent with \citet{OVMZ13}. The solar Be abundance does not show an obvious depletion, indicating that the mixing cannot extend to $r=0.6R_\odot$. In this paper, the overshoot mixing is the mechanism to deplete lithium and the parameter $C_{\rm{X}}$ of the overshoot mixing is based on the required lithium depletion. If other mixing mechanisms are taken into account (e.g., rotational mixing \citep{bi11,yang16,yang19}, internal wave mixing \citep{Arnett10}), the value of $C_{\rm{X}}$ should be downward revised. However, any mixing mechanism below the BCZ should be restricted by the above results indicated by the solar Li and Be abundances. Therefore they should lead to a similar strength of the adopted overshoot mixing and should not change the main results on the improved solar models.

\acknowledgments

We thank the anonymous referee for providing comments which improved the original version. Fruitful discussions with Prof. J.~H. Guo and W.~M. Yang are highly appreciated. We are grateful to Maria Pia Di Mauro for providing the helioseismic inversion code used in this project. This work is co-sponsored by the National Natural Science Foundation of China through grant No. 11303087, 11773064, 11333006 and 11521303, and the foundation of Chinese Academy of Sciences (Light of West China Program and Youth Innovation Promotion Association). Funding for the Stellar Astrophysics Centre is provided by The Danish National Research Foundation (Grant DNRF106). This research was partially conducted during the Exostar19 program at the Kavli Institute for Theoretical Physics at UC Santa Barbara, which was supported in part by the National Science Foundation under Grant No. NSF PHY-1748958.


\end{document}